\documentclass[useAMS,usenatbib]{mn2e}

\usepackage{graphicx}
\usepackage{epsfig}
\usepackage{amssymb,amsmath}
\usepackage{multirow}
\usepackage{mathrsfs}
\usepackage{times}
\usepackage{setspace}
\usepackage{color}
\usepackage{amssymb}

\usepackage[colorlinks, breaklinks, citecolor=blue]{hyperref}

\newcommand{\rhomean}{\rho_{\rm mean}}
\newcommand{\varpiv}{\boldsymbol{\varpi}}

\defcitealias{thompson16a}{TK16}

\title[Observable Properties of Cool Winds]{The Observable Properties of Cool Winds from Galaxies, AGN, and Star Clusters. I. Theoretical Framework}

\author[Krumholz et al.]{Mark R. Krumholz$^1$\thanks{mark.krumholz@anu.edu.au},
Todd A. Thompson$^2$, Eve C. Ostriker$^3$, and Crystal L. Martin$^4$
\\ \\
$^1$ Research School of Astronomy \& Astrophysics, Australian National University, Canberra, ACT, Australia\\
$^2$ Department of Astronomy, The Ohio State University, Columbus, OH, USA\\
$^3$ Department of Astrophysical Sciences, Princeton University, Princeton, NJ, USA\\
$^4$ Department of Physics, University of California, Santa Barbara, CA, USA
}

\begin{document}
\maketitle
\label{firstpage}
\begin{abstract}
Winds arising from galaxies, star clusters, and active galactic nuclei are crucial players in star and galaxy formation, but it has proven remarkably difficult to use observations of them to determine physical properties of interest, particularly mass fluxes. Much of the difficulty stems from a lack of a theory that links a physically-realistic model for winds' density, velocity, and covering factors to calculations of light emission and absorption. In this paper we provide such a model. We consider a wind launched from a turbulent region with a range of column densities, derive the differential acceleration of gas as a function of column density, and use this result to compute winds' absorption profiles, emission profiles, and emission intensity maps in both optically thin and optically thick species. The model is sufficiently simple that all required computations can be done analytically up to straightforward numerical integrals, rendering it suitable for the problem of deriving physical parameters by fitting models to observed data. We show that our model produces realistic absorption and emission profiles for some example cases, and argue that the most promising methods of deducing mass fluxes are based on combinations of absorption lines of different optical depths, or on combining absorption with measurements of molecular line emission. In the second paper in this series, we expand on these ideas by introducing a set of observational diagnostics that are significantly more robust that those commonly in use, and that can be used to obtain improved estimates of wind properties.
\end{abstract}


\section{Introduction}

Outflows of gas from galaxies, active galactic nuclei (AGN), and embedded star clusters are ubiquitous phenomena in astrophysics. Wherever gas flows converge to form stars or accrete onto a black hole, the subsequent release of energy appears to expel a portion of the inflowing gas in a high speed wind. On cosmological scales, outflows from galaxies are likely responsible for explaining both the relatively small baryon fractions of dark matter halos and the ubiquity of metals in the intergalactic medium \citep[and references therein]{veilleux05a}. On the scales of AGN, winds are a major candidate for explaining the observed correlation between black hole mass and the properties of galactic bulges \citep[e.g.,][]{king03a}. On the scales of stars, outflows driven by a variety of feedback mechanisms are the most likely explanation for why the typical outcome of the star formation process is an unbound association rather than a bound cluster \citep[and references therein]{krumholz14c, krumholz14e}.

Despite their ubiquity and importance, however, outflows are very difficult to measure. Some outflowing gas can be distinguished by its high temperatures and thus its emission at X-ray wavelengths, and outflows of hot gas are indeed observed wherever outflows are present. However, this hot component is generally thought to carry only a small fraction of the outflow mass flux, with the bulk in a cooler component\footnote{In this paper, ``cool" refers to any gas at a temperature $\lesssim 10^5$ K, including both the traditional ``warm" ($T\sim 10^3-10^4$ K) and ``cold" ($T\sim 10-100$ K) components of the interstellar medium.} that is either entrained by the hot gas or driven out by some other mechanism. This cooler component is very difficult to separate from the usually much brighter emission of the region responsible for launching the outflow, particularly because, while the cool component is much denser than the hot, X-ray emitting portion of the wind, it is generally much more diffuse than the gas found in the launching region. Observations therefore usually rely on spectroscopy in either absorption \citep[e.g.,][]{heckman00a, martin05a, rupke05a, rupke05b, steidel10a, werk14a} or emission \citep[e.g.,][]{alatalo11a, genzel11a, janssen16a}, which allows the outflowing component to be separated based on its velocity, or by observing the tenuous off-plane emission of the flow \citep[e.g.,][]{leroy15b}.

Each of these techniques has serious limitations. Absorption spectroscopy using background sources is limited by the availability of such sources, and thus is usually available for at most a handful of lines of sight. ``Down the barrel" absorption measurements, which use the galaxy itself as a backlight, must contend with the lower signal to noise that having a weaker backlight implies, along with the unknown distance between the absorbing systems and the launching galaxy, which influences the conclusions that one draws about the mass outflow rate. Emission spectroscopy is limited by confusion with the wind launching region, which masks any low-velocity component of the outflow behind the much brighter emission of non-wind material moving at similar velocities. This masking is not problematic if it is assumed that such low-velocity material is doomed to fall back rather than enter the wind, but there is little physical reason to make this assumption, and, as we will show in this paper, it is often unjustified. Off-plane emission observations are possible only for the most nearby systems at close to edge-on orientation, where our resolution is high enough to separate the plane and off-plane regions, and the off-plane emission is bright enough to be seen. Moreover, in such configurations the velocity of the outflowing material is poorly constrained. Due to these limitations, each type of measurement is very difficult to interpret, and quantitative estimates of quantities such as mass outflow rates usually derived from simple heuristic arguments.

The goal of this work is to significantly improve this situation by introducing a simple analytic model that self-consistently couples the launching and kinematics of the cool component of the wind with a calculation of its observable absorption and emission of light. Our approach here differs from the more common tactic of conducting a full numerical simulation of a wind and then post-processing the results to produce synthetic emission or absorption data \citep[e.g.,][]{fujita09a, shen12a, stinson12a, hummels13a, suresh15a}. While such an approach is more accurate than any purely analytic model can hope for, because simulations sample only a tiny part of parameter space they cannot easily be used to extract physical quantities from a given set of observed data. Moreover, because one cannot explore parameter space with simulations, it is hard to draw general conclusions from them.

Given these limitations of the fully numerical approach, analytic models that allow rapid and efficient computation are clearly required. However, the few analytic models in the literature are extremely simple, and tend to adopt prescriptions for quantities like the wind velocity, density, and covering fraction -- e.g., homologously expanding spherical shells, or flows where only a single velocity is present at any radial distance -- that are chosen more for numerical simplicity than on the basis of a detailed physical model \citep[e.g.,][]{steidel10a, prochaska11a, scarlata15a}. In contrast, here we develop a model for computing the observable emission and absorption properties of winds that, while still idealised enough to be amenable to analytic treatment, is based on a physical model for wind launching that naturally and deterministically links the wind mass flux, density distribution, and velocity distribution, and allows for partial covering. We have implemented the software required to carry out these computations as an extension of the open source code Derive the Energetics and SPectra of Optically Thick Interstellar Clouds (\textsc{despotic}) \citep{krumholz14b}.\footnote{\textsc{despotic} is available from \url{https://bitbucket.org/krumholz/despotic/}. The scripts that use \textsc{despotic} to generate all the figures in this paper are available from \url{https://bitbucket.org/krumholz/despotic_winds/}.}.

This paper is the first in a series. Here we develop our analytic theory for the cool components of the wind, and illustrate the power of our model with some examples. We draw some general conclusions about what physical properties can and cannot be determined robustly from different types of observations, but do not tackle the full problem of constraining physical quantities from a set of observables. This inversion problem forms the basis for the second paper in this series.

In the remainder of this paper, we first present our physical model for the properties of winds in \autoref{sec:phys}. We calculate how such winds absorb (\autoref{sec:absorb}) light, and how they emit it in two limiting cases (\autoref{sec:emit_LTE} and \autoref{sec:emit_subcrit}). In \autoref{sec:example} we bring this formalism together to compute the observable properties of starburst galaxy whose properties are inspired by those of M82. In \autoref{sec:discussion} we discuss what we have learned from this exercise, and we summarise and conclude in \autoref{sec:conclusion}.

\section{Physical Model}
\label{sec:phys}

In order to compute the observable emission and absorption properties of a cool wind, we must begin from a model for the physical properties of that wind -- density, velocity, filling factor, etc. For this purpose we will adopt the \citet{thompson16a} (hereafter \citetalias{thompson16a}) model for wind launching, and extending it following an approach that combines elements of this model with elements of the models presented by \citet{thompson15a} and \citet{krumholz17a}. In all our models we will assume that the properties of the wind are a function only of the spherical radius $r$ (though the wind need not cover all possible directions from the source -- see below), and that the wind is in steady state.

\subsection{Summary of the \citetalias{thompson16a} Model}

We begin with a brief conceptual overview of the wind model proposed in \citetalias{thompson16a}, in order set the stage for our calculation here. This model treats the gas from which the wind is launched as an isothermal, turbulent medium with a lognormal probability distribution function (PDF) for volume and column densities. The medium is characterised by a mean surface density $\overline{\Sigma}_0$ at scale $r_0$, and is confined by a gravitational potential. Gravity is opposed by some form of feedback, the operation of which is described as exerting a constant force per unit solid angle, or equivalently injecting momentum into the gas, in a direction opposite that of gravity. The rate of momentum injection is $\dot{p}$, and ratio of the feedback force to the gravitational force for the regions at the mean surface density $\overline{\Sigma}_0$ is $\Gamma$, the generalised Eddington ratio. 

This characterisation of the feedback as exerting a constant force per unit solid angle is reasonable for many possible feedback mechanisms, including the pressure of direct starlight and the ram pressure exerted by a hot flow past cold clouds (assuming, in both cases, a basic spherical geometry). The primary requirement for this model to be applicable is that the material being ejected not trap the ``working fluid" (hot gas or radiation) for the feedback so effectively that it converts to an energy driven flow (e.g., trapped infrared photons or confined hot gas). We state this requirement more precisely below.

The central argument in \citetalias{thompson16a} is that, if the Eddington ratio $\Gamma < 1$, then the feedback mechanism cannot eject the bulk of the material in a dynamical time. However, there is a lognormal distribution of gas surface densities, and the force of gravity per unit solid angle is proportional to the local gas surface density $\Sigma$. This means that, in sufficiently under-dense regions gravity may exert less force per unit solid angle than the feedback mechanism, and as a result material will be accelerated outward into a wind. To be precise, gas is accelerated outward if its logarithmic column density $x = \ln \Sigma/\overline{\Sigma}_0$ satisfies $x < x_{\rm crit} = \ln \Gamma$. Thus even in a gas where feedback cannot expel the bulk of the material, it can still produce a wind by expelling the lower tail of the gas column density PDF with $x < x_{\rm crit}$. Since turbulence in the still-confined bulk of the gas will re-fill the tail on a dynamical timescale, the result will be continual loss of mass. The bulk of \citetalias{thompson16a} is concerned with calculating the wind mass flux produced by this process. Our goal in the remainder of this section is to compute the kinematics and structure of the resulting wind.

Before proceeding, we note that several aspects of the \citetalias{thompson16a} model have been verified in the numerical radiation hydrodynamic simulations of \citet[and 2017, submitted]{raskutti16a}.  In particular, these simulations, which focus on the effects of radiation pressure forces from forming star clusters on the surrounding gas in GMCs, demonstrate that the gas has a lognormal distribution, and only structures with sufficiently low surface density are ejected. 

\subsection{Wind Acceleration Laws}
\label{ssec:velocity_structure}

\subsubsection{Ideal Momentum-Driven Winds}
\label{sssec:ideal_winds}

The first step in our calculation is to determine the velocity structure of the wind material. Consider a region where the total mass interior to some radius $r$ is $M_r$, and there is some constant, isotropic momentum injection rate $\dot{p}$ at the origin, which is deposited in the surrounding material and accelerates it outward. The assumption of constant momentum deposition per unit solid angle is a simplification that we shall drop in subsequent sections, but which we adopt now for illustrative purposes. We also adopt the simplifying assumption that on each line of sight all of the material is collected into a single structure that intercepts and absorbs the injected momentum. The equation of motion for a thin shell accelerating outward as part of a time-steady wind, or a segment thereof (generically a ``cloud"), with column density $\Sigma_c$ when viewed along a ray from the origin is
\begin{equation}
\label{eq:eom}
\frac{dv_r}{dt} = v_r \frac{dv_r}{dr} = -\frac{GM_r}{r^2} + \frac{\dot{p}}{{4\pi r^2}} \frac{1}{\Sigma_c},
\end{equation}
where $v_r$ is the radial velocity. In writing this equation, we assume that the wind is quasi-ballistic, in the sense that individual ``clouds" accelerate under the forces applied to them without regard to the behaviour of other clouds.

Next we non-dimensionalise the system. Let $r_0$ be the radius from which the wind is launched, i.e., we take $v_r=0$ at $r=r_0$. We define a dimensionless position $a$ by $a=r/r_0$ and a dimensionless velocity $u$ by $u_r = v_r/v_0$, where $v_0 = \sqrt{2 G M_0/r_0}$; here $M_0$ is $M_r$ evaluated at $r_0$, and thus $v_0$ is just the escape velocity from radius $r_0$ in the case of a point mass potential. Similarly, we write the mass interior to radius $r$ as $M_r = m M_0$, where $m$ is a dimensionless function of $a$ that obeys $m=1$ at $a=1$. An isothermal mass distribution would have $m=a$, while a point mass distribution has $m=1$.

Let the gas at $r_0$ have a mean column density $\overline{\Sigma}_0$ when viewed along a radial ray outward from the origin, and let $\Sigma_{0,c}$ be the local column density of the cloud we are considering at $r=r_0$. Following the approach of \citetalias{thompson16a}, we define $x = \ln (\Sigma_{0,c} /\overline{\Sigma}_0)$ as the logarithmic under- or over-density of a given cloud or segment of a shell.

As the shell segment or cloud with which we are concerned moves outward, it may change its cross-sectional area to the driving mechanism. For pressureless dust, we expect the material to move purely radially, and thus maintain a constant solid angle as seen from the centre of the outflow, $A_c\propto r^2$, where $A_c$ is the cloud area. For a cloud that is pressure-confined, and has time as it moves outward for pressure forces to act, the expansion will be dictated by pressure balance: for a quasi-spherical cloud, the area scales with density as $A_c\propto \rho^{-2/3}$, and the density scales with external pressure as $\rho \propto P^{1/\gamma_{\rm cl}}$, where $\gamma_{\rm cl}$ is the cloud's adiabatic index. If the mechanism driving the cool gas is a hot wind with roughly constant velocity, the density of the hot gas drops as $\rho_{\rm w} \propto r^{-2}$, and its pressure therefore drops as $P\propto r^{-2\gamma_{\rm w}}$, where $\gamma_{\rm w}$ is the adiabatic index of the hot wind. Combining these scalings, the cloud's area varies as $A_c\propto r^{4\gamma_{\rm w}/3\gamma_{\rm cl}}$. For an adiabatic wind with $\gamma_{\rm w} = 5/3$ and a cool cloud whose pressure is predominantly turbulent, giving $\gamma_{\rm cl} \approx 5/3$ in many circumstances (\citealt{robertson12a}; Birnboim, Federrath, \& Krumholz 2017, in preparation), this produces $A_c\propto r^{4/3}$. If the cool gas pressure is predominantly thermal and the gas is kept isothermal (e.g., by radiation), then $\gamma_{\rm cl}\approx 1$, and $A_c\propto r^{20/9}$, though in this case the expansion would likely by limited by the fact that the cloud cannot expand faster than its sound speed. Finally, numerical simulations of magnetised clouds swept up by hot winds by \citet{mccourt15a} suggest that magnetic fields significantly reduce the rate of cloud expansion, though the amount by which they do so is difficult to determine from \citeauthor{mccourt15a}'s numerical experiments, which have uniform pressure and a wind that is idealised as plane parallel rather than spherically divergent. In summary, depending on the nature of the cool material and the driving mechanism, a wide range of behaviours for $A_c(r)$ are possible.

For this reason, we choose to parameterise our ignorance by introducing a function $y$ such that the surface density $\Sigma_c = \Sigma_{0,c}/y$, where $y$ is a function of $a$ that has the property $y=1$ at $a=1$; the constant area case corresponds to $y=1$, while the constant solid angle case is $y=a^2$. We refer to $y$ as the expansion function. Intuitively, it is simply the cross sectional area of the cloud versus radius, normalised by the area the cloud starts with at radius $r_0$. Below we will consider both the limiting constant area and constant solid angle cases, as well as an intermediate one, $y = a$.

Given these definitions, with some algebra \autoref{eq:eom} may be rewritten as
\begin{equation}
\label{eq:eom_nondim}
2 u_r \frac{du_r}{da} = \frac{1}{a^2}\left(y\Gamma e^{-x} - m\right),
\end{equation}
where
\begin{equation}
\label{eq:Gamma}
\Gamma = \frac{\dot{p}}{4\pi G M_0 \overline{\Sigma}_0}
\end{equation}
is the generalised Eddington ratio at radius $r_0$. The boundary condition is $u_r = 0$ at $a=1$, where the wind is launched. Only material with column density $x < x_{\rm crit} = \ln \Gamma$ has an outward acceleration at $a=1$ and becomes part of the wind. In the bulk of this paper we will assume that all such material is part of the wind, but we note that for many wind driving mechanisms and expansion geometries the velocity of the wind approaches 0 as $x\rightarrow x_{\rm crit}$. We therefore consider in \autoref{app:fcrit} how the conclusions of our paper would change if we assumed that only material with starting logarithmic surface density smaller than $x_{\rm crit}$ by some finite amount were launched into the wind. We refer to a solution to \autoref{eq:eom_nondim} as a ballistic wind acceleration law, and formally write it out as $U_a(x)$, i.e., $U_a(x)$ is the function that maps between initial surface density $x$ of a cloud and its radial velocity $u_r$ at a particular radius $a$.

\begin{table*}
\caption{
\label{tab:eom_solutions}
Wind acceleration laws $U_a(x)$, obtained by solving \autoref{eq:eom_nondim} for various potentials and cloud expansion behaviours. Note that what is reported in the table is $U_a^2(x)$ rather than $U_a(x)$.}
\begin{tabular}{l||cc}
\hline\hline
& \multicolumn{2}{c}{Mass Distribution} \\ 
Cloud expansion & Point, $m=1$ & Isothermal, $m=a$ \\
\hline\hline
Fixed area, $y=1$ & $U_a^2(x) = \left(\Gamma e^{-x}-1\right)\left(\frac{a-1}{a}\right)$ &
$U_a^2(x) = \Gamma e^{-x} \left(\frac{a-1}{a}\right) - \ln a$ \\
Intermediate, $y=a$ & $U_a^2(x) = \Gamma e^{-x} \ln a - \frac{a-1}{a}$ & $U_a^2(x) = \left(\Gamma e^{-x} - 1\right) \ln a$ \\
Fixed solid angle, $y=a^2$ & $U_a^2(x) = \left(\Gamma e^{-x}-\frac{1}{a}\right)\left(a-1\right)$ &
$U_a^2(x) = \Gamma e^{-x}\left(a-1\right)-\ln a$
\\
\hline
\end{tabular}
\end{table*}

\begin{figure}
\includegraphics[width=\columnwidth]{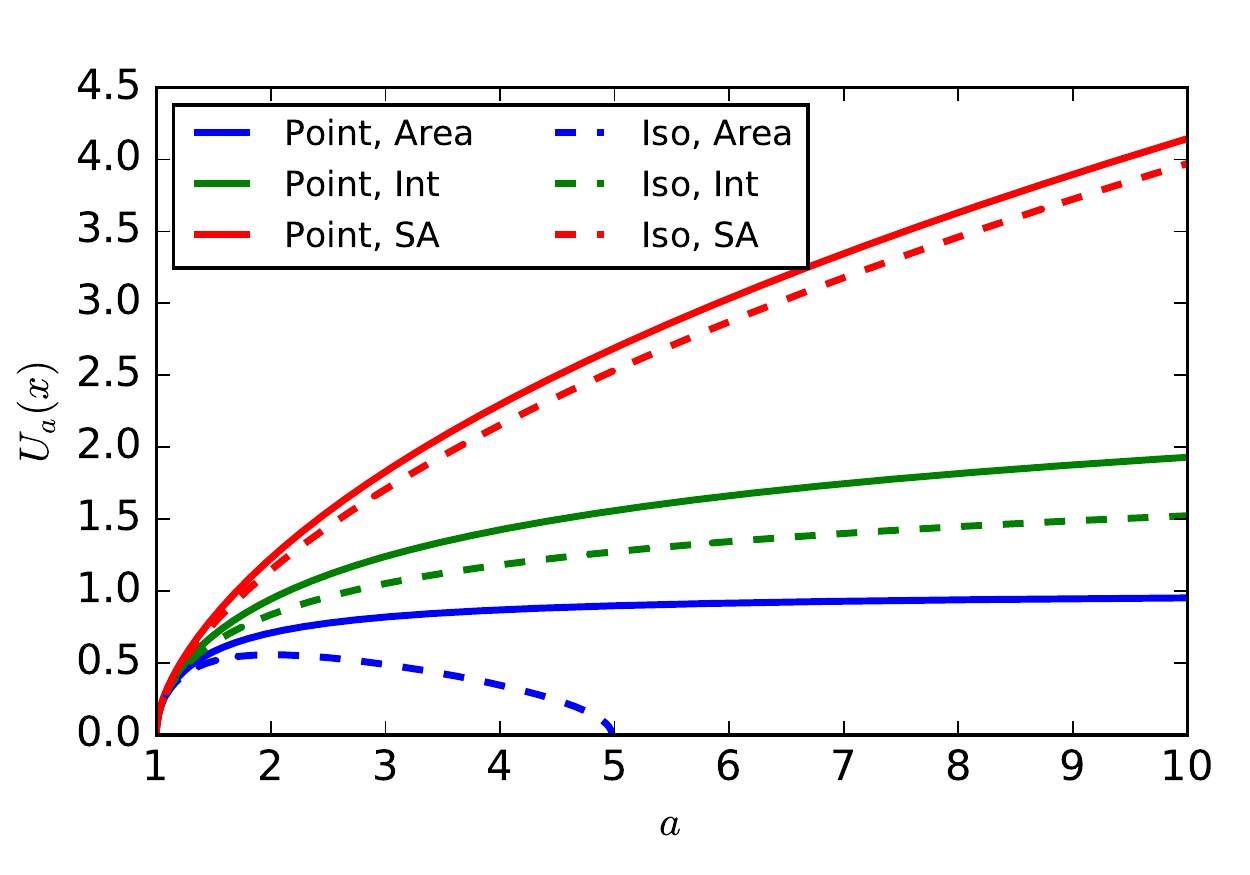}
\caption{
\label{fig:acc_law_ideal}
Example wind acceleration laws $U_a(x)$, plotted for gas with starting surface density $x=-3$ and winds with $\Gamma = 0.1$ and a variety of expansion behaviour $y$ and potentials $m$. The expansion behaviours shown are constant area $A_c$ ($y=1$, blue), constant solid angle $\Omega$ ($y=a^2$, red), and intermediate between the two ($y=a$, green), and the potentials shown are point ($m=1$, solid) and isothermal ($m=a$, dashed).
}
\end{figure}

It is instructive to solve \autoref{eq:eom_nondim} in a few representative cases. In \autoref{tab:eom_solutions} we give solutions for expansion functions $y=1$ (constant area), $y=a$ (intermediate), and $y=a^2$ (constant solid angle), and potentials $m=1$ (point mass) and $m=a$ (isothermal). We plot some examples in \autoref{fig:acc_law_ideal}. We can classify possible solutions as wind solutions, which have the property that $U_a(x)$ is real and non-zero for all combinations of $x < x_{\rm crit}$ and $a>1$, and fountain solutions, which do not have that property. For wind solutions a finite mass flux reaches infinity, while for fountain solutions no mass reaches infinity. Wind solutions arise for combinations of potential and lateral expansion rate such that $y$ increases with $a$ faster than $m$ does, and fountains arise when the reverse is true. For winds, the velocity approaches asymptotically constant values as $a\rightarrow \infty$ if $y$ increases with $a$  linearly or more slowly, and diverges to arbitrarily high velocities if $y$ increases with $a$ superlinearly. This divergence is not physically realistic, since $\dot{p}$ cannot remain constant as $u_r \rightarrow \infty$, and does not occur for the more realistic wind models we shall consider next.

\subsubsection{Radiatively-Driven Winds}

\begin{table*}
\caption{
\label{tab:eom_solutions_rad}
Same as \autoref{tab:eom_solutions}, but for a radiation pressure driven wind where the momentum input rate depends on the optical depth $\tau_0$ (\autoref{eq:eom_rad}). In the formulae below, $\tau\equiv \tau_0 e^x$, and $\mathrm{Ei}(x)$ is the exponential integral of $x$. We also give the maximum velocity $u_{\rm max}$ for each case; this maximum depends only on $m$ and not on $y$, so a single value is given for each column}
\begin{tabular}{l||cc}
\hline\hline
Cloud Expansion & \multicolumn{2}{c}{Mass Distribution} \\ 
& Point, $m=1$ & Isothermal, $m=a$ \\
\hline\hline
Fixed area, $y=1$ &
$\begin{array}{lcl}
U_a^2(x) & = & \left[\Gamma e^{-x}\left(1-e^{-\tau}\right)-1\right]\left(\frac{a-1}{a}\right)
\end{array}$ &
$\begin{array}{lcl}
U_a^2(x) = \Gamma e^{-x} \left(1-e^{-\tau}\right) \left(\frac{a-1}{a}\right) - \ln a
\end{array}
$ \\[1ex]
%%%%
Intermediate, $y=a$ &
$\begin{array}{lcl}
U_a^2(x) & = & \Gamma e^{-x} \left[\mathrm{Ei}\left(-\frac{\tau}{a}\right) - \mathrm{Ei}(-\tau) + \ln a\right] \nonumber \\
& & \, {} - \frac{a-1}{a}
\end{array}$
&
$\begin{array}{lcl}
U_a^2(x) & = & \Gamma e^{-x} \left[\mathrm{Ei}\left(-\frac{\tau}{a}\right) - \mathrm{Ei}(-\tau) + \ln a\right] \nonumber \\
& & \, {} - \ln a
\end{array}$
\\[1ex]
%%%%
Fixed solid angle, $y=a^2$ & 
$\begin{array}{lcl} U_a^2(x) & = & \Gamma e^{-x}\left[a\left(1-e^{-\tau/a^2}\right)-1+e^{-\tau} + {}\right.
\\ & & \, \left.
\sqrt{\pi\tau}\left(\mbox{erf}\sqrt{\tau}-\mbox{erf}\sqrt{\tau/a^2}\right)\right]-\frac{a-1}{a}
\end{array}$ &
$\begin{array}{lcl} U_a^2(x) &= & \Gamma e^{-x}\left[a\left(1 - e^{-\tau/a^2}\right) -1+e^{-\tau} + {}\right. \\
& & \, \left.\sqrt{\pi \tau}\left(\mbox{erf}\sqrt{\tau}-\mbox{erf}\sqrt{\tau/a^2}\right)\right] - \ln a
 \end{array}$
\\
\hline 
\parbox{2.5cm}{\raggedright Maximum velocity \\(Optically thin)}
& $u^2_{\rm max} =\Gamma\tau_0 - 1$ & $u^2_{\rm max} =\Gamma\tau_0 - \ln\Gamma\tau_0 - 1$ \\
\hline
\end{tabular}
\end{table*}

The computation in \autoref{sssec:ideal_winds} is for an idealised wind with $\dot{p}$ constant, independent of cloud surface density $\Sigma_c$. Clearly this is not fully realistic. We therefore now consider more realistic driving mechanisms where $\dot{p}$ is not purely constant and derive acceleration laws $U_a(x)$ for them.

One potentially important mechanism for driving winds is the momentum of starlight photons leaving a galaxy, interacting with dust grains suspended in the gas \citep{murray05a, murray10a, thompson15a}. The grains then transmit this momentum to the gas via collisions or magnetic forces. For a direct radiation field (i.e., one that is traveling radially outward from the stars, rather than a reprocessed one produced by the dust grains themselves), the momentum carried by the radiation field is $\dot{p}=L/c$, where $L$ is the luminosity. We will assume that $L$ is constant with radius, so that a negligible portion of the radiation field is absorbed by the wind; we check this assumption explicitly in \autoref{ssec:validity}. 

The momentum per unit area applied to the gas by the radiation field is
\begin{equation}
\frac{d^2 p_{\rm gas}}{dt\, dA} = \frac{\dot{p}}{4\pi r^2} \left(1 - e^{-\kappa_F \Sigma_c}\right)
\end{equation}
where $\kappa_F$ is the flux mean opacity of the dusty gas in the shell under consideration. The corresponding non-dimensional equation of motion (c.~f.~\autoref{eq:eom_nondim}) is
\begin{equation}
\label{eq:eom_rad}
2 u_r \frac{du_r}{da} = \frac{1}{a^2} \left\{y \Gamma e^{-x} \left[1 - \exp\left(-\frac{e^x\tau_0}{y}\right)\right] - m\right\},
\end{equation}
where $\tau_0 \equiv \kappa_F \overline{\Sigma}_0$. Material is launched into a wind only at overdensities $x<x_{\rm crit}$, where $x_{\rm crit}$ is given implicitly by
\begin{equation}
\label{eq:rad_ejec_condition}
\Gamma e^{-x_{\rm crit}} \left[1-\exp\left(-\tau_0 e^{x_{\rm crit}}\right)\right] - 1 = 0.
\end{equation}
This equation has a solution for $x_{\rm crit}$ only if $\Gamma\tau_0 > 1$, and so if $\Gamma \tau_0 < 1$ there is no wind at all. This is equivalent to the requirement $\kappa_F > 4 \pi G M_0/\dot{p}$, i.e., there is a lower limit on $\kappa_F$ for wind launching that depends on the mass and momentum flux. In the case of young star clusters this limit is $\kappa_{F} = 4 \pi G c/\langle L_*/M_*\rangle$, where $\langle L_*/M_*\rangle$ is the light to mass ratio of the newborn stars \citep{murray10a, krumholz10b, skinner15a}. From \autoref{eq:rad_ejec_condition}, an upper limit on the
surface density that can be accelerated is  $\Sigma_{\rm max} =
\langle L_*/M_* \rangle/4\pi c G$. For the value of $\langle L_*/M_*\rangle$
appropriate for a fully-sampled IMF, this limit is $\approx 400$
$M_\odot$ pc$^{-2}$ \citep{raskutti16a}.

\begin{figure}
\includegraphics[width=\columnwidth]{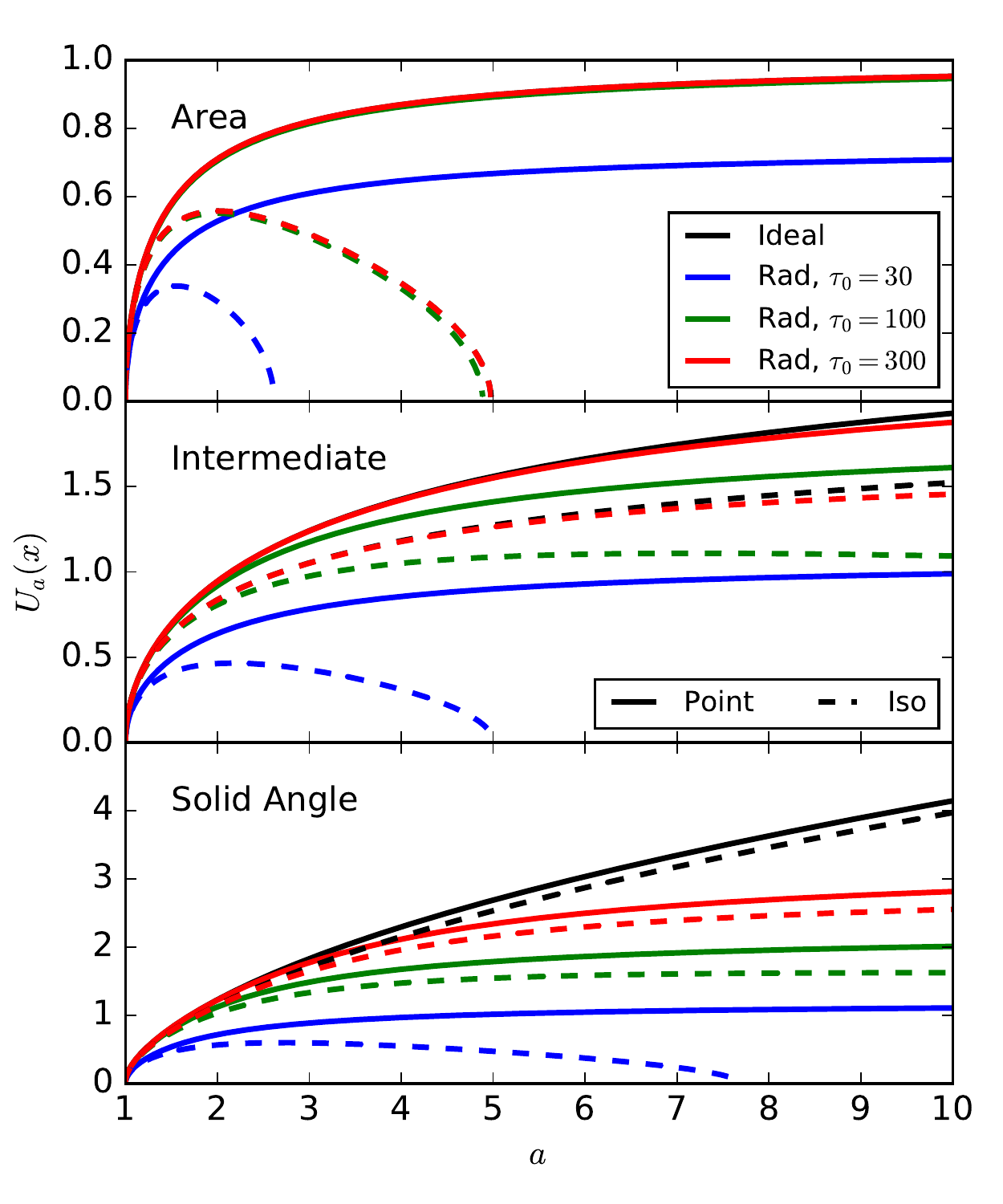}
\caption{
\label{fig:acc_law_rad}
Example wind acceleration laws $U_a(x)$, plotted for gas with starting surface density $x=-3$ and winds with $\Gamma = 0.1$. Black lines show ideal momentum-driven winds, while other colours show radiatively driven winds with a variety of $\tau_0$ values, as indicated in the legend. Solid lines are for point potentials, dashed lines for isothermal potentials. Panels show, from top to bottom, constant area ($y=1$), intermediate ($y=a$), and constant solid angle ($y=a^2$) clouds. Note that, in the top panel, the black line for ideal is completely hidden by the red line for $\tau_0=300$.
}
\end{figure}

In cases where a wind is launched, solutions for $U_a(x)$ for various
choices of $y$ and $m$ are given in
\autoref{tab:eom_solutions_rad}.\footnote{The solution for $y=a^2$,
  $m=1$ has previously been given in Raskutti, Ostriker, \& Skinner
  (2017, submitted).} We plot some examples, and compare to an
idealised wind, in \autoref{fig:acc_law_rad}. For a radiatively driven
wind, a point potential ($m=1$) yields wind solutions, with a finite
mass flux reaching infinity. An isothermal potential ($m=a$) produces
fountain solutions, with no mass reaching infinity, regardless of the
expansion factor $y$.

From the standpoint of observables, the main difference between a purely ideal momentum-driven wind and one driven by direct photon pressure is that force imparted by the radiation field drops off as the material becomes optically thin, i.e., in the case that $y$ increases with $a$. As a result, the velocity does not diverge to infinity as the surface density drops to zero. Instead, because the radiative force is proportional to the surface density for optically thin material, this exactly compensates for the factor $y$ in the first term of \autoref{eq:eom_rad}, and the radiative acceleration becomes proportional to $1/a^2$. When this gas travels outward, the velocity will reach a finite maximum $u_{\rm max}$. If the potential is point-like this maximum velocity is achieved in the asymptotic limit as $a\rightarrow \infty$, while for an isothermal potential it occurs at a finite radius $a$ (which depends on the initial surface density $x$), and the velocity declines logarithmically at larger radii. We give the value of $u_{\rm max}$ along with the acceleration law $U_a(x)$ in \autoref{tab:eom_solutions_rad}. 

\subsubsection{Winds Driven by a Volume-Filling Hot Gas}

In addition to radiation pressure, another potentially important mechanism for driving cool gas out of galaxies is the ram pressure and hydrodynamic drag associated with a hot wind of supernova heated gas. The entrainment process is complex and poorly-understood. Hydrodynamic simulations suggest that it is quite destructive, with little if any cold material surviving acceleration to significant speeds, and the effective cross-sectional area of that material that does survive being significantly reduced \citep[e.g.,][]{williamson14a, scannapieco15a, bruggen16a, schneider17a}. However, the presence of magnetic fields may allow cold material to survive acceleration \citep{mccourt15a}. It is also possible that cool material is initially destroyed near where the wind is launched, but re-forms at larger radii where it is more stable, and is subsequently accelerated outward \citep[e.g.,][]{martin15a, thompson16b}. Given our goal of producing a simple model that can be evaluated analytically and used to fit obsevations, we will not attempt to incorporate all of the complex physics captured in this mostly-numerical work, and we will instead make a number of simplifying assumptions.

Consider a volume-filling hot wind with a mass flux $\dot{M}_h$ and a velocity $v_h$, which will be of order the hot gas sound speed. The density of the hot wind is $\rho_h = \dot{M}_h/(4\pi r^2 v_h)$, and its momentum flux is $\dot{M}_h v_h$. A hot wind will not have exactly constant velocity, but since thermal pressure-driven winds accelerate only very slowly with radius (for example, an supersonic wind in an isothermal potential has $v_h \propto \sqrt{\ln r}$), we will treat $v_h$ as constant for mathematical simplicity.

\begin{figure}
\includegraphics[width=\columnwidth]{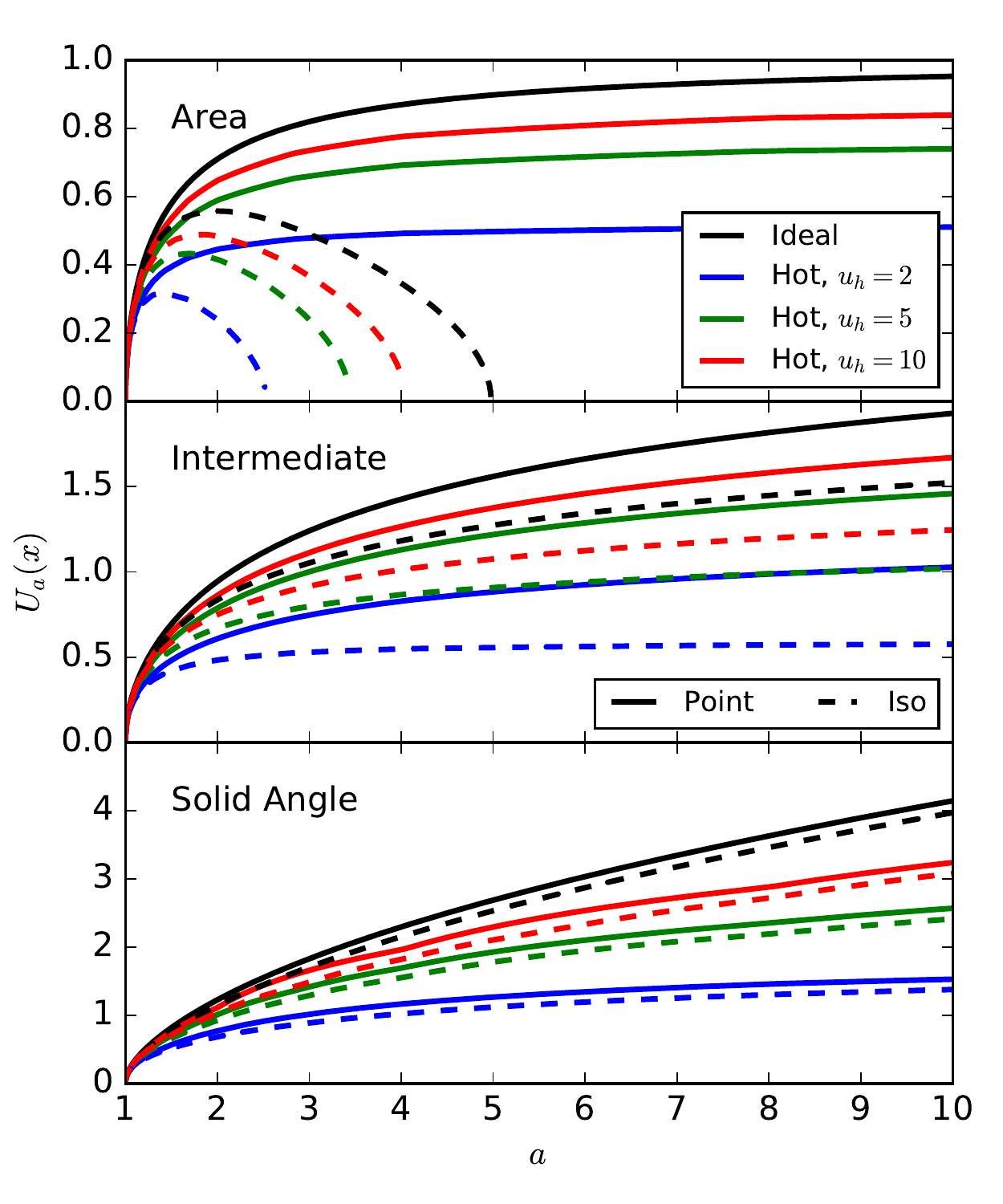}
\caption{
\label{fig:acc_law_hot}
Same as \autoref{fig:acc_law_rad}, except that now the coloured lines show hot gas-driven winds with a variety of values for $u_h$, as indicated in the legend.
}
\end{figure}

As this wind impacts and flows by cold clouds, it will deposit outward momentum into them at a rate
\begin{equation}
\frac{d^2 p_{\rm gas}}{dt\, dA} = C_t \rho_h (v_h-v_r)^2,
\end{equation}
where $v_r$ is the velocity of the cold gas and $C_t \geq 1$ is a coefficient of order unity that describes the extra momentum transferred to the cloud as a result of the turbulent wake created by the flow of hot gas around the cold cloud. Using this in our equation of motion \autoref{eq:eom}, and non-dimensionalising in analogy with \autoref{eq:eom_nondim}, gives
\begin{equation}
\label{eq:eom_hot}
2 u_r \frac{du_r}{da} = \frac{1}{a^2} \left[ \Gamma y e^{-x}\left(1 - \frac{u_r}{u_h}\right)^2 - m\right],
\end{equation}
where $\Gamma = C_t \dot{M}_h v_h/(4\pi G M_0 \overline{\Sigma}_0)$ is the momentum flux imparted by the wind normalised to the strength of gravity, and $u_h = v_h/v_0$. The critical density for wind launching is $x_{\rm crit} = \ln \Gamma$, the same as for an idealised wind. There is no closed-form analytic solution for \autoref{eq:eom_hot}, but it is trivial to evaluate numerically. \autoref{fig:acc_law_hot} shows some sample solutions for $U_a(x)$ for a hot gas-driven wind. The solutions are qualitatively similar to those for the idealised case, except that the velocity increases more slowly as $u_r\rightarrow u_h$, and the velocity eventually asymptotes to a maximum of $u_h$ rather than diverging as $x\rightarrow-\infty$ or $a\rightarrow\infty$. Thus the hot wind case is qualitatively similar to the radiation pressure case, except that the maximum radial velocity $u_{\rm max}$ is always $u_h$. As in the radiative case, the influence of a finite value of $u_h$ is largest for more rapidly-expanding clouds, and is more modest for $y=1$ or even $y=a$.

\subsection{Density Structure}\label{ssec:meandensity}

Given a wind acceleration law $U_a(x)$, our next step is to calculate
the density structure of the wind. To do so, we begin from the
\citetalias{thompson16a} ansatz that at any time the surface density of the
gas in the wind launching region follows a lognormal distribution.  We
also assume that gas at starting column densities $x$ that satisfy the
condition for wind launching ($x < x_{\rm crit}$) will be ejected on a
dynamical timescale of the wind launching region, so that
the distribution of mass flux with $x$ obeys
\begin{equation}\label{eq:massflux}
\frac{d\dot{M}}{dx} = 4 \pi r_0^2 v_0 \rho_{\rm norm}  p_M,
\end{equation}
where $r_0$ and $v_0$ are the launch radius and escape speed from this radius as defined in \autoref{sssec:ideal_winds},
\begin{equation}
p_M = \frac{1}{\sqrt{2\pi \sigma^2_x}}\exp\left[-\frac{\left(x-\sigma_x^2/2\right)^2}{2\sigma_x^2}\right]
\end{equation}
is the mass-weighted lognormal distribution,
and $\rho_{\rm norm}$ is an overall normalisation constant. Following \citetalias{thompson16a}, we take the dispersion of the lognormal to be
\begin{equation}
\sigma_x = \sqrt{\ln \left(1 + R \mathcal{M}^2/4\right)},
\end{equation}
where $\mathcal{M}$ is the Mach number of the turbulence,
\begin{equation}
R = \frac{1}{2}\left(\frac{3-\alpha}{2-\alpha}\right)\left[\frac{1-\mathcal{M}^{2(2-\alpha)}}{1-\mathcal{M}^{2(3-\alpha)}}\right],
\end{equation}
and $\alpha = 2.5$.

The material at distance $a$ is clumpy, but we can define the
contribution to the mean density from material with initial column
$x$, $d\rhomean(x)$, in terms of the mass in a shell
$d\dot M(x) \Delta t$ divided by the volume of that shell, $4 \pi r_0^2
v_0 a^2 U_a(x) \Delta t$; here $\Delta t$ is the time required for material
with initial surface density $x$ to traverse the distance from radius $a$ to $a+\Delta a$.  This yields
\begin{equation}
\label{eq:rhomean}
\frac{d\rhomean}{dx}  = \rho_{\rm norm} \frac{p_M}{a^2 U_a}.
\end{equation}
It is convenient to write the normalisation factor $ \rho_{\rm norm}$ in terms of the overall isotropic mass loss rate $\dot{M}$, evaluated immediately outside the wind launch point (i.e., before any significant portion of the gas has turned around, in fountain cases where the mass flux does not reach infinity). Doing so gives
\begin{equation}
\rho_{\rm norm} = \frac{\dot{M}}{4\pi r_0^2 v_0} \left(\frac{1}{\zeta_M}\right),
\end{equation}
where
\begin{equation}
\zeta_M = \frac{1}{2} \left[1 - \mbox{erf}\left(\frac{-2 x_{\rm crit} + \sigma_x^2}{2\sqrt{2}\sigma_x}\right)\right]
\end{equation}
is the fraction of the mass with column densities $x<x_{\rm crit}$. We pause to emphasise that the $\dot{M}$ that appears here is the mass flux assuming the wind subtends $4\pi$ sr, rather than the true mass flux, which may be smaller if the wind covers a smaller solid angle. We will consider geometries where the wind only escapes in certain directions (e.g., only through a conical region) in \autoref{ssec:geometry}.

\begin{figure}
\includegraphics[width=\columnwidth]{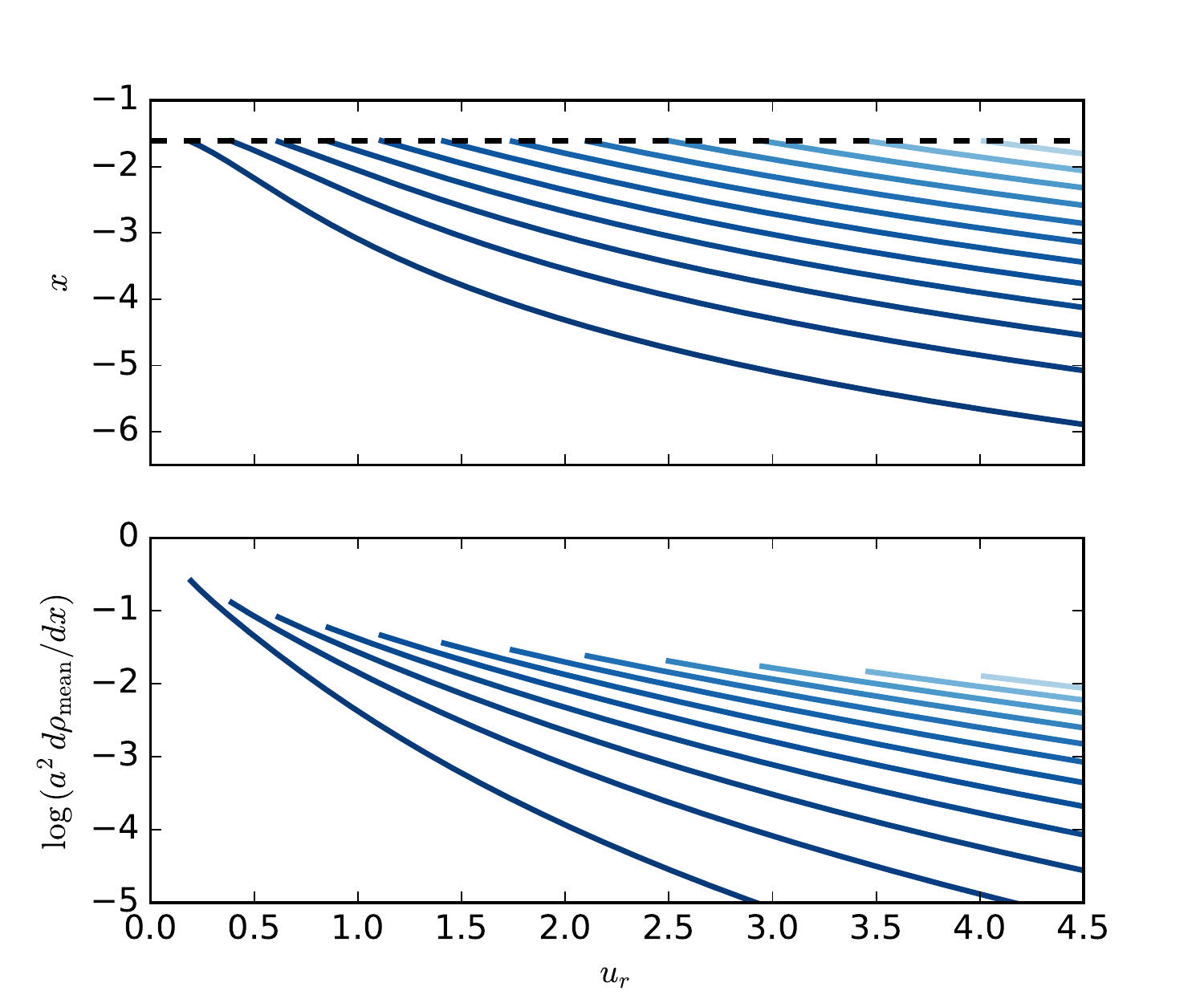}
\caption{
\label{fig:den_struct_cut}
Example calculation of the density structure of a wind; the example shown is an ideal wind with $\Gamma=0.2$, $\mathcal{M} = 50$, $y=a^2$, and $m=a$. Lines show the logarithmic surface density $x$ (top panel) and the normalised density per unit $x$ scaled by shell volume, $a^2 (d\rho_{\rm mean}/dx)$ (bottom panel) as a function of radial velocity $u_r$. The different lines correspond to different radii $a$, with the lines starting and $\log a = 0.25$ (darkest line) and increasing by $0.25$ per line. In the top panel, the dashed black line shows $x=x_{\rm crit} = \log\Gamma$.
}
\end{figure}

\begin{figure}
\includegraphics[width=\columnwidth]{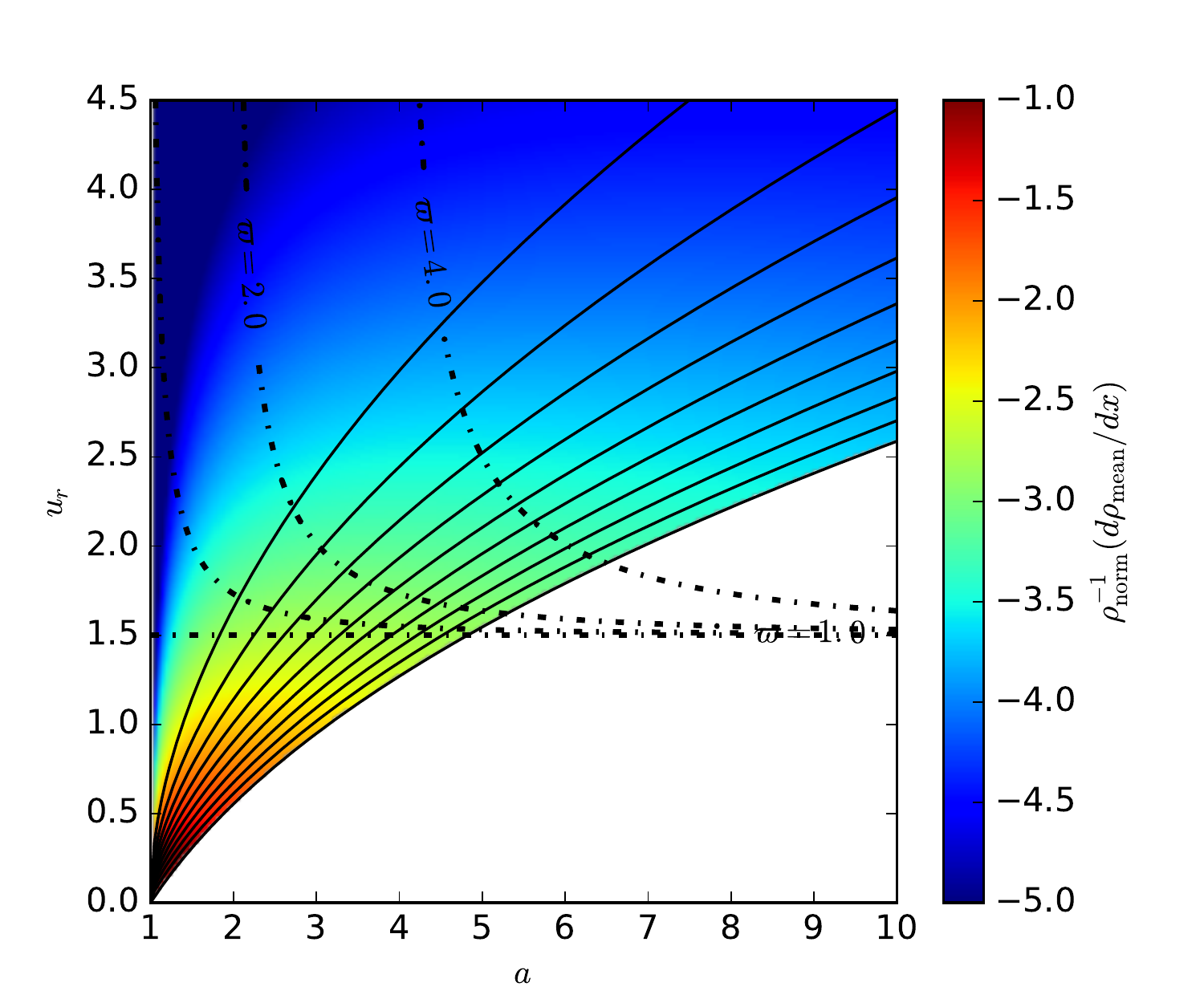}
\caption{
\label{fig:den_struct}
Mass flux and differential density $d\rho_{\rm mean}/dx$ for the same wind shown in \autoref{fig:den_struct_cut}. Colour indicates $d\rho_{\rm mean}/dx$ as a function of radius $a$ and radial velocity $u_r$. Solid lines show the acceleration law for material at a range of starting surface densities $x$, i.e., material born with a particular $x$ corresponding to one of the lines will move along that line as it flows outward. The lowest line corresponds to $x=x_{\rm crit}$, and subsequent lines are spaced so that 10\% of the wind mass flux at each radius is contributed by gas with velocities between each pair of lines. The dot-dashed lines show lines of constant line of sight velocity $u=1.5$ for observations with total impact parameter $\varpi = 0$, 1, 2, and 4, as indicated (see \autoref{ssec:geometry}).
}
\end{figure}

We note that, since $p_M$ is a function of $x$ alone, and $U_a$ maps between $x$, $a$, and $u_r$, \autoref{eq:rhomean} implicitly gives $d\rhomean/dx$ as a function of $u_r$ and $a$, something we
will exploit below. We illustrate the nature of this relation in \autoref{fig:den_struct_cut} and \autoref{fig:den_struct}.

In \autoref{fig:den_struct_cut}, the upper panel shows the relationship between radial velocity $u_r$ and initial logarithmic surface density $x$ at different radii, with each line corresponding to a different radius. That is, at each radius $a$, there is a one-to-one mapping between cloud velocity and the surface density it had when first entrained into the wind, and this relationship is different at different radii. The dashed line represents the maximum possible column density $x = x_{\rm crit}$ -- clouds with column densities larger than this value are too strongly held by gravity to accelerate outward. The lower panel of \autoref{fig:den_struct_cut} is similar, but instead it shows the relationship between $u_r$ and $a^2 d\rho_{\rm mean}/dx$ (\autoref{eq:rhomean}). Recall that $d\rho_{\rm mean}/dx$ describes the differential contribution made by material of surface density $x$ to the mean density in a given radial shell; we multiply by $a^2$ to normalise out the geometric drop in mean density coming from the fact that shells at larger radii have larger volumes. Thus each line in the lower panel of \autoref{fig:den_struct_cut} can be read as showing how much material at a given velocity contributes to the total mass budget in a given radial shell. The upper envelope of the lines shown in the figure corresponds to the acceleration law $U_a$ for a cloud with $x=x_{\rm crit}$, i.e., the highest surface density cloud that can be ejected into the wind at all.

In \autoref{fig:den_struct}, we show how the acceleration law shown in \autoref{fig:den_struct_cut} translates into masses and mass fluxes. Solid lines show wind acceleration laws for different starting surface densities, i.e., they show velocity versus radius for particular clouds. The lines are spaced so that an equal amount of mass lies between each pair of them, and they form a sequence in $x$, starting with $x=x_{\rm crit}$ (the lowest velocity line) and then moving to lower $x$ (higher velocity) with each subsequent line. The colour shows the differential density of wind material at any combination of radius $a$ and velocity $u_r$. Thus for example one can read off from the colour scale that the densest material at any given radius is at low velocity, and started with surface densities $x$ close to $x_{\rm crit}$. Similarly, at any given velocity, the mass budget will tend to be dominated by material at the largest radius where there is any wind at all. However, we emphasise that these statements are true for the particular example shown. Different wind acceleration mechanisms and expansion laws would produce different distributions in \autoref{fig:den_struct}.

\subsection{Domain of Validity for Momentum-Driven Winds}
\label{ssec:validity}

Before proceeding further, we pause to consider under what circumstances our simple kinematic model for wind acceleration is reasonable. In particular, we assume that the force $\dot{p}$ applied to the wind does not explicitly depend on the distance $a$ from the wind launching region, though, as in the case of a radiatively- or hot gas-driven wind, it may indirectly depend on $a$ through its dependence on the wind density, velocity, or some other property. This is reasonable for a momentum-driven wind only if the momentum transferred to the wind constitutes a relatively small fraction of the total momentum budget available to the driving mechanism. For example in the case of radiation driving, our assumption of constant $\dot{p}$ and thus constant $\Gamma$ is reasonable only for winds where the optical depth averaged over scales much larger than an individual cloud is small, so that the typical cloud that might be accelerated is not shadowed by one closer to the source.\footnote{Clouds may be shadowed at frequencies that are resonant with some line of interest, and we explicitly allow for this possibility below. We only require that such shadowing not be so broad in frequency as to substantially affect the total momentum budget of the radiation field.} If this were not the case, then photons injected at small radii would not reach large radii, and $\Gamma$ would become a strong function of position. This limitation does not apply to energy-driven mechanism such as the pressure exerted by hot gas, since for these mechanisms different parts of the expanding shell of cool material are in causal contact, and momentum added to one region of the wind can be cancelled by momentum added to the opposite region, or to the central driving object.

To investigate under what circumstances our models are appropriate for momentum-limited winds, we can compute the momentum flux $\dot{p}_{\rm sh}(a)$ passing through a thin shell at radius $a$, and compare this to the total momentum budget $\dot{p}$. We can write the momentum flux as
\begin{equation}
\dot{p}_{\rm sh}(a) = \int_{-\infty}^{x_{\rm cr}} \frac{d\dot{p}}{dx} \, dx
\end{equation}
where
\begin{equation}
\frac{d\dot{p}}{dx} = v_0 U_a(x) \frac{d\dot{M}}{dx}
\end{equation}
is the momentum flux carried by material with starting surface densities in the range $x$ to $x+dx$. Inserting our expression of $d\dot{M}/dx$ (\autoref{eq:massflux}), we obtain
\begin{eqnarray}
\frac{\dot{p}_{\rm sh}(a)}{\dot{p}} & = & \frac{2}{3f_g \Gamma} \left(\frac{r_0}{v_0}\right) \left(\frac{\dot{M}}{M_0}\right) \frac{1}{\zeta_M} \int_{-\infty}^{x_{\rm cr}} p_M U_a \, dx \\
& = & \frac{2}{3f_g \Gamma} \left(\frac{t_c}{t_w}\right) \frac{1}{\zeta_M} \int_{-\infty}^{x_{\rm cr}} p_M U_a \, dx,
\end{eqnarray}
where $f_g$ is the gas fraction, and  in the second step we have defined $t_c = r_0/v_0$ and $t_w = M_0/\dot{M}$ as the crossing time and the wind mass removal time, respectively. We can further simplify by noting that the \citetalias{thompson16a} wind model predicts a wind mass flux $\dot{M} \approx f_g M_0/\zeta_M t_c$, i.e., the mass flux is simply the gas mass $f_g M_0$ divided by the crossing time (which is roughly the same as the free-fall time for a virialised system), reduced by a factor of $1/\zeta_M$ representing the fraction of the mass that is super-Eddington. Inserting this expression for $\dot{M}$, we have
\begin{equation}
\frac{\dot{p}_{\rm sh}(a)}{\dot{p}} \approx \frac{1}{\Gamma} \int_{-\infty}^{x_{\rm cr}} p_M U_a \, dx.
\end{equation}
For our model to be valid for a momentum-limited driving mechanism, this ratio should be $\ll 1$.

\begin{figure}
\includegraphics[width=\columnwidth]{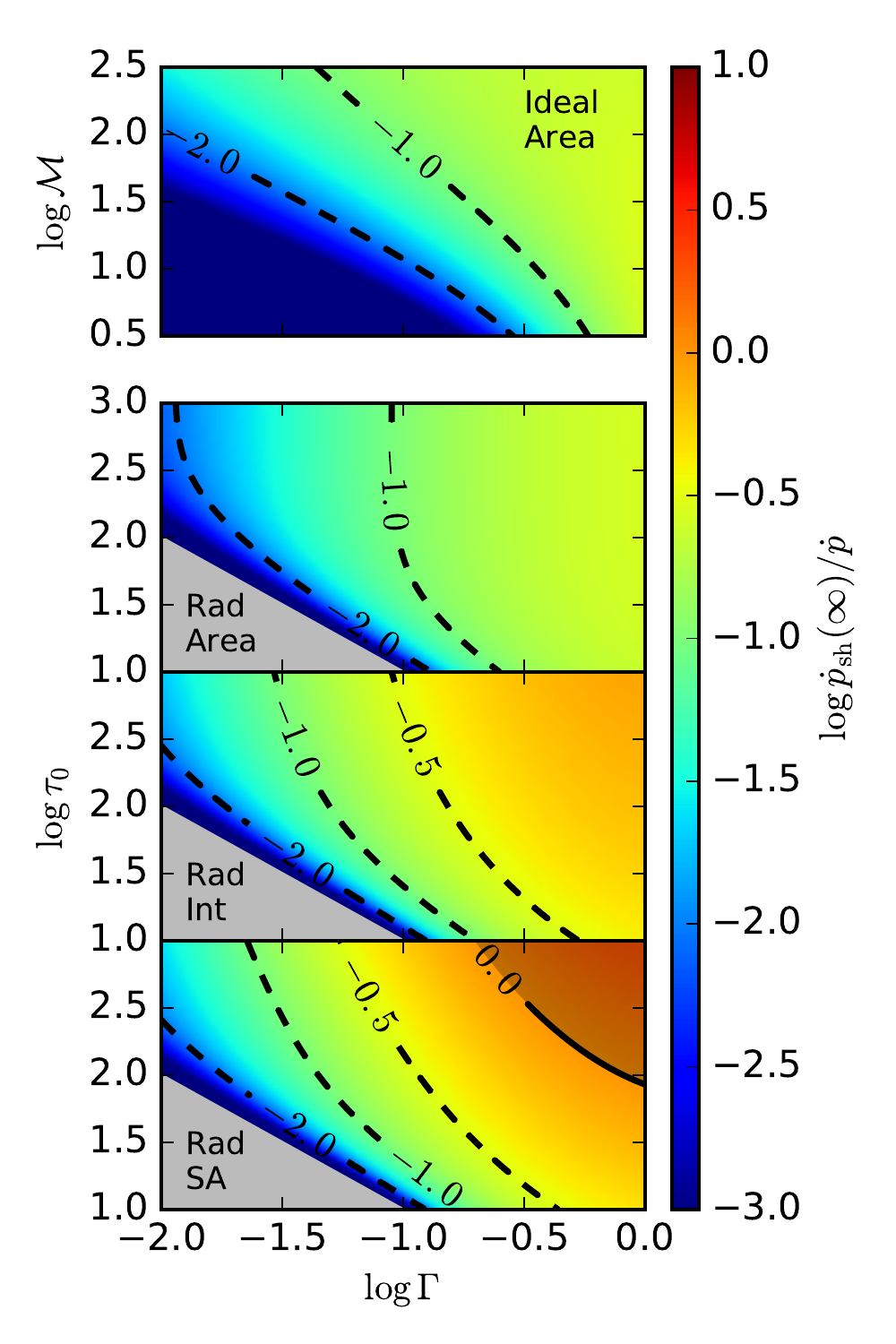}
\caption{
\label{fig:momentum}
Wind momentum flux through the shell at infinity, $\dot{p}_{\rm sh}(\infty)$, normalised by the driving momentum flux $\dot{p}$. The top panel shows this quantity as a function of Eddington ratio $\Gamma$ and Mach number $\mathcal{M}$ for an ideal wind with constant-area clouds (i.e., a wind following the acceleration law given in the top left entry in \autoref{tab:eom_solutions}). Recall that $\mathcal{M}$ is the Mach number of the region from which the wind is being accelerated, not the Mach number of the wind gas (which we have implicitly assumed is $\gg 1$, since we treat thermal pressure as unimportant). The bottom three panels show this quantity for radiatively-driven winds with constant area, intermediate, and constant solid angle clouds, as a function of $\Gamma$ and the mean optical depth $\tau_0$; each of these calculations used $\mathcal{M}=100$. In each panel, colour shows the value of $\dot{p}_{\rm sh}/\dot{p}$ at each parameter combination, while contours show loci of constant $\log \dot{p}_{\rm sh}/\dot{p}$. The contours shown are $\log \dot{p}_{\rm sh}/\dot{p} = -2, -1, -0.5$, and $0$, as indicated in the plot. The grey region in the lower left corner of the panels for radiatively-driven winds are combinations of $\Gamma\tau_0 < 1$, which are forbidden because they do not result in a wind being driven. The shadowed region in the upper right corner of the bottom panel indicates $\dot{p}_{\rm sh}/\dot{p} > 1$, where the model is invalid.
}
\end{figure}

It is immediately clear that an ideal wind cannot fully satisfy this constraint if the material expands as $y=a$ or faster, since in this case the velocity diverges as $a\rightarrow \infty$, and the momentum flux therefore does as well. For all other driving mechanisms, and for ideal driving with constant area, $U_a$ remains finite as $a\rightarrow \infty$, and we can evaluate the maximum value of $\dot{p}_{\rm sh}(\infty)/\dot{p}$ by substituting in the limiting value of $U_a$. In \autoref{fig:momentum} we plot the momentum flux at infinity carried by an ideal wind with constant area clouds, and by a radiatively-driven wind for any expansion factor. We do not make an analogous plot for hot gas-driven winds, because these are energy- rather than momentum-limited, and thus can have any value of $\dot{p}_{\rm sh}/\dot{p}$.

From the Figure, it is clear that the constraint that $\dot{p}_{\rm sh}(\infty)/\dot{p} \ll 1$ is well satisfied over most of parameter space for either radiatively-driven or constant-area ideal winds. The only cases where $\dot{p}_{\rm sh}(\infty)/\dot{p} \sim 1$ for radiatively-driven winds are for clouds with constant solid angle, very high initial optical depth ($\tau_0 \gtrsim 100$), and very high Eddington ratio ($\Gamma\gtrsim 0.3$).  This conforms with our intuition about momentum-driven spherical shells, where we expect the shell to gather all of the available momentum produced by the source, as long as it remains optically thick \citep{thompson15a}. Over most of the parameter space of interest --- e.g., $\log \Gamma\sim-1$ --- we can therefore safely use our models even for momentum-limited winds. The physical reason $\dot{p}_{\rm sh}(\infty)/\dot{p}\ll1$ for most of the region of parameter space with $\Gamma<1$ is fundamentally because only low column density sightlines are accelerated, and these do not block a significant fraction of the source.

\subsection{Wind Geometry}
\label{ssec:geometry}

\begin{figure}
\vspace{-0.4in}
\includegraphics[width=\columnwidth]{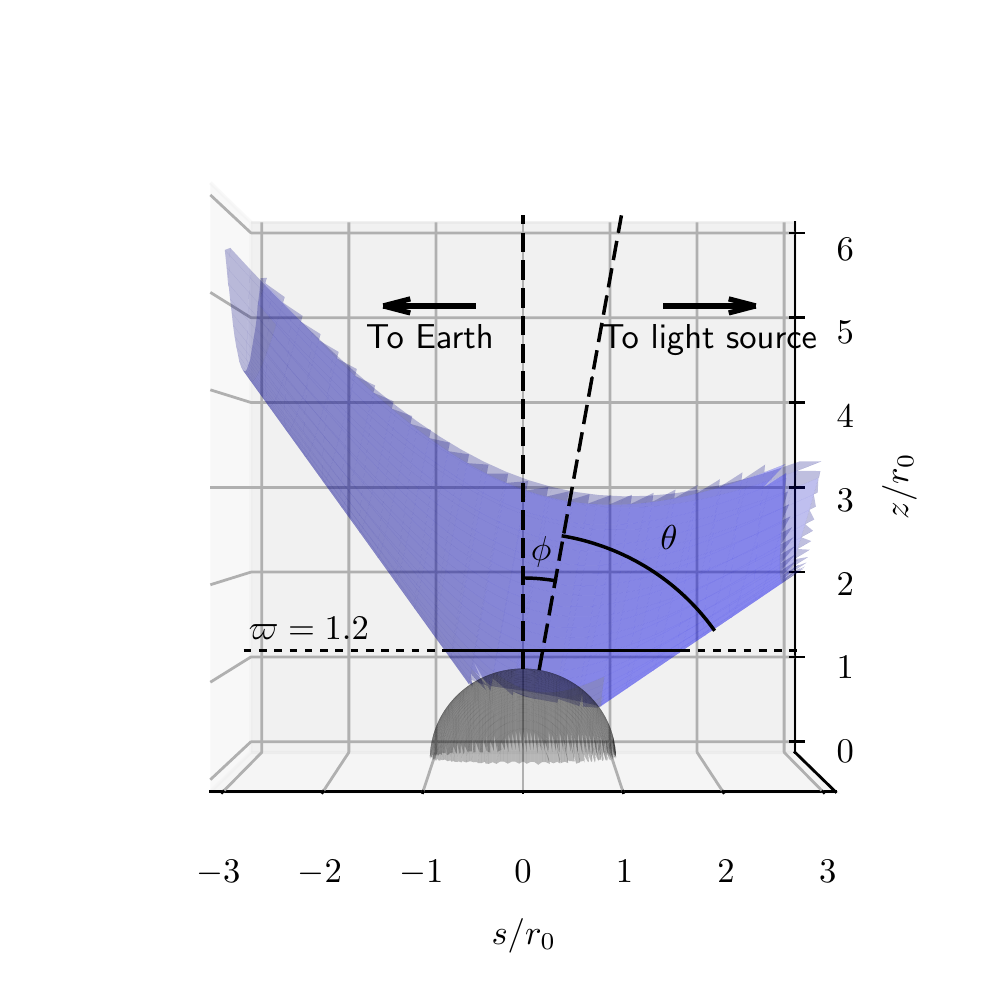}
\includegraphics[width=\columnwidth]{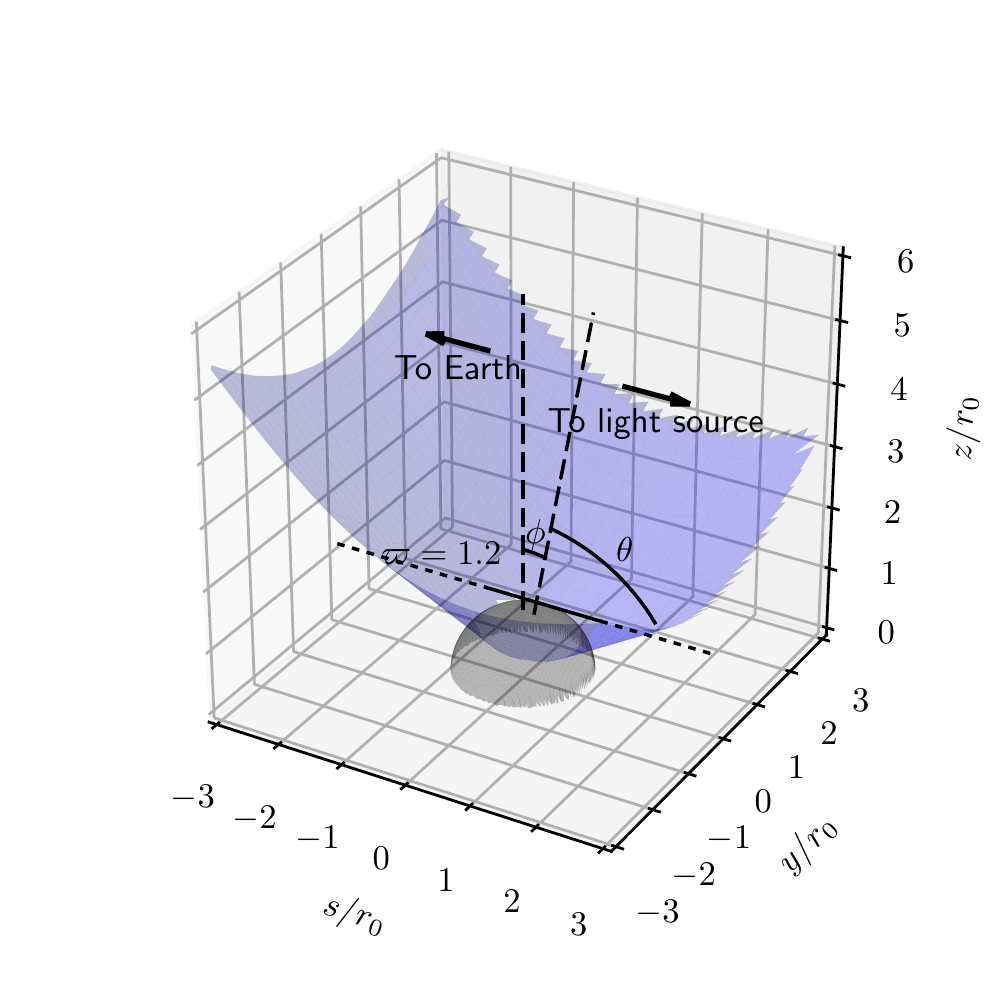}
\caption{
\label{fig:geom}
Geometry of our winds, showing only the upper half volume $\varpi_a>0$ to minimise confusion. In this example, we show a wind that is a fully-filled cone. The two panels show the same structures from two different perspectives. The grey hemisphere shows the wind launching region, while the blue cone shows the outer edge of the wind region, with opening angle $\theta = \theta_{\rm out} = 45^\circ$. The black dashed line shows the central axis of the wind cones, which is inclined at an angle $\phi = 10^\circ$ relative to the line of sight. The dotted and solid line shows an example line of sight with $\varpi_a=1.2, \varpi_t = -0.3$. The portions of this line of sight that lie within the wind are drawn as solid, while the remainder is drawn as dotted.
}
\end{figure}

Thus far we have considered spherical, isotropic winds. However, for winds driven from galaxies and AGN, then outflow is usually blocked by the presence of large quantities of dense material over solid angles subtended by the galactic or AGN accretion disk. Moreover, different chemical components of the wind (e.g., an ionised component and a molecular component) may not always be mixed, but may instead subtend different parts of the sky as viewed from the driving object. In the context of our simple model, we idealise this by considering that the wind may be limited to a certain range of angle around some central axis. The wind is characterised by an outer opening angle $\theta_{\rm out} \in (0, \pi/2]$ and an inner opening angle $\theta_{\rm in}\in [0, \theta_{\rm out})$. The central axis of the wind is inclined by an angle $\phi \in [-\pi/2,\pi/2]$ relative to our line of sight, with $\phi=0$ corresponding to the central axis lying in the plane of the sky and $\phi = \pi/2$ corresponding to the axis pointing directly away from  the Earth. Thus we can consider isotropic spherical winds, winds that are filled cones, and winds that are sheaths bounded by an inner and an outer cone.

Within this geometry, we consider some line of sight of interest, and we characterise it by three coordinates, all of which we non-dimensionalise by scaling to $r_0$: $\varpi_a$ is the impact parameter of the line of sight along the central axis of the wind, $\varpi_t$ is the impact parameter transverse to the wind axis, and $s$ is position along the line of  sight, with $s=0$ corresponding to the position where the line of sight crosses the plane of the sky in which the wind centre ($a=0$) lies. Values of $s<0$ correspond to the near side of the wind to Earth, while $s>0$ is the far side. We illustrate our adopted geometry and coordinate system in \autoref{fig:geom}.

The line of sight intersects the outer cone that defines the edges of the wind region at any point $(s,\varpi_a, \varpi_t)$ satisfying
\begin{equation}
(s\cos\phi-\varpi_a \sin\phi)^2 + \varpi_t^2 - (\varpi_a \cos\phi+s\sin\phi)^2 \tan^2\theta_{\rm out} = 0,
\end{equation}
subject to the constraint $s^2 + \varpi_a^2 + \varpi_t^2 > 1$, and similarly for the inner cone with opening angle $\theta_{\rm in}$. It intersects the spherical region within which there is no wind at any point satisfying $s^2 + \varpi_a^2 + \varpi_t^2 = 1$. Depending on the values of $\phi$, $\theta_{\rm out}$, $\theta_{\rm in}$, $\varpi_a$, and $\varpi_t$, this system of non-linear equations may have a variable number of solutions for positions along the line of sight $s$, which define where the line of sight enters and exits the wind region. We refrain from writing out all the possible cases, as they are somewhat tedious, but it is straightforward to solve the equations to obtain them, and the software required to do so is included in the public release as part of \textsc{despotic}.

For the purposes of the remainder of this paper, we define $s_{0,i}$ and $s_{1,i}$ as the positions $s$ along the line of sight that define where the line of sight enters and exits the wind, i.e., wind material exists at any $s$ satisfying $s_{0,i} < s < s_{1,i}$. If a given line of sight never exits the wind on the near or far side, then $s_{0,i}$ and $s_{1,i}$ is $-\infty$ or $\infty$, respectively. We further denote by $s^+_{0,i}$ and $s^+_{1,i}$ the wind entry and exit points only on the far side, $s > 0$, and we similarly denote by $s^-_{0,i}$ and $s^-_{1,i}$ points on the near side. If a given line of sight it still within the wind when it passes through the plane $s=0$, then we will have $s^-_{1,i} = 0$ or $s^+_{0,i} = 0$. Finally, we define the radii corresponding to these points by $a^\pm_{0,i} = \sqrt{\left(s^\pm_{0,i}\right)^2 + \varpi_a^2 + \varpi_t^2}$, and similarly for $a^\pm_{1,i}$; for notational simplicity we take the radii to always be ordered so that $a^\pm_{0,i} < a^\pm_{1,i} < a^\pm_{0,i+1}$, and we define $\varpiv = (\varpi_a, \varpi_t)$, and $\varpi = \sqrt{\varpi_a^2+\varpi_t^2}$.

\section{Line Absorption by Winds}
\label{sec:absorb}

We are now in a position to compute the observable absorption profile for light passing through the wind. In this calculation, we assume that over the great majority of the wind the thermal velocity dispersion is small compared to the spread in wind velocities associated with variations in the bulk velocity caused by variations in the initial surface density $x$. We also assume that the solid angle that clouds subtend is at most constant with radius, so $y(a) \leq a^2$. Finally, we focus on the case of absorption of light from a background source at infinity, rather than light from the source galaxy or star cluster located at the origin; the latter can be treated trivially simply by neglecting absorption from any position $s>0$, i.e., those on the far side of the source.

Now consider a transition with oscillator strength $\Omega$, produced by a species for which the mass per member of the species is $m_X$, i.e., in gas of mass density $\rho$ the number density of members of species $X$ is $n_X = \rho/m_X$. This is related to the abundance $[\mathrm{X}/\mathrm{H}]$ of the absorbing element relative to hydrogen, the ionisation correction $f_{\rm ion}$, and the depletion factor $f_{\rm dep}$, by $m_X = \mu_{\rm H} m_{\rm H}/(f_{\rm ion} f_{\rm dep} [\mathrm{X}/\mathrm{H}])$, where $\mu_{\rm H}$ is the gas mass per H nucleus in units of $m_{\rm H}$ ($\mu_{\rm H} = 1.4$ for standard cosmic composition), $f_{\rm ion}$ is the fraction of the absorbing element in the ionisation state that produces the absorption, and $f_{\rm dep}$ is the fraction of the element that is in the gas phase rather than depleted onto dust grains. The transition is between a lower state $\ell$ and an upper state $u$, which have degeneracies $g_\ell$ and $g_u$, respectively. The fractions of absorbers in the lower and upper states are $f_\ell$ and $f_u$, and the excitation temperature $T_{\rm ex}$ is defined as usual, $\exp(-h\nu_0/k_B T_{\rm ex}) = g_\ell f_u / g_u n_\ell$, where $\nu_0$ is the line centre frequency. We are interested in computing the optical depth $\tau_\nu$ produced by this transition for a measurement at line of sight velocity $v$ (corresponding to an observed frequency $\nu = \nu_0(1+v/c)$) along a line of sight that has a specific impact parameter. We define $u = v/v_0$ as the dimensionless line of sight velocity.

For this situation, the absorption coefficient is given by
\begin{equation}
\kappa_\nu = \frac{\pi e^2}{m_e c} \Omega n_X f_\ell \left(1 - e^{-h\nu_0/k_B T_{\rm ex}}\right) \phi_\nu,
\end{equation}
where $\phi_\nu$ is the line shape function, which measures the distribution of the absorbers in frequency. For optical lines the fraction in the lower state $f_\ell$ and the stimulated emission correction $1-e^{-h\nu_0/k T_{\rm ex}}$ are generally both unity, but we retain them for generality, because we shall discuss radio lines below. It is convenient to rewrite the absorption coefficient in terms of the absorber velocity distribution, as
\begin{equation}
\label{eq:kappa_nu}
\kappa_\nu = \frac{\pi e^2}{m_e c} \Omega \frac{\lambda_0}{v_0} f_\ell \left(1 - e^{-h\nu_0/k_B T_{\rm ex}}\right) \frac{dn_X}{du}
\end{equation}
where $\lambda_0 = c/\nu_0$ is the line centre wavelength, and we use vacuum rather than air wavelengths here and throughout. Note that, because we have written $\kappa_\nu$ as proportional to the gas bulk velocity distribution $dn_X/du$ rather than the convolution of the bulk and thermal velocity distributions, we have implicitly adopted a large velocity gradient approximation whereby we neglect  thermal broadening in comparison to bulk motion. Since we are concerned with the cool components of the wind (temperature $\lesssim 10^4$ K), this approximation is quite reasonable: the thermal broadening for a species with atomic mass $\mu$ in gas of temperature $T$ is $9.1 \mu^{-1/2} (T/10^4\,{\rm K})^{1/2}$ km s$^{-1}$, while the winds with which we are concerned generally have velocities of hundreds to thousands of km s$^{-1}$.

To compute $dn_X/du$, we need to compute the number density absorbers within our observed beam. At a distance $a$ from the origin, using \autoref{eq:rhomean} the mean number density is
\begin{equation}
n_X = \frac{1}{m_X} \int_{-\infty}^{x_{\rm crit}} \frac{d\rhomean}{dx}\, dx.
\end{equation}
Since we are interested in the distribution of absorbers with respect to velocity, it is helpful at this point to make a change of variables from $x$ to $u_r$. Doing so gives
\begin{equation}
n_X = \frac{1}{m_X} \int_{0}^{\infty} \sum_i \frac{d\rhomean/dx}{|dU_a/dx|}\, du_r,
\end{equation}
where $d\rhomean/dx$ and $dU_a/dx$ are to be evaluated at overdensities $x=x_i$ satisfying $u_r = U_a(x)$ with $x_i < x_{\rm crit}$. The summation is over all such solutions, and the integrand is zero at values of $(u_r,a)$ such that this equation has no real roots satisfying $x < x_{\rm crit}$. We now make one more change of variables: the quantity of interest to us is not the radial velocity $u_r$, but the line of sight velocity $u$. The two are related by
\begin{equation}
\label{eq:u_ur}
u = \pm u_r \sqrt{1-\frac{\varpi^2}{a^2}}.
\end{equation}
Here the positive solution corresponds to the material at $s > 0$ and the negative solution material at $s < 0$. Using \autoref{eq:u_ur}, the density is
\begin{equation}
\label{eq:n_u}
n_X = \frac{1}{m_X} \frac{a}{\sqrt{a^2-\varpi^2}} \int_{-\infty}^{\infty} \sum_i
\frac{d\rhomean/dx}{|dU_a/dx|} \, du,
\end{equation}
where the summation is over solutions $x_i$ to the implicit equation
\begin{equation}
\label{eq:u_x_map}
u^2 = U_a^2(x) \left(1 - \frac{\varpi^2}{a^2}\right),
\end{equation}
subject to the constraint $x < x_{\rm crit}$. The density contribution $d\rhomean/dx$ and velocity derivative $dU_a/dx$ are to be evaluated at these solutions. As above, if for a particular combination of $u$, $\varpi$, and $a$ \autoref{eq:u_x_map} has no real solutions with $x < x_{\rm crit}$, then the integrand is zero. With the volume density rewritten in the form given by \autoref{eq:n_u}, it is trivial to extract the differential contribution of the material at observed velocity $u$, which is simply
\begin{equation}
\label{eq:dndu}
\frac{dn_X}{du} = \frac{1}{m_X} \frac{a}{\sqrt{a^2-\varpi^2}} \sum_i
\frac{d\rhomean/dx}{|dU_a/dx|}.
\end{equation}

We can understand the procedure graphically via \autoref{fig:den_struct}. In the Figure, the dot-dashed lines show loci of constant $u=1.5$ through the $(a, u_r)$ plane for a range of impact parameters $\varpi$. For an observation with a particular impact parameter $\varpi$ at a particular line of sight velocity $u$, one can use the dot-dashed curve to read off the value of $x$ and the differential density $d\rhomean/dx$ at each radius $a$. This in turn defines the absorption coefficient at that radius. In the parts of the figure where the dot-dashed line is outside the coloured region, there is no material with line of sight velocity $u$, and thus no absorption.

To proceed we must now make some assumptions about geometry. The need for a geometric assumption is easiest to understand if we consider the analogous but simpler problem of computing the amount of continuum light absorbed by dust along a line of sight. For a fixed mass of dust, we will find very different amounts of light are blocked if the dust is arranged into very small particles, each of which is individually optically thin, than if we arrange the dust into large boulders, each of which is individually opaque. The analogous question in our line absorption case is whether the material in different radial shells is coherent over size scales such that individual clouds become opaque. We will consider two limiting cases.

\subsection{The Uncorrelated Case}

First suppose that the clouds being ejected in the wind have such tiny radial extents that each individually is optically thin, and that the gas on different radial shells is completely uncorrelated -- in effect, the wind is like a mist, and each droplet in the mist is individually transparent. This case is analogous to the case of dust consisting of tiny particles. Since clouds are randomly placed, each one sees the same average flux emerging from the direction of the source. Thus if a flux $F_\nu$ propagates a distance $ds$ along a line of sight at a distance $a$ from the origin, it is reduced by an amount $dF_\nu = -\kappa_\nu F_\nu \, ds$. Since $\kappa_\nu$ is an explicit function of radius $a$, it is convenient to re-express this as
\begin{equation}
\label{eq:dflux_uncorr}
dF_\nu = -\kappa_\nu \left(\frac{a}{\sqrt{a^2-\varpi^2}}\right) F_\nu\, r_0 \, da,
\end{equation}
where the factor in parentheses is a geometric correction to account for the angle between the radial length element $da$ and the length element along the line of sight $ds$. Note that this expression assumes that scattering into the beam from other directions is negligible, and thus is inapplicable for lines (e.g., Ly$\alpha$) where this process is important. If we let $F_\nu^{(0)}$ be the flux at the point where the light enters the wind, and insert \autoref{eq:kappa_nu} and \autoref{eq:dndu} for $\kappa_\nu$, we can integrate the equation to find the flux that reaches the observer at infinity,
\begin{equation}
F_\nu = F_\nu^{(0)} e^{-\tau_\nu^{\rm (uc)}},
\end{equation}
with
\begin{eqnarray}\label{eq:tauuc_def}
\tau_\nu^{\rm (uc)} & = & \frac{\pi e^2}{m_e c} \frac{\Omega}{m_X}  f_\ell \left(1 - e^{-h\nu_0/k_B T_{\rm ex}}\right)\frac{\lambda_0}{v_0} r_0
\cdot {}
\nonumber \\
& & \qquad
\sum_j \int_{a^{\pm}_{0,j}}^{a^{\pm}_{1,j}} 
\frac{a^2}{a^2-\varpi^2} \sum_i \frac{d\rhomean/dx}{ \left|dU_a/dx\right|}\, da.
\end{eqnarray}
The summation here is over the ranges of radii when the line of sight is within the wind; the positive sign applies to velocities $u>0$, for which only the far side of the wind can absorb, while the negative sign applies to velocities $u<0$, for which only the near side absorbs. The superscript (uc) is to indicate that this is for the uncorrelated case, and the integral here is
$d\rhomean/dx$ integrated along one of the dot-dashed curves
in \autoref{fig:den_struct}. Inserting \autoref{eq:rhomean} for 
$d\rhomean/dx$ and simplifying gives
\begin{equation}
\label{eq:tau_v_uc}
\tau_\nu^{(\rm uc)} =  \frac{t_X}{t_w}f_\ell \left(1 - e^{-h\nu_0/k_B T_{\rm ex}}\right)
\Phi^{\rm (uc)}(u, \varpiv, 1, \infty),
\end{equation}
where we have defined the quantities
\begin{eqnarray}
\label{eq:tX}
t_X & = & \frac{\Omega \lambda_0}{4G m_X} \frac{e^2}{m_e c} \\
t_w & = & \frac{M_0}{\dot{M}},
\end{eqnarray}
and
\begin{eqnarray}
\lefteqn{\Phi^{\rm (uc)}(u,\varpiv, a_0, a_1) =}
\nonumber \\
& & \frac{1}{\zeta_M} \sum_j \int_{\max(a_0, a^\pm_{0,j})}^{\min(a_1, a^\pm_{1,j})} 
\sum_i \frac{p_M}{\left|dU_a^2/dx\right|}
\frac{1}{a^2-\varpi^2}\, da.
\end{eqnarray}
As a reminder, the integrand in the integral that defines $\Phi^{\rm(uc)}(u,\varpiv,a_0,a_1)$ is to be evaluated at the solutions $x_i$ to \autoref{eq:u_x_map}, and is taken to be zero if no real solution with $x < x_{\rm crit}$ exists. The quantity $t_X$ (which has units of time) depends only on fundamental constants and on the abundance of the absorbers, and can be thought of as a parameter that describes the intrinsic strength of the absorption feature in question. The quantity $t_w$ is a characteristic timescale for the wind; it is the time required for the wind to carry away a mass $M_0$. The function $\Phi^{\rm(uc)}(u,\varpiv,a_0,a_1)$ can be thought of as a dimensionless version of the line shape function for material in the radial range from $a_0$ to $a_1$. In general it cannot be evaluated analytically, but is straightforward to compute numerically.\footnote{In this case, since $a_0$ is always unity and $a_1$ is always infinity, there is no need to give the arguments explicitly. However, we retain them because we will need them in subsequent discussions.}

If there are multiple transitions that are closely-enough spaced in frequency that they must be considered together (for example doublets such as Mg~\textsc{ii} $\lambda\lambda 2796, 2804$), then in the uncorrelated case we can generalise \autoref{eq:tau_v_uc} simply by adding the optical depths contributed by each transition. Specifically, the optical depth for multiple transitions becomes
\begin{eqnarray}
\label{eq:tau_v_uc_multiple}
\lefteqn{\tau_\nu^{(\rm uc)} = \sum_k \frac{t_{X,k}}{t_w}
f_{\ell,k} \left(1 - e^{-h\nu_{0,k}/k_B T_{{\rm ex},k}}\right)
}
\nonumber \\
& & 
{} \cdot \Phi^{\rm (uc)}(u - u_k, \varpiv, 1, \infty),
\end{eqnarray}
where the sum is over the $k$ transitions under consideration, $t_{X,k}$, $f_{\ell,k}$, $\nu_{0,k}$, and $T_{{\rm ex},k}$ are the timescale for each transition, lower level population, rest frequency, and excitation temperature for each transition, and $u_k \equiv (1-\nu_{k,0}/\nu_{0,0}) (c/v_0)$ is the dimensionless velocity shift that corresponds to the wavelength separation of the transitions.

\begin{table*}
\begin{tabular}{lcccccc}
\hline\hline
Line(s) & $f_{\rm ion}$ & $f_{\rm dep}$ & $[\mathrm{X}/\mathrm{H}]$ & $m_X$ [g] & $ \Omega$ & $t_X$ [Gyr] \\
\hline\hline
Mg~\textsc{i} $\lambda 2853$ & 0.1 & 0.64 & $4.4\times 10^{-5}$ & $8.2\times 10^{-19}$ & $1.0$ & 34 \\
Mg~\textsc{ii} $\lambda\lambda 2796, 2804$ & 0.9 & 0.64 & $4.4\times 10^{-5}$ & $9.3\times 10^{-20}$ & $0.62, 0.31$ & $190$, $95$ \\
Na~\textsc{i} $\lambda\lambda 5892, 5898$ & 0.01$^\dagger$ & 1.0$^\dagger$ & $2\times 10^{-7}$ & $1.2\times 10^{-16}$ & 0.64, 0.32 & $0.32$, $0.16$ \\
Fe~\textsc{ii} $\lambda 2383$ & 0.5 & 0.14 & $3.5\times 10^{-5}$ & $9\times 10^{-19}$ & 0.32 & 8.5 
\\ \hline
\end{tabular}\\
\parbox{4.5in}{$^\dagger$The ionisation fraction and depletion factor for Na are substantially uncertain, due to the fact that the dominant interstellar ionisation state, Na$^+$, has no transitions at energies $<13.6$ eV. Only Na$^0$ can be observed, and the ionisation and depletion factor are determined from its abundance coupled with an uncertain ionisation correction. The depletion factor given in the Table comes from \citet{weingartner01b}. However, the product $f_{\rm ion} f_{\rm dep}$ is better constrained than either one alone, and the value of 0.01 for this quantity that we give in the Table is consistent with these constraints \citep[e.g.,][]{rupke15a}. More discussion of the ionisation state of Na is given in \citet{murray07a}.}
\caption{
\label{tab:tA}
Selected absorption line parameters. For each line or group of lines, we give a typical ionisation correction (defined as the fraction of that element in the ionisation state that produces the absorption line) for low ionisation sources such as star-forming galaxies or LINERs, a depletion factor, an abundance relative to H, the mass per absorber / emitter $m_X$, the oscillator strength $\Omega$, and the strength parameter $t_X$ (\autoref{eq:tX}). Abundances and depletion factors for are taken from Table 9.5 of \citet{draine11a}, using their WIM values. Oscillator strengths for atomic lines are from Table 9.4 of \citet{draine11a} or Table 2 of \citet{morton03a}.
}
\end{table*}

\autoref{tab:tA} gives typical values of $t_X$ for frequently-observed low ionisation species. We see that for many low ionisation lines $t_X/t_w \gg 1$ for any value of $t_w$ small enough (i.e., for any wind mass loss rate $\dot{M}$ large enough) to be interesting from the standpoint of galaxy formation. Depending on the transition selected and the source in question, typical values of $t_X/t_w$ are likely to be in the range $\sim 1 - 100$. Since we rarely observe complete absorption by winds in these transitions, which is what we generally expect for large $t_X/t_w$ in the uncorrelated case, this is strong evidence that reality is unlikely to be close to the uncorrelated case. We therefore turn next to the correlated case.

\subsection{The Correlated Case}
\label{ssec:correlated}

If ``flights'' of clouds that are launched at different times are correlated
in their angular coordinates, then
\autoref{eq:dflux_uncorr} is not valid, because all the material at a
given radial position does not see the same flux emerging from the
direction of the source. Instead, the flux incident to a given cloud may be reduced
by absorption of material closer to the source.  This shadowed material absorbs no additional
light, but the correlation of material along some lines of sight
creates other lines of sight that are clear of gas and the source is
completely unabsorbed.  As a consequence,
the effective mean optical depth for a particular wind is less than
what it would have been for uncorrelated clouds.

We can account for this effect as follows. First, we consider a given
``flight'' of clouds, ejected from the source over its dynamical time.
We assume that, at the point where it is launched, the wind material is
confined within a fraction $f_w$ of the total available solid angle. At
a distance $a$ from the launching point, the fraction of the solid angle
covered is therefore
\begin{equation}
f_c = f_w \frac{y}{a^2}.
\end{equation}
Within the covered region, clouds
are more closely crowded together, which increases the mean volume
density by a factor $1/f_c$ compared to \autoref{eq:rhomean}, and 
therefore the mean optical depth within this solid angle increases
by a factor $1/f_c$ compared to \autoref{eq:tauuc_def}.  
However, only a fraction $f_c$ of the light is attenuated at all.

What is the value of $f_w$? We note that gas in the launching region 
should have a roughly lognormal distribution of area
as well as mass, so that the differential area fraction occupied by material with
surface density between $x$ and $x+dx$ is given by
\begin{equation}
p_A = \frac{1}{\sqrt{2\pi\sigma_x^2}} 
\exp\left[-\frac{\left(x+\sigma_x^2/2\right)}{2\sigma_x^2}\right].
\end{equation}
In this case the total fraction of area subtended by regions from which
the wind is launched is
\begin{equation}
\zeta_A = \int_{-\infty}^{x_{\rm crit}} p_A \, dx
= \frac{1}{2}\left[1+\mbox{erf}\left(\frac{2 x_{\rm crit}+\sigma_x^2}{2\sqrt{2}\sigma_x}\right)\right],
\end{equation}
so as a rough guess one might set $f_w = \zeta_A$. However,
Raskutti, Ostriker, \& Skinner (2017, submitted) find from their numerical
simulations that, while
the distribution of mass with circumcluster surface density follows a
lognormal, the distribution of solid angle with circumcluster
surface density has an excess at low surface density compared to a lognormal;
this is equivalent to a lognormal distribution that holds only within a
fraction of the total available solid angle. For this reason, we choose to leave
$f_w$ as a free parameter, for which $\zeta_A$ is a rough estimate, but one that
is probably a bit too large.

In the limiting case in which all flights of clouds are aligned in their
angular distribution, and $f_w$ is the same for all flights of clouds, 
the flux that reaches infinity in the case where $y=a^2$ and thus $f_c=f_w$ is
constant with radius will be
\begin{equation}
\label{eq:f_nu_const_sa}
F_\nu = F_\nu^{(0)} \left\{1 - f_c + f_c \exp\left[-\tau_\nu(\infty)\right]\right\},
\end{equation}
where
\begin{eqnarray}
\tau_\nu(a) & = & r_0 \sum_j \int_{\min(a, a^\pm_{0,j})}^{\min(a,a^\pm_{1,j})}  \frac{a'}{\sqrt{a'^2-\varpi^2}} \frac{\kappa_\nu}{f_c} \, da' \\
& = & \frac{t_X}{t_w}f_\ell \left(1 - e^{-h\nu_0/k_B T_{\rm ex}}\right) \Phi^{\rm (c)}(u,\varpiv,1, a),
\label{eq:tau_cor}
\end{eqnarray}
and we have defined
\begin{eqnarray}\label{eq:Phi_cor}
\lefteqn{\Phi^{\rm (c)}(u,\varpiv, a_0, a_1) = }
\nonumber \\
& & \frac{1}{\zeta_M} \sum_j \int_{\max(a_0,a^\pm_{0,j})}^{\min(a_1,a^\pm_{1,j})} 
\frac{1}{f_c} \sum_i \frac{p_M}{\left|dU_a^2/dx\right|}
\frac{1}{(a^2-\varpi^2)}\, da.
\end{eqnarray}
The quantity $\tau_\nu(a)$ is the optical depth from radius $a^\pm_{0,0}$ (i.e., the minimum radius at which the line of sight is within the wind) to radius $a$, accounting for the increase in attenuation that comes from the absorbing material being concentrated into a fraction $f_c$ of the available solid angle. Note that the quantity $\Phi^{\rm (c)}$ on which $\tau_\nu$ depends differs from $\Phi^{\rm (uc)}$ only due to the extra factor $1/f_c$. Also, we caution that, as for the correlated case, \autoref{eq:f_nu_const_sa} is not valid for transitions where scattering into the beam from other directions is significant.\footnote{
If there are multiple ``flights" of clouds that are internally correlated but that were launched at different epochs and are not correlated with each other, then \autoref{eq:f_nu_const_sa} can be generalised to $F_\nu = F_\nu^{(0)} \prod_{i=1}^N \left\{1 - f_{c,i} + f_{c,i} \exp\left[-\tau_{\nu,i}(\infty)\right]\right\},$ where there are $N$ distinct flights, and $f_{c,i}$ and $\tau_{\nu,i}(\infty)$ are the covering fraction and optical depth for flight $i$. In the limit $N\rightarrow \infty$, and where the launching time interval and thus the radial range that contributes to each $\tau_{\nu,i}(\infty)$ becomes infinitesimally small, this approaches the result for the uncorrelated case, \autoref{eq:tau_v_uc}. However, since $N$ is not generally known, we will for the rest of this paper limit ourselves to the two limiting cases of a single correlated flight or purely uncorrelated gas.
}

In \autoref{eq:f_nu_const_sa}, the term $1-f_c$ is the fraction of the solid angle that is not covered, and thus from which light escapes unattenuated. The term $f_c e^{-\tau_\nu(\infty)}$ is the product of the fraction of the solid angle that is attenuated with the fraction of light that is transmitted through this solid angle. The corresponding optical depth that would be inferred by an observer at infinity who could not resolve individual clouds within the beam is
\begin{equation}
\label{eq:tau_c_const_sa}
\tau_\nu^{\rm (c)} = \ln \frac{F_\nu^{(0)}}{F_\nu} = -\ln \left\{1 - f_c + f_c \exp\left[-\tau_\nu(\infty)\right]\right\}.
\end{equation}
Comparing this to \autoref{eq:tau_v_uc}, one can see with minimal algebra that they are identical in the limit $\tau_\nu(\infty)\rightarrow 0$. However, they behave very differently in the limit $\tau_\nu(\infty) \rightarrow \infty$, with $\tau_\nu^{\rm (uc)}$ approaching $\infty$ as well in this case, while $\tau_\nu^{\rm (c)}\rightarrow-\ln(1-f_c)$, and thus remains finite.

The generalisation of \autoref{eq:f_nu_const_sa} to multiple overlapping transitions is the same as in the uncorrelated case,
\begin{eqnarray}
\label{eq:tau_cor_multiple}
\lefteqn{\tau_\nu(a) = \sum_k \frac{t_{X,k}}{t_w}
f_{\ell,k} \left(1 - e^{-h\nu_{0,k}/k_B T_{{\rm ex},k}}\right)
}
\nonumber \\
& & 
{} \cdot  \Phi^{\rm (c)}(u - u_k, \varpiv, 1, a).
\end{eqnarray}
Physically, this amounts to saying that we simply add the contributions to the optical depth from each transition before considering the effects of partial covering.

If $y$ increases with radius more slowly than $y=a^2$, then the covering fraction decreases as clouds move away from the origin. Consider light at a frequency corresponding to a velocity $u < 0$, so it is attenuated only on the near side of the wind. The covering fraction is at a maximum when this light begins to be attenuated at a small radius, but it falls as the light moves further from the origin toward the Earth. Thus some light will be attenuated only out to a certain radius, and will not be attenuated at larger radii because there will be no wind material covering it. The converse is true if we consider the case $u>0$. The covering fraction at $s \gg 1$ is small, so no light is attenuated, but as the beam of light approaches the plane $s=0$ where the wind is being launched, the covering fraction rises, so some light is again attenuated only within a certain distance of the wind launching plane.

For either near- or far-side attenuation, the flux that reaches infinity will be
\begin{equation}
\label{eq:f_nu_var_sa}
F_\nu = F_\nu^{(0)} \left\{1 - f_c(a^\pm_{0,0}) + \int_{a^\pm_{0,0}}^{\infty} \left|\frac{df_c}{da}\right| \exp\left[-\tau_\nu(a)\right] \, da\right\},
\end{equation}
where $|df_c/da| = f_w |a y' - 2y| / a^2$, and $a^\pm_{0,0}$ is the smallest radius at which the line of sight is within the wind and the covering fraction is largest.\footnote{This expression is valid only for $f_c \rightarrow 0$ as $a\rightarrow\infty$. Since this holds for all the wind expansion laws with variable $f_c$ that we consider in this paper, we refrain from giving the more general expression to avoid unnecessary complexity.} As in \autoref{eq:f_nu_const_sa}, the term $1-f_c(a^\pm_{0, 0})$ is the fraction of the area that is never covered by wind material at any radius, and thus is unattenuated. The integral in \autoref{eq:f_nu_var_sa} is the integrated transmission fraction for covered areas, and can be thought of as a weighted average. Specifically, it is the fraction $e^{-\tau_\nu(a)}$ of light transmitted if attenuation occurs only out to radius $a$, weighted by the fraction $|df_c/da| \, da$ of the area that is attenuated only out to radius $a$. The corresponding inferred optical depth will be
\begin{equation}
\label{eq:tau_c_var_sa}
\tau_\nu^{\rm (c)} = -\ln \left\{1 - f_c(a^\pm_{0,0}) + \int_{a^\pm_{0,0}}^{\infty} \left|\frac{df_c}{da}\right| \exp\left[-\tau_\nu(a)\right] \, da\right\}.
\end{equation}

Extending this solution to the case of multiple transitions is only slightly more complex than in the case of fixed $f_c$. The added complication is that different transitions may have different signs of $u-u_k$, so some transitions produce absorption on the near side, while other produce it on the far side. We cannot in general assume that $a_{0,0}^{+} = a_{0,0}^{-}$, and if $f_c$ is not constant then the fraction of area that is not covered is different for the near and far sides, i.e., $f_c(a_{0,0}^{+}) \neq f_c(a_{0,0}^{-})$. However, this is easy to handle. Suppose that $a_{0,0}^{+} < a_{0,0}^-$, which under our assumption that $f_c$ is decreasing with $a$ (if it is not constant) implies that $f_c(a_{0,0}^{+}) > f_c(a_{0,0}^-)$, i.e., that the maximum covering fraction on the far side is higher than on the near side. In this case a fraction $1-f_c(a_{0,0}^-)$ of the light is not absorbed by gas on either the near or the far side, a fraction $f_c(a_{0,0}^+) - f_c(a_{0,0}^-)$ is absorbed only on the far side, and a fraction $f_c(a_{0,0}^{-})$ is absorbed on both sides. If $a_{0,0}^+ > a_{0,0}^-$, the converse is true: some light is absorbed on neither side, some is absorbed only on the near side, and some is absorbed on both sides. Considering both possibilities, the corresponding effective optical depth is
\begin{eqnarray}
\tau_\nu^{\rm (c)} & = & -\ln \left\{1 - f_c(a^\mp_{0,0})
+ \int_{a^\mp_{0,0}}^{a^\pm_{0,0}} \left|\frac{df_c}{da}\right| \exp\left[-\tau^\pm_\nu(a)\right] \, da 
\right.
\nonumber \\
& & \quad
\left. {}
+  \int_{a^\pm_{0,0}}^\infty \left|\frac{df_c}{da}\right| \exp\left[-\tau^+_\nu(a)-\tau^-_\nu(a)\right] \, da\right\},
\label{eq:tau_c_var_sa_multiple}
\end{eqnarray}
where the upper sign in \autoref{eq:tau_c_var_sa_multiple} applies if $a_{0,0}^{+} < a_{0,0}^-$, the lower if $a_{0,0}^{+} > a_{0,0}^-$, and we define $\tau_\nu^+(a)$ and $\tau_\nu^-(a)$ as in \autoref{eq:tau_cor_multiple}, except that $\tau_\nu^+(a)$ sums only over transitions for which $u-u_k > 0$, and $\tau_\nu^-(a)$ only over transitions for which $u-u_k < 0$.

\begin{figure}
\includegraphics[width=\columnwidth]{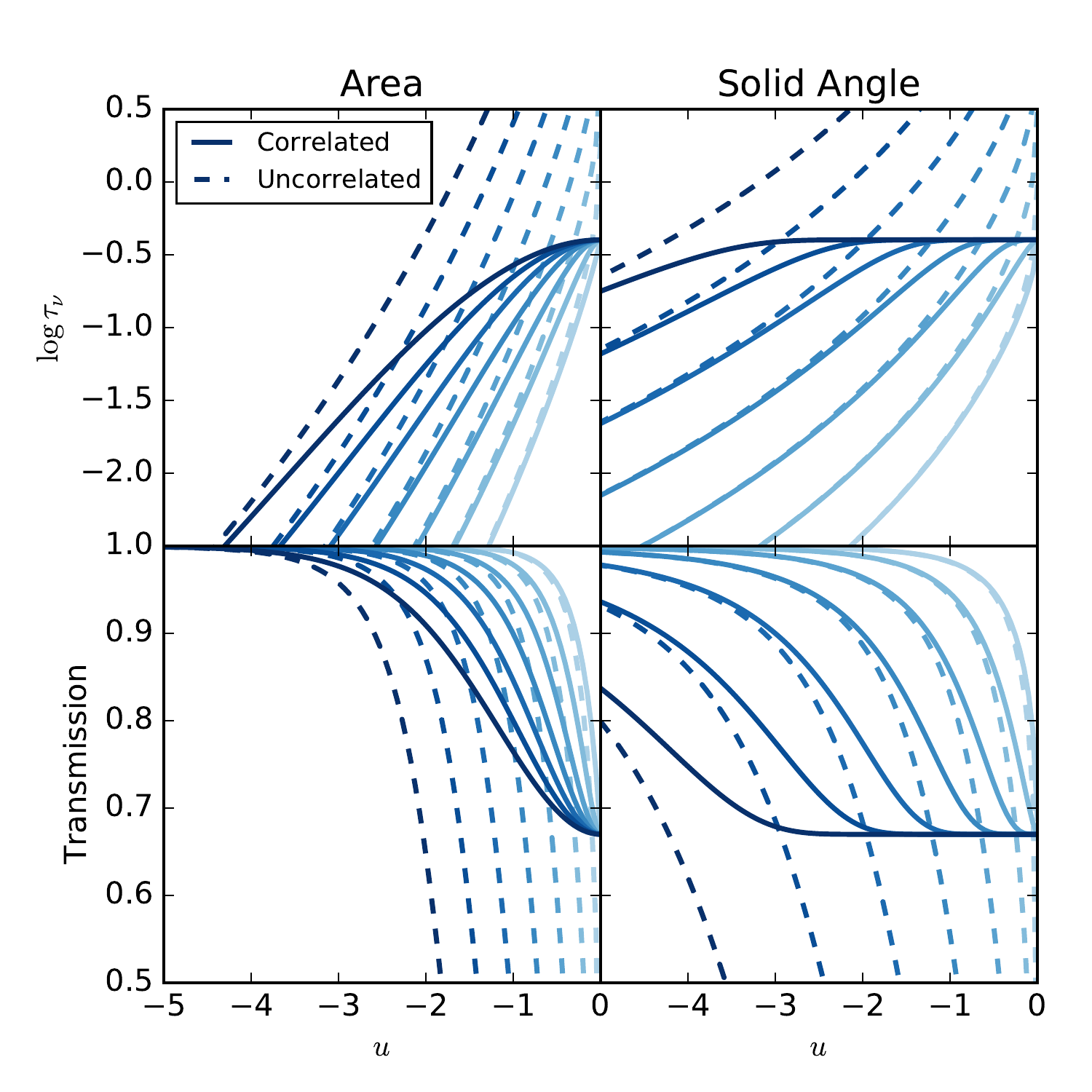}
\caption{
\label{fig:absorption}
Profiles of optical depth $\tau_\nu$ (top panel) and transmission fraction $e^{-\tau_\nu}$ (bottom panel) as a function of velocity $u$ for an ideal, spherical wind in an isothermal potential, observed down the barrel ($\varpi=\varpi_t=0$). All the lines are for winds with $\Gamma=0.2$, $\mathcal{M} = 50$, and the correlated case uses $f_w = \zeta_A = 0.33$. The left column shows a wind with constant area clouds, and the right column shows one with constant solid angle clouds. Solid lines show the correlated case, and dashed lines show the uncorrelated case. Colours indicate the value of $t_X/t_w$, from $t_X/t_w = 0.1$ (lightest) to $t_X/t_w = 100$ (darkest) in steps of 0.5 dex. All calculations use lower state fraction $f_\ell = 1$ and excitation temperature $T_{\rm ex} = 0$.
}
\end{figure}

We show example optical depths and absorption profiles for single transitions in \autoref{fig:absorption}. The plot shows both perfectly correlated  (\autoref{eq:tau_cor}) and uncorrelated (\autoref{eq:tau_c_const_sa} or \autoref{eq:tau_c_var_sa}) cases, and also compares absorption for a constant area case with a constant solid angle case. For the correlated case, in this plot we use the approximation $f_w = \zeta_A$. First comparing the uncorrelated and correlated cases, we see that they are identical at large velocities, where the wind becomes optically thin, but, as expected, they are very different at velocities of order $u\approx 1$, where even a moderate strength transition, one with $t_X/t_w \approx 10$, yields a wind that is completely opaque out to line of sight velocities $u\approx 2$. As noted above, this is not particularly consistent with observations, which suggests that the correlated case is more realistic. For the correlated case, the transmission saturates at a minimum value of $1-\zeta_A = 67\%$ for low velocities. For constant solid angle clouds, this saturation level is reached at line of sight velocities that can range from $\approx 20\%$ of to $\approx 3$ times the escape speed, depending on the strength of the transition and the wind mass flux.

\section{Line Emission by Winds: Species in Local Thermodynamic Equilibrium}
\label{sec:emit_LTE}

We next compute the emission profile for the wind. For a completely general emission mechanism, one where the emissivity is a complex function of density and temperature and where radiative transfer effects substantially modify level populations, such a calculation can only be accomplished by brute force numerical integration. However, we can obtain a result analytically (at least up to a straightforward numerical integral) in two limiting cases, both of which are relevant for a wide range of observational diagnostics. In this section we consider the first of these: transitions with critical densities low enough that we can approximate the level populations as close to local thermodynamic equilibrium (LTE). The most prominent example of this is emission in the low $J$ lines of CO.

\subsection{Line Profiles}
\label{ssec:LTE_lineprof}

As for absorption, consider a line produced by an emitting species for which there is one of that species per mass $m_X$ of gas. The species has energy levels $E_i$ with multiplicity $g_i$, and we are concerned with emission of a line corresponding to a transition between an upper level $u$ and a lower level $\ell$, for which the spontaneous emission rate is $A_{u\ell}$. In LTE, the fraction of the population in level $i$ is given by the usual Boltzmann distribution
\begin{equation}
f_i = \frac{g_i e^{-E_i/k_B T}}{\sum_j g_j e^{-E_j/k_B T}},
\end{equation}
where $T$ is the gas kinetic temperature, which is identical to the excitation temperature $T_{\rm ex}$ for a species in LTE. The energy emission rate per unit gas mass is
\begin{equation}
\frac{\mathcal{L}}{\rho} = \frac{1}{m_X} f_u E_{u\ell} A_{u\ell},
\end{equation}
where $E_{u\ell} = h \nu_0$ is the energy difference between the two levels, and $\nu_0$ is the line centre frequency. Thus the energy emission rate per unit volume for gas at radial distance $a$ is
\begin{equation}
\mathcal{L} = \frac{1}{m_X} f_u E_{u\ell} A_{u\ell} \int_{-\infty}^{x_{\rm crit}} \frac{d\rhomean}{dx} \, dx,
\end{equation}
and making the same change of variables from $x$ to line of sight velocity $u$ as in \autoref{sec:absorb} gives
\begin{equation}
\mathcal{L} = \frac{1}{m_X} f_u E_{u\ell} A_{u\ell} \frac{a}{\sqrt{a^2-\varpi^2}}
\int_{-\infty}^\infty \sum_i \frac{d\rhomean/dx}{|dU_a/dx|} \, du,
\end{equation}
where the sum as usual is over solutions $x_i$ to \autoref{eq:u_x_map}. The specific luminosity at frequency $\nu$ is therefore
\begin{equation}
\mathcal{L}_\nu = \frac{f_u}{m_X} \frac{hc}{v_0} A_{u\ell} \frac{a}{\sqrt{a^2-\varpi^2}}
\sum_i \frac{d\rhomean/dx}{|dU_a/dx|},
\end{equation}
where the final term is to be evaluated at a velocity $u = (c/v_0) (1 - \nu/\nu_0)$.

\begin{table*}
\begin{tabular}{lcccccc}
\hline\hline
Line & $[\mathrm{X}/\mathrm{H}]$ & $m_X$ & $ \Omega$ & $t_X$ & $X^{\rm (thin)}$ & $\alpha^{\rm (thin)}$ \\
& & [g] & & [Myr] & [cm$^{-2}$ / K km s$^{-1}$] & [$M_\odot$ pc$^{-2}$ / K km s$^{-1}$] \\
\hline\hline
CO & $1.1\times 10^{-4}$ & $2.1\times 10^{-20}$ \\ 
$\ldots J=1-0$ & & & $2.19\times 10^{-8}$ & 270 & $3.3\times 10^{18}$ & 0.036 \\
$\ldots J=2-1$ & & & $2.92\times 10^{-8}$ & 180 & $1.4\times 10^{18}$ & 0.015\\
$\ldots J=3-2$ & & & $3.94\times 10^{-8}$ & 160 & $8.5\times 10^{17}$ & 0.0094\\
HCN & $4.0\times 10^{-9}$ & $5.9\times 10^{-16}$ \\
$\ldots J=1-0$ & & & $1.24\times 10^{-5}$ & 7.1 & $1.6\times 10^{20}$ & 1.8 \\
$\ldots J=2-1$ & & & $1.65\times 10^{-5}$ & 4.8 & $6.6\times 10^{19}$ & 0.75\\
$\ldots J=3-2$ & & & $2.23\times 10^{-5}$ & 4.3 & $4.1\times 10^{19}$ & 0.47\\
CS & $6.3\times 10^{-9}$ & $3.7\times 10^{-16}$ \\
$\ldots J=1-0$ & & & $2.95\times 10^{-6}$ & 4.9 & $4.2\times 10^{20}$ & 4.7 \\
$\ldots J=2-1$ & & & $3.93\times 10^{-6}$ & 3.3 & $1.8\times 10^{20}$ & 2.0 \\
$\ldots J=3-2$ & & & $5.30\times 10^{-6}$ & 2.9 & $1.1\times 10^{20}$ & 1.2 \\

\hline
\end{tabular}
\caption{
\label{tab:emission}
Selected emission line parameters; note that $t_X$ here is the same quantity as in \autoref{tab:tA}, but for convenience we use different units in this Table. The CO abundance is set equal to the WIM C abundance taken from Table 9.5 of \citet{draine11a}; other abundances are the M82 values given in Table 7 of \citet{martin06a}. Einstein coefficients and wavelengths are taken from the Leiden Atomic and Molecular Database \citep[\url{http://home.strw.leidenuniv.nl/~moldata}]{schoier05a}.
}
\end{table*}

Because the wind can self-absorb, not all of the light that is emitted at a given point will escape to infinity. Let $\tau_\nu(a)$ be the optical depth at frequency $\nu$ from a given emitting radius $a$ to the observer at $s=-\infty$. With this definition, we can compute the specific intensity of light that an observer receives from the wind by integrating along the line of sight:
\begin{equation}
I_\nu = \frac{r_0}{4\pi} \sum_j \int_{a_{0,j}^\pm}^{a_{1,j}^\pm} \mathcal{L}_\nu e^{-\tau_\nu(a)} \frac{a}{\sqrt{a^2-\varpi^2}}\, da,
\end{equation}
where the factor $a/\sqrt{a^2-\varpi^2}$ accounts for the ratio of path length $ds$ to radial distance $da$. With some algebra, and invoking the relation $\Omega = (g_u/g_\ell) m_e c \lambda_0^2 / (8\pi^2 e^2) A_{u\ell}$ to convert between Einstein coefficient and oscillator strength, we can rewrite this as
\begin{equation}
\label{eq:Inu_LTE}
I_\nu = B_\nu(T) f_\ell \left(1-e^{-E_{u\ell}/k_B T}\right)\left(\frac{t_X}{t_w}\right) \eta(u,\varpiv)
\end{equation}
where
\begin{equation}
\eta(u,\varpiv) = \frac{1}{\zeta_M} \sum_j \int_{a^\pm_{0,j}}^{a^\pm_{1,j}} \frac{1}{a^2-\varpi^2} e^{-\tau_\nu(a)} \sum_i  \frac{p_M}{|dU_a^2/dx|} \, da
\end{equation}
and $B_\nu(T) = (2h\nu^3/c^2)[1/(e^{E_{u\ell}/k_B T}-1)]$ is the usual Planck function. As for absorption of a background source, the characteristic strength of the line is dictated primarily by a timescale $t_X$ that depends only on the abundance of the species and on quantum mechanical constants. We give values of $t_X$ for some lines of interest in \autoref{tab:emission}. 

The remaining step in computing the line emission is to determine the self-absorption optical depth of the wind. Following our calculation in \autoref{sec:absorb}, in the uncorrelated case the optical depth at frequency $\nu$ for an emitting region at radius $a$ to the observer located at $s=-\infty$ is
\begin{equation}
\label{eq:emit_tau_uc}
\tau_\nu^{\rm(uc)} = 
\frac{t_X}{t_w} f_\ell \left(1 - e^{-\frac{E_{u\ell}}{k_B T}}\right) \cdot
\left\{
\begin{array}{ll}
\Phi^{\rm(uc)}(u,\varpiv,a,\infty), & u < 0 \\
\Phi^{\rm(uc)}(u,\varpiv,1,a), & u>0
\end{array}
\right..
\end{equation}
Here the case $u<0$ represents emission coming from the near side of the wind, which is absorbed by material at larger radii, so $\Phi^{\rm(uc)}$ is evaluated from $a$ to $\infty$. The case $u>0$ corresponds to emission from the far side, so absorption is by material at smaller radii and the optical depth is computed from $1$ to $a$.\footnote{Light emitted on the far side must of course pass through the near side too, but since all the near side gas is at velocities $u<0$, and we are neglecting the velocity dispersion of the wind material in comparison to its bulk motion, gas on the near side cannot absorb light from the far side, which has $u>0$. Thus only the far side contributes to the optical depth.}

While self-absorption is exactly the same as absorption of light from a background source in the uncorrelated case, they are not identical in the correlated case. If clouds maintain constant solid angle so that the covering fraction $f_c$ is constant, then in the case of self-absorption the entire emitting region is always covered, rather than only a fraction $f_c$ of the emission as we assumed for absorption of a background source -- the absorbers line up perfectly with the emitting region. Thus for constant $f_c$ (meaning cloud expansion as $y=a^2$) the optical depth is simply
\begin{equation}
\label{eq:emit_tau_c_const_sa}
\tau_\nu^{\rm(c)} = 
\frac{t_X}{t_w} f_\ell \left(1 - e^{-\frac{E_{u\ell}}{k_B T}}\right) \cdot
\left\{
\begin{array}{ll}
\Phi^{\rm(c)}(u,\varpiv,a,\infty), & u < 0 \\
\Phi^{\rm(c)}(u,\varpiv,1,a), & u>0
\end{array}
\right.,
\end{equation}
i.e., the same as in the uncorrelated case (\autoref{eq:emit_tau_uc}), but with $\Phi^{\rm(c)}$ in place of $\Phi^{\rm(uc)}$. Since $\Phi^{\rm(c)}$ is larger than $\Phi^{\rm(uc)}$ by a factor of $1/f_c$, this will produce higher absorption than in the uncorrelated case, which is the opposite of the effect correlation has on background absorbers. However, this is exactly as we should expect: if the gas is clumped together in angle, it will be less effective at blocking light emitted by a randomly-placed background source, but more effective at blocking light emitted by the gas itself.

If clouds do not maintain constant solid angle, then there is an asymmetry between emission from the near and far sides. Recall that we have assumed that the covering factor is non-increasing with radius. This means that light emitted on the far side ($u>0$) is emitted in a region of lower covering factor and then moves through a region of higher covering factor on its way to the observer. If the wind is perfectly correlated, this means that the emitting region will be fully covered at all radii, and so the optical depth is the same as for the case of constant $f_c$, i.e., it is the value given by \autoref{eq:emit_tau_c_const_sa}. In contrast, light emitted on the near side ($u<0$) is emitted in a region of high covering factor and then moves through a region of lower covering factor as it propagates. In analogy to \autoref{eq:tau_c_var_sa}, this produces an attenuation factor $e^{-\tau}$ that is an area-weighted average of the attenuation factors experienced by different portions of the emitting region,
\begin{eqnarray}
\lefteqn{
\tau_\nu^{\rm(c)} = -\ln \left\{\frac{1}{f_c(a)}
\int_a^\infty \left|\frac{df_c}{da'}\right| \cdot {}
\right.
}
\nonumber \\
& &
\left.
\exp\left[-\frac{t_X}{t_w} f_\ell \left(1 - e^{-\frac{h\nu_0}{k_B T}}\right)
\Phi^{\rm(c)}(u,\varpiv,a,a')\right]
\right\} \, da'.
\label{eq:emit_tau_c_var_sa}
\end{eqnarray}

\begin{figure}
\includegraphics[width=\columnwidth]{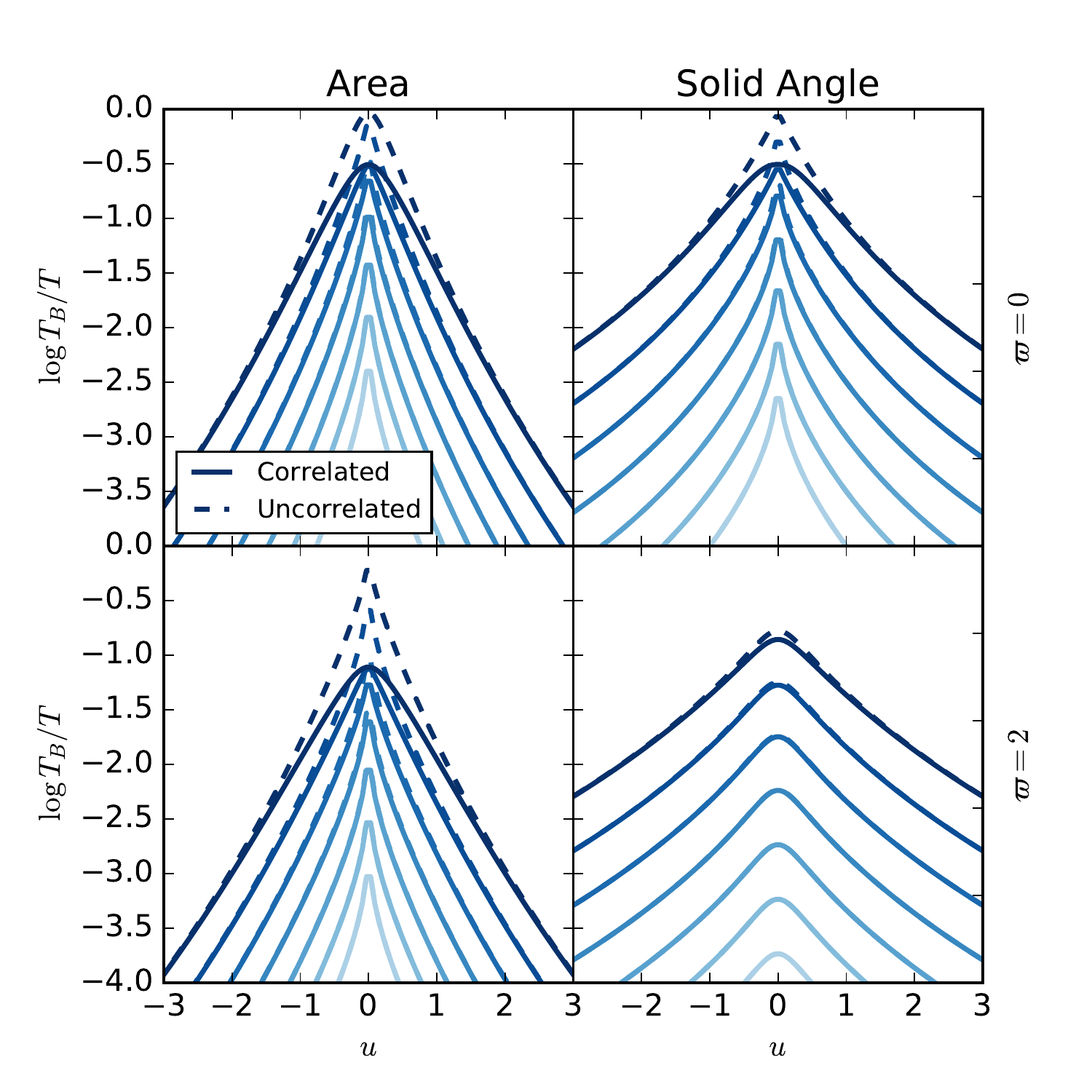}
\caption{
\label{fig:CO_line}
Ratio of brightness temperature $T_B$ to gas kinetic temperature $T$ versus normalised velocity $u$ for CO $J=1-0$ emission from a spherical, ideal wind with $\Gamma = 0.2$, $\mathcal{M} = 50$. The top two panels show impact parameter $\varpi=0$, and the bottom two show $\varpi=2$. The left column shows winds with constant area clouds, while the right shows winds with constant solid angle. Lines show results for values of $t_X/t_w = 0.1$ (lightest) to $t_X/t_w = 100$ (darkest) in steps of 0.5 dex. As in \autoref{fig:absorption}, solid lines show the correlated case, and dashed lines show the uncorrelated case.
}
\end{figure}

We show example LTE line emission profiles in \autoref{fig:CO_line}; rather than plotting intensity directly, we plot brightness temperature $T_B$, the temperature of a blackbody that produces the specified intensity at a given frequency. The figure shows the results for the CO $J=1-0$ line for gas at a temperature of $T=50$ K, which gives $f_\ell = 0.05$ and $E_{u\ell}/k_B T = 0.11$. As in \autoref{fig:absorption}, we adopt $f_w = \zeta_A$ for the correlated case.  Note that there is a small asymmetry between the near ($u<0$) and far ($u>0$) side line emission, as expected based on the above discussion, but for the parameters shown here it is small enough that it is not easily visible on the plot.

\subsection{Integrated Intensity and the ``$X$" Factor}
\label{sssec:xfactor}

\begin{figure}
\includegraphics[width=\columnwidth]{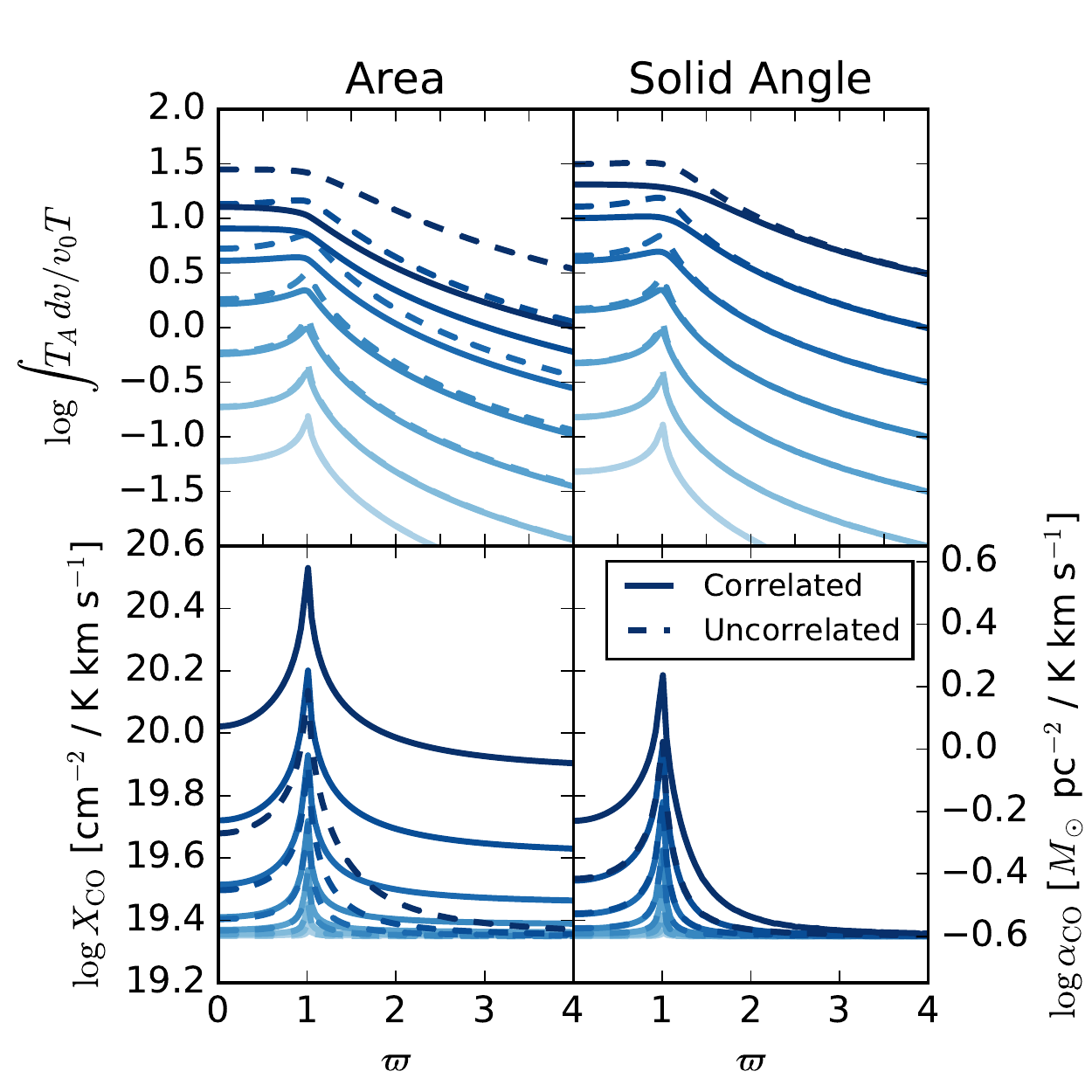}
\caption{
\label{fig:CO_line_int}
Velocity-integrated brightness temperature normalised to $v_0 T$ (top row) and CO to H$_2$ conversion factor $X_{\rm CO}$ or $\alpha_{\rm CO}$ (bottom row), both plotted as a function of projected radius $\varpi$. All quantities plotted are for CO $J=1-0$ emission from a spherical, ideal wind with $\Gamma = 0.2$, $\mathcal{M} = 50$ and a temperature of 50 K. The left column shows winds with constant area clouds, while the right shows winds with constant solid angle. Lines show results for values of $t_X/t_w = 0.1$ (lightest) to $t_X/t_w = 100$ (darkest) in steps of 0.5 dex. As in \autoref{fig:CO_line}, solid lines show the correlated case, and dashed lines show the uncorrelated case.
}
\end{figure}

It is also of interest to examine the velocity- or frequency-integrated intensity, and the closely related ``$X$" factor used to convert an integrated intensity to a gas column (or equivalently an $\alpha$ factor used to convert to a mass per unit area). We can obtain the integrated intensity by integrating the line profile (\autoref{eq:Inu_LTE}) over frequency, giving
\begin{equation}
\label{eq:inu_int}
\int I_\nu\, d\nu = B_\nu(T) f_\ell \left(1-e^{-E_{u\ell}/k_B T}\right) \left(\frac{t_X}{t_w}\right) \left(\frac{v_0}{\lambda_0}\right) \Psi(\varpiv)
\end{equation}
where
\begin{equation}
\Psi(\varpiv) = \int_{-\infty}^\infty \eta(u,\varpiv)\, du.
\end{equation}
We may write the velocity-integrated antenna temperature (with respect to which $X$ is more commonly defined) corresponding to \autoref{eq:inu_int} as
\begin{equation}
\int T_A \, dv = T_{u\ell} v_0 f_\ell \left(1-e^{-E_{u\ell}/k_B T}\right) \left(\frac{t_X}{t_w}\right) \Psi(\varpiv),
\end{equation}
where $T_{u\ell} = E_{u\ell}/k_B$. This quantity can be compared to the total gas column density along a line of sight to compute an $X$ factor. The total hydrogen column per unit area is given by
\begin{equation}
N_{\rm H} = \frac{v_0}{4\pi G \mu_{\rm H} m_{\rm H}} \frac{1}{t_w} \Psi^{\rm (thin)}(\varpiv)
\end{equation}
where $\mu_{\rm H}$ is the mean mass per H nucleus in the gas, and we have defined $\Psi^{\rm(thin)}(\varpiv)$ as $\Psi(\varpi)$ with $\tau_\nu = 0$, i.e., $\Psi^{\rm(thin)}(\varpiv)$ is simply the function $\Psi(\varpiv)$ in the limit that all photons escape. Combining the antenna temperature and the column, the $X$ factor for any given transition is
\begin{equation}
X = \frac{N_{\rm H}}{\int T_A\, dv} = \frac{X^{\rm (thin)}}{f_u} \frac{\Psi^{\rm (thin)}(\varpiv)}{\Psi(\varpiv)},
\end{equation}
where $f_u$ is the fraction of the population in state $u$, and
\begin{equation}
X^{\rm (thin)} = \frac{8\pi}{A_{u\ell} T_{u\ell} \lambda_0^3 \left[\mathrm{X}/\mathrm{H}\right]}
\end{equation}
depends only on the abundance $[\mathrm{X}/\mathrm{H}]$ of the species in question and the quantum mechanical constants for the line. The combination $X^{\rm (thin)}/f_u$ is the $X$ factor if the gas is optically thin. The equivalent $\alpha$ factor is
\begin{equation}
\alpha = \frac{\alpha^{\rm (thin)}}{f_u} \frac{\Psi^{\rm (thin)}(\varpiv)}{\Psi(\varpiv)}
\end{equation}
with
\begin{equation}
\alpha^{\rm (thin)} = \frac{8\pi m_X}{A_{u\ell} T_{u\ell} \lambda_0^3}.
\end{equation}
We give values of $X^{\rm (thin)}$ and $\alpha^{\rm (thin)}$ for some commonly-observed lines in \autoref{tab:emission}. \autoref{fig:CO_line_int} shows example calculations of integrated intensity and $X_{\rm CO}$ for CO $J=1-0$ emission, using the same parameters as shown in \autoref{fig:CO_line}. Note that formally $X_{\rm CO}$ diverges at $\varpi = 1$ exactly; this is an artefact of assuming that the gas velocity is exactly 0 at $a = 1$.

\section{Line Emission by Winds: Optically Thin, Subcritical Lines}
\label{sec:emit_subcrit}

The other case for which we can compute emission analytically is for lines with critical densities significantly higher than the mean density for in the wind, and where radiative trapping effects are unimportant; examples of this include collisionally excited emission in the C~\textsc{ii} 158 $\mu$m line and recombination radiation in Balmer and Paschen lines.

\subsection{Emissivity and the Microscopic Density}

For a subcritical line produced by collisions between two partners with number densities $n_1$ and $n_2$, the emission rate per unit volume is proportional to $n_1 n_2$.  Let the total rate of energy emission per unit volume by the gas be
\begin{equation}
\label{eq:Lambda_thin}
\mathcal{L} = \Lambda\left(\frac{\rho}{m_{\rm H}}\right)^2,
\end{equation}
where $\rho$ is the gas density, $\Lambda$ is a cooling constant for the line, with units of energy per unit volume per unit time, and $m_{\rm H}$ is the gas mass per H nucleus. With this definition, the cooling rate constant $\Lambda$ is defined such that it is the rate of radiative emission per H nucleus squared. It depends on the abundances of the species in question and on the rate coefficients describing line emission; specifically, for a line produced by a collision or reaction between two species with abundances $[\mathrm{X}_1/\mathrm{H}]$ and $[\mathrm{X}_2/\mathrm{H}]$ and a rate coefficient $k_{12}$, the cooling constant is
\begin{equation}
\Lambda = \left[\frac{\mathrm{X}_1}{\mathrm{H}}\right] \left[\frac{\mathrm{X}_2}{\mathrm{H}}\right] k_{12} \frac{hc}{\lambda_0},
\end{equation}
where $\lambda_0$ is the wavelength for the transition in question. We limit our calculation to cases where we can treat $\Lambda$ as at least approximately constant over the radiating region of the wind. Examples of $\Lambda$ for some lines of interest are given in \autoref{tab:low_ncrit_lines}. Note that our assumption of constant $\Lambda$ is not valid in cases where phase changes within the wind significantly influence its emission; for example, if cool clouds capable of producing C~\textsc{ii} emission only condense out of a hot phase at radii $a \gg 1$, as happens for example in the model of \citet{thompson16b}, our simple calculation cannot capture this effect without including the additional phase information.

\begin{table}
\begin{tabular}{lccc}
\hline\hline
Line & Species & $[\mathrm{X}_{1,2}/\mathrm{H}]$ & $\Lambda$ [erg cm$^3$ s$^{-1}$] \\ 
\hline\hline
[C~\textsc{ii}] 158 $\mu$m & C$^+$, H & $1.1\times 10^{-4}$, 1 & $1.4\times 10^{-28}$ \\
$[\mathrm{O~\textsc{i}}]$
& O, H & $4.6\times 10^{-4}$, 1 \\
$\ldots$ 63 $\mu$m & & &  $1.5\times 10^{-28}$ \\
$\ldots$ 145 $\mu$m  & & & $1.0\times 10^{-29}$ \\
H$\alpha$ & H$^+$, $e^{-}$ & 1, 1.1 & $3.9\times 10^{-25}$ \\
H$\beta$ & H$^+$, $e^-$ & 1, 1.1 & $1.4\times 10^{-25}$ \\
\hline
\end{tabular}
\caption{
\label{tab:low_ncrit_lines}
Parameters for selected low critical density lines. Species lists the species whose collision produces the line. Abundances for C~\textsc{ii} and O~\textsc{i} are the CNM C and O abundances given in Table 9.5 of \citet{draine11a}, while those for H$\alpha$ and H$\beta$ assume fully ionised hydrogen and singly-ionised helium. Rate coefficients are taken from the Leiden Atomic and Molecular Database \citep[\url{http://home.strw.leidenuniv.nl/~moldata}]{schoier05a} at a temperature of 200 K for the low far-IR forbidden lines; the underlying atomic data are from \citet{launay77a} and \citet{barinovs05a} for C$^+$ and \citet{abrahamsson07a} for O. For H$\alpha$ and H$\beta$ we use the case B effective $\alpha$ values at a temperature of $10^4$ K and a density of $10^3$ cm$^{-3}$ taken from Table 14.2 of \citet{draine11a}.
}
\end{table}

The density $\rho$ that appears in \autoref{eq:Lambda_thin} is in general not the same as the mean density $\rho_{\rm mean}$ computed in \autoref{ssec:meandensity}. Knowledge of the mean density is sufficient to compute absorption and LTE  emission profiles, because for both of these processes the absorption and emission rates per unit mass are constant. For subcritical lines, however, the emissivity per unit mass is not constant, and depends on the local, microscopic density. This is
much more uncertain than the mean density, because it depends on the
detailed interaction between clouds and their surroundings in the wind. For example,
as clouds are accelerated by a hot wind, they may also be crushed. This will not
change the mean density in the wind, but it will change the local density
and thus enhance the rate of emission in subcritical lines.

Given this complexity, our approach is to parameterise the problem. We can
set a lower limit on the microscopic density by considering the case where
the clouds are not compressed at all. This amounts to computing the density
for clouds that have initial column
$\Sigma_0/\bar\Sigma_0 =e^x$ and area ratio $A/A_0 = y$
(where $A_0$ is the area occupied by the cloud when it is at radius $r_0$,
so $y = \Sigma_{0,c}/\Sigma_c = A/A_0$, assuming constant cloud mass)
by assuming that the volume filling fraction at $r/r_0=a$ is equal to the solid
angle filling fraction.  If the material has a log-normal distribution
of area at radius $r_0$, the initial differential area occupied 
by material with column density in the range $x$ to $x+dx$ is
\begin{equation}
\frac{dA_0}{dx} = 4\pi r_0^2 p_A
\end{equation}
where
\begin{equation}
p_A = \frac{1}{\sqrt{2\pi \sigma^2_x}}\exp\left[-\frac{\left(x+\sigma_x^2/2\right)^2}{2\sigma_x^2}\right].
\end{equation}
The cross-sectional area of the same material when it reaches radius $r$ is simply $dA = y\, dA_0$, and the fraction of the total solid angle is
$df_\Omega = dA/(4\pi r^2)=(y p_A/a^2)dx$.  The
solid angle PDF is then 
\begin{equation}\label{eq:pOmega}
\frac{df_\Omega}{dx}\equiv p_\Omega=\frac{y}{a^2} p_A ,
\end{equation}
where this represents the fraction of the total solid angle containing clouds
with this $x$.
Since we have the mapping $U_a$ between $u$, $x$, and $a$, \autoref{eq:pOmega}
implicitly defines the differential covering fraction
as a function of $u$ and $a$.

The lower limit on microphysical density comes from
assuming that $df_\Omega/dx$ also characterises the volume filling factor of gas at
density $x$. In this case, the microphysical density of gas characterised by $x$,
when it reaches radius $a$, is then given by the
ratio of the contribution of that gas to the mean density,
$d\rho_{\rm mean}/dx$ (\autoref{eq:rhomean}), to its contribution to the volume
 $df_\Omega/dx=p_\Omega$:
\begin{equation}\label{eq:rhomicromin}
\rho_{\rm micro,min}(x,a)= \rho_{\rm norm} \left(\frac{p_M}{p_A}\right) \frac{1}{U_a y}.
\end{equation}
Given this result, we choose to write the actual microphysical density as
\begin{equation}
\rho_{\rm micro}(x,a) =  c_\rho \rho_{\rm norm} \left(\frac{p_M}{p_A}\right) \frac{1}{U_a y},
\end{equation}
where $c_\rho$ is a clumping factor whose value is $\geq 1$. If $c_\rho$ does not
vary strongly with position or velocity, then it only provides an overall scaling to the
emission profile, without changing its shape.

\subsection{Line Profiles}

Now that we have computed the microphysical density, we are prepared to compute the line
profile. Following the same approach as in \autoref{ssec:LTE_lineprof}, the energy emission
rate per unit volume for gas at radial distance $a$ is
\begin{equation}
\mathcal{L} = \frac{\Lambda}{m_{\rm H}} \int_{-\infty}^{x_{\rm crit}} \frac{\rho_{\rm micro}}{m_{\rm H}} \frac{d\rho_{\rm mean}}{dx}\, dx
\end{equation}
and the specific luminosity at frequency $\nu$ is therefore
\begin{equation}
\mathcal{L}_\nu = \frac{\Lambda}{m_{\rm H}} \frac{\lambda_0}{v_0} \frac{a}{\sqrt{a^2-\varpi^2}}
\sum_i \frac{\rho_{\rm micro}}{m_{\rm H}} \frac{d\rho_{\rm mean}/dx}{|dU_a/dx|}.
\end{equation}
The intensity received by an observer at $s=-\infty$, assuming that the gas is optically thin, is
\begin{equation}
\label{eq:I_nu_thin}
I_\nu = \left[\frac{c_\rho}{18 \pi} \Lambda \left(\frac{\rho_0}{m_{\rm H}}\right)^2 r_0\right]
 \frac{\lambda_0}{v_0} \left(\frac{t_c}{t_w}\right)^2 \Xi(u, \varpi)
\end{equation}
with
\begin{eqnarray}
\rho_0 & = & \frac{3 M_0}{4\pi r_0^3} \\
t_c & = & \frac{r_0}{v_0}
\end{eqnarray}
and
\begin{eqnarray}
\lefteqn{\Xi(u,\varpi) =  \frac{1}{\zeta_M^2} \cdot {}}
\nonumber \\
& & \sum_j  \int_{a^\pm_{0,j}}^{a^\pm_{1,j}} \frac{1}{ya(a^2-\varpi^2)}
\sum_i \frac{p_M^2}{p_A} \frac{1}{U_a \left|dU_a^2/dx\right|} \, da.
\end{eqnarray}
We can provide a physical interpretation to \autoref{eq:I_nu_thin} as follows. The first term, in square brackets, is up to factors of order unity the characteristic intensity we would expect to be produced in the region from which the wind is launched, since it involves the emissivity multiplied by the square of the number density multiplied by the size scale $r_0$. The term $\lambda_0/v_0$ is the characteristic frequency width of the emission. Thus the product of these two terms should roughly describe the intensity of emission at $u = 0$, where emission is presumably dominated not by the wind but by gas in the emitting region.\footnote{In fact, for some wind acceleration laws $\Xi(u,\varpi)$ diverges as $u\rightarrow 0$. This divergence is a result of a breakdown in the large velocity gradient approximation we have adopted. One cannot neglect the thermal velocity dispersion of the wind material at velocities near zero. However, this issue only affects velocities smaller than the thermal velocity dispersion, i.e., $\lesssim 10$ km s$^{-1}$.} The term $t_c/t_w$ is the ratio of the crossing time to the wind ejection time; it provides a dimensionless measure of how quickly the wind is evacuating the launching region, and thus how much mass it contains at any given time. Finally, $\Xi(u,\varpi)$ is a dimensionless function that determines the precise shape of the emission in frequency. This term together with $t_c/t_w$ determines the brightness of the line wings, which come from the wind, will be in comparison to the core of the line, which is dominated by the wind-launching region.

\begin{figure}
\includegraphics[width=\columnwidth]{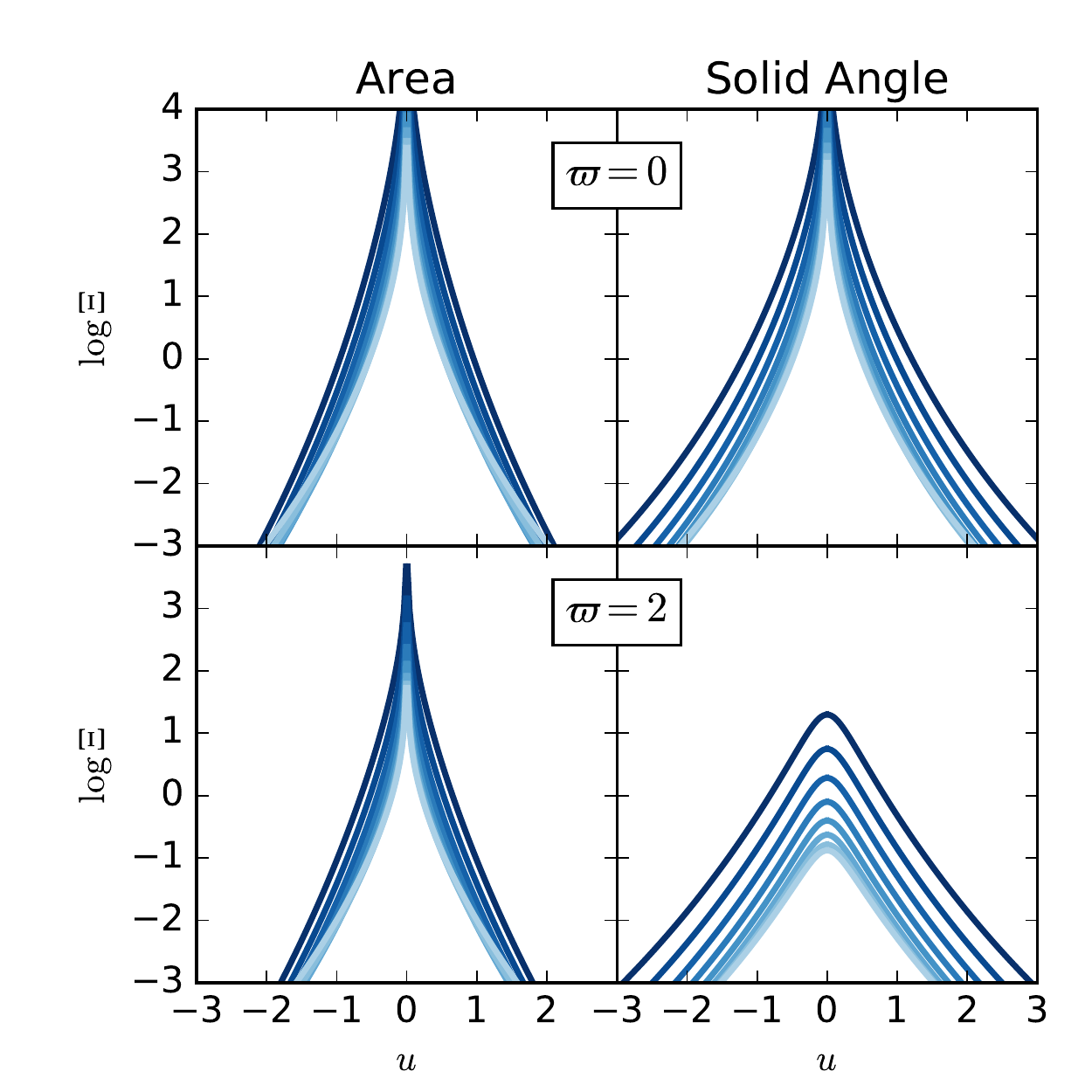}
\caption{
\label{fig:subcritical}
The optically thin subcritical emission function $\Xi(u,\varpi)$ (\autoref{eq:I_nu_thin}) versus line of sight velocity $u$. All calculations show an ideal, spherical wind with $\mathcal{M} = 50$. The top two panels show impact parameter $\varpi=0$, and the bottom two show $\varpi=2$. The left column shows winds with constant area clouds, while the right shows winds with constant solid angle. Lines show a range of values of $\Gamma$, from $\log\Gamma = 0$ (darkest) to $\log\Gamma = -2$ (lightest), in steps of 0.25 dex.
}
\end{figure}

We show an example calculation of optically thin line emission in \autoref{fig:subcritical}. Note that we plot only the shape function $\Xi(u,\varpi)$, because for an optically thin subcritical wind the intensity is simply this function multiplied by a dimensional constant.

The corresponding frequency-integrated intensity is
\begin{equation}
\label{eq:int_intensity}
\int I_\nu \, d\nu = \frac{r_0}{4\pi}\sum_j \int_{a_{0,j}}^{a_{1,j}} \int_{-\infty}^{x_{\rm crit}}
\mathcal{L} \frac{a}{\sqrt{a^2-\varpi^2}} \, dx \, da
\end{equation}
With a bit of algebra, this reduces to
\begin{equation}
\int  I_\nu \, d\nu = \left[\frac{c_\rho}{36 \pi} \Lambda \left(\frac{\rho_0}{m_{\rm H}}\right)^2 r_0\right] \xi(u, \varpi)
\end{equation}
where
\begin{equation}
\xi(\varpi) =  \frac{1}{\zeta_M^2} \sum_j  \int_{a^\pm_{0,j}}^{a^\pm_{1,j}} \frac{1}{ya\sqrt{a^2-\varpi^2}}
\int_{-\infty}^{x_{\rm crit}} \frac{p_M^2}{p_A} \frac{1}{U_a^2}\, dx \, da.
\end{equation}
However, we caution that this quantity is undefined for any location in the flow where $U_a^2 = 0$ at finite $x$. This is because the assumption to steady state must break down in such cases. However, this divergence does not prevent $\Xi$ from being well-defined at velocities $u > 0$.

\section{Case Study: An M82-Like Starburst}
\label{sec:example}

In this section we demonstrate the applications of our formalism by computing a range of observable properties for the cool wind from a galaxy whose bulk physical parameters are chosen to match those of the dwarf starburst M82. In the process we will be able to make a number of remarks about how the observable properties of wind absorption and emission can be connected to the physical properties of the underlying wind. We caution that in this section we are not attempting to match the observed emission of M82 directly; such an effort would require significant model fitting, a task we leave to the next paper in this series. Our goal here is merely to demonstrate that our model, though simple, can reproduce the major qualitative features of a real wind.

\subsection{Physical Parameters for the Model}

The starbursting centre of M82 has a diameter of $\approx 29''$ \citep{kennicutt98a}, corresponding to a radius $r_0\approx 250$ pc. The circular speed at the edge of this region is $\approx 120$ km s$^{-1}$ \citep{greco12a}, so we take $v_0$ to be a factor of $\sqrt{2}$ larger, $v_0 = 170$ km s$^{-1}$. (Recall that $v_0$ is the escape speed just considering material interior to radius $r_0$ -- the true escape speed for a realistic halo will be substantially larger.) The corresponding mass is $M_0 = 8.2\times 10^8$ $M_\odot$, which agrees well with the dynamical mass measured by \citet{forster-schreiber01a}. We assume that the potential is isothermal. The wind may be driven either by radiation pressure \citep{coker13a} or by the hot gas, which has a terminal velocity $v_h \approx 1400 - 2200$ km s$^{-1}$ \citep{strickland09a}, corresponding to $u_h \approx 10$. For the purposes of this example we assume the latter and therefore use hot wind model with $u_h = 10$. For our fiducial model we assume constant solid angle clouds.

To make a model of this system, we require knowledge of the Mach number $\mathcal{M}$ and the Eddington ratio $\Gamma$. For the former, we note that the gas velocity dispersion in the disc is measured to be $\approx 40$ km s$^{-1}$ \citep{leroy15b}, and the molecular gas temperature within the galaxy is $\approx 50$ K or more \citep{wild92a}; combining these gives $\mathcal{M} \sim 100$. For the latter, we assume that the isotropic mass loss rate $\dot{M}$, fraction of area from which the wind escapes $f_A$, and the Eddington ratio $\Gamma$ are related as predicted by the  \citetalias{thompson16a} model, which gives the mass loading factor $\eta$ as a function of the Eddington ratio. Specifically
\begin{equation}
\eta = \frac{f_A \dot{M}}{\dot{M}_*} = \frac{\zeta_M}{\epsilon_{\rm ff}},
\label{eq:eta}
\end{equation}
where $\epsilon_{\rm ff}\approx 0.01$ is the star formation rate per free-fall time \citep[and references therein]{krumholz14c} and $\dot{M}_* = 4.1$ $M_\odot$ yr$^{-1}$ is M82's star formation rate \citep{kennicutt98a} adjusted to a \citet{chabrier05a} IMF. Since $\zeta_M$ is a function of $\Gamma$ and $\mathcal{M}$, this relation enables us to derive $\Gamma$ for any value of $\dot{M}$. As a fiducial parameter we adopt an isotropic mass loss rate $\dot{M} = 100$ $M_\odot$ yr$^{-1}$ and an area fraction $f_A = 0.22$ (see below), which corresponds to a total mass loss rate of 22 $M_\odot$ yr$^{-1}$, consistent with the recent estimate of \citet{leroy15b}. Formally, we note that this should be interpreted as the mass loss rate only in a single chemical phase, either ionised or molecular, so that the actual mass loss rate incorporating all phases is twice as large.

Finally, we require knowledge of the geometry. Observations of line splitting strongly suggest that the wind geometry is biconical, at least for the cool phase, with a moderate opening angle and a small inclination relative to the plane of the sky \citep[e.g.,][]{heckman90a, mckeith93a, mckeith95a, shopbell98a, coker13a, leroy15a}. For the purposes of our example we use $\phi = -5^\circ$, $\theta_{\rm in} = 30^\circ$, and $\theta_{\rm out} = 50^\circ$, which are within the range of the published estimates. The fraction of available area through which the wind is ejected is $f_A = \cos \theta_{\rm in} - \cos \theta_{\rm out} = 0.22$. Again, we caution that our goal is not to reproduce the observations precisely, but rather to reproduce them qualitatively, so that a precise match can be used to constrain the parameters of the model.

\subsection{The Fiducial Model}

\begin{figure*}
\includegraphics[width=0.8\textwidth]{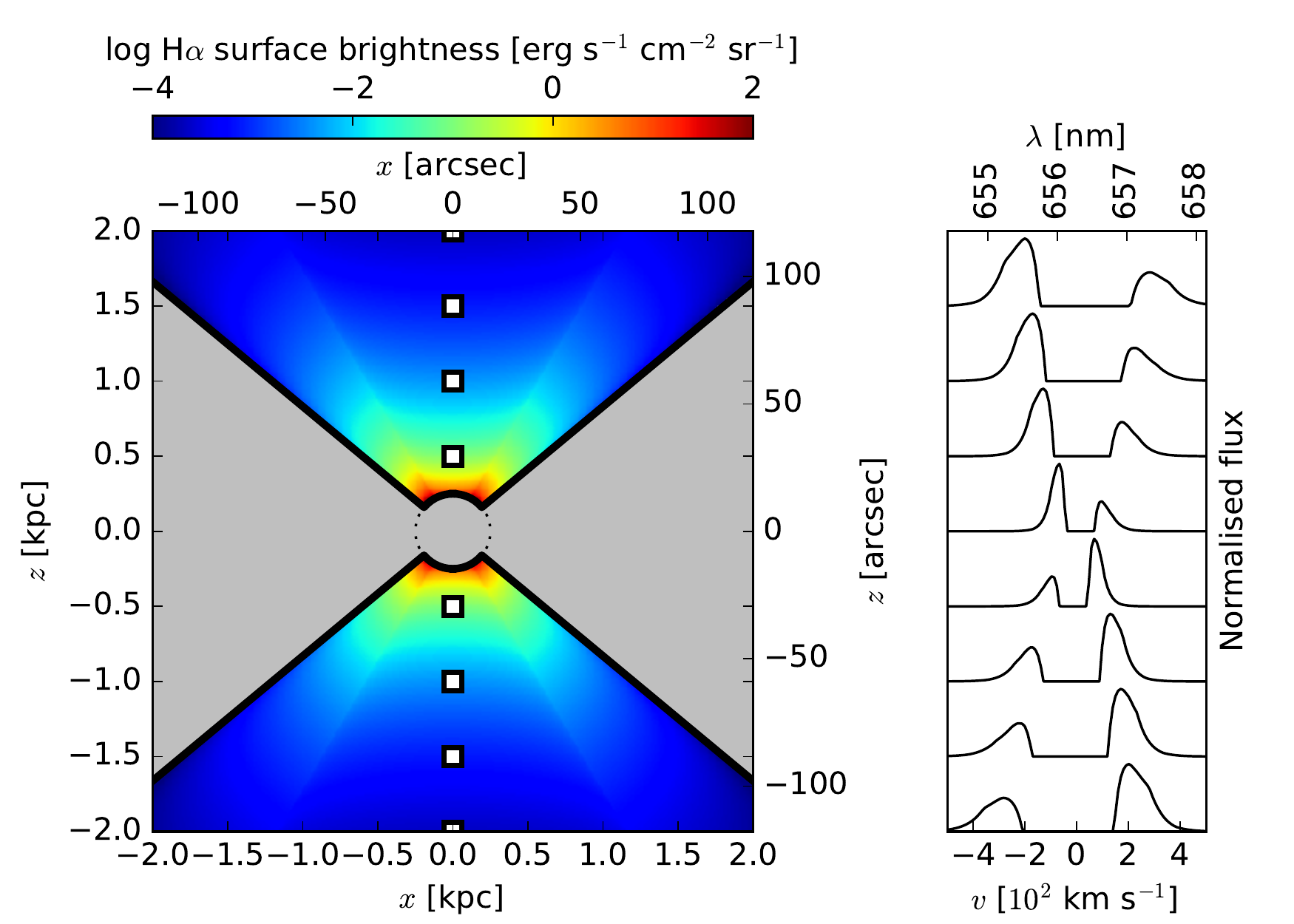}
\caption{
\label{fig:m82_Ha}
Computed H$\alpha$ emission from our M82-like galaxy. The left panel shows the velocity-integrated surface brightness of the line, as a function of projected distance from the wind axis, which lies in the $xz$-plane. The angular distances given on the top and right axes are for an assumed distance of 3.5 Mpc. The grey region indicates lines of sight that do not intersect the wind, and the central circle is the wind launching region. The right panel shows H$\alpha$ line spectra measured at eight positions evenly along the wind axis, starting at $z=-2$ kpc (bottom) and increasing to $z=2$ kpc (top) in steps of 0.5 kpc (skipping $z=0$, where there is no wind). These positions marked with white squares in the left panel.
}
\end{figure*}

We first examine a fiducial model with the parameters described in the previous section, before considering how the results would change if we were to alter some of the more uncertain of those parameters. We show the H$\alpha$ emission expected from our wind in \autoref{fig:m82_Ha}. This computation uses the value of $\Lambda$ given in \autoref{tab:low_ncrit_lines}, and a clumping factor $c_\rho = 1$, and thus should be regarded as a lower limit on the true emission. We note that the model produces clear line splitting and limb-brightening features as a result of the biconical geometry. The median velocity shifts from negative to positive as one moves upward along the wind cone, and there is an asymmetry between the positive and negative velocity portions of the spectra. Both of these features are a result of the tip in the wind cone, i.e., the fact that $\phi \neq 0$.

\begin{figure*}
\includegraphics[width=0.8\textwidth]{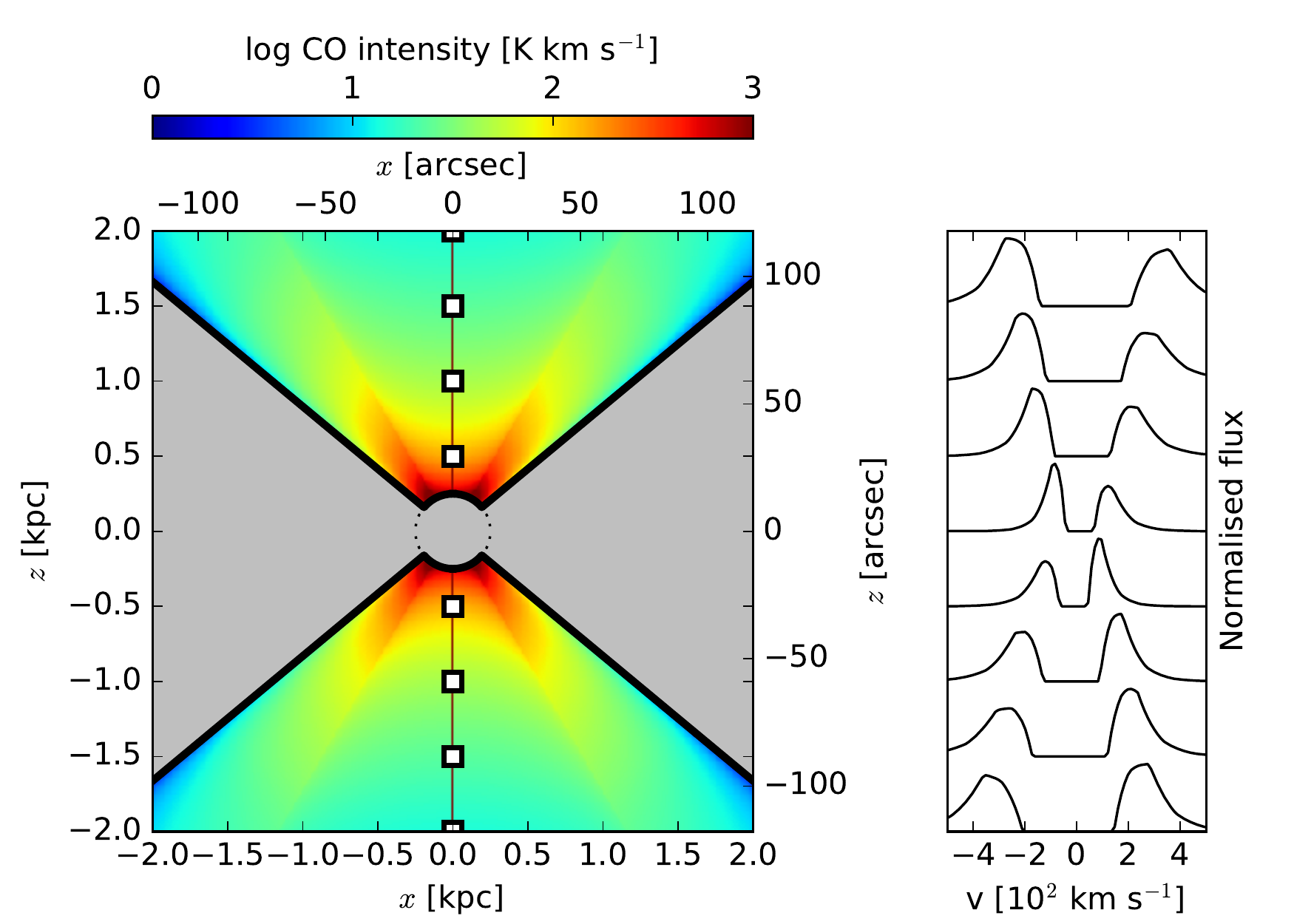}
\caption{
\label{fig:m82_CO}
Same as \autoref{fig:m82_Ha}, for for the CO $J=1-0$ line rather than the H$\alpha$ line. The intensity shown is the velocity-integrated antenna temperature.
}
\end{figure*}

\autoref{fig:m82_CO} shows a map of CO $J=1-0$ emission for our model, in the same form as that for H$\alpha$ shown in \autoref{fig:m82_Ha}.  For this computation we use a gas temperature of 50 K, and the value of $t_X$ shown in \autoref{tab:emission}. Qualitatively the result is similar to that we obtain for the H$\alpha$, including the effects of limb brightening and line splitting. The overall intensity is comparable to that measured for M82 by \citet{leroy15b}.

\begin{figure}
\includegraphics[width=\columnwidth]{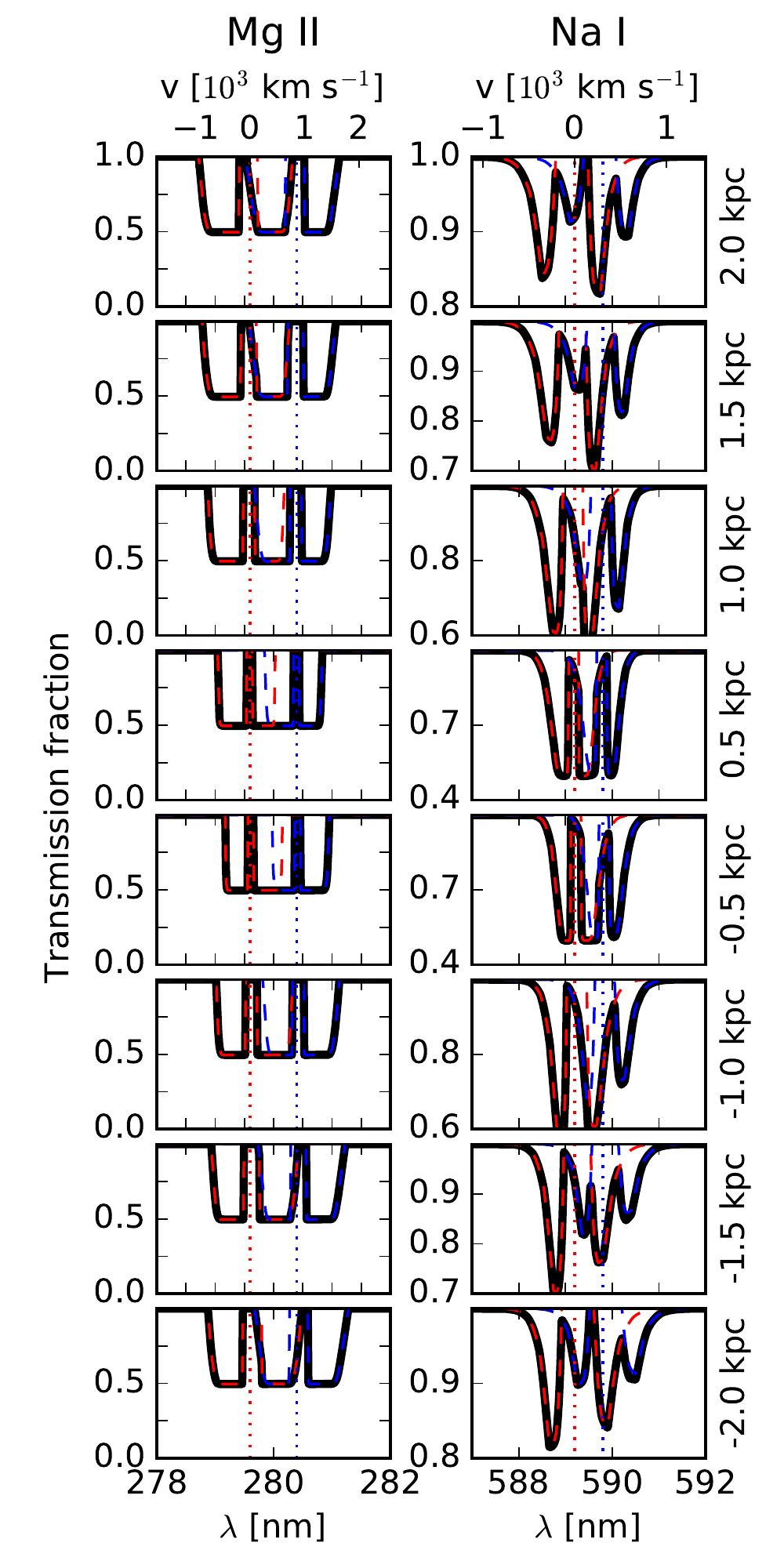}
\caption{
\label{fig:m82_odepth}
Transmission fraction versus wavelength for our M82-like galaxy in the Mg~\textsc{ii} $\lambda\lambda 2796, 2804$ and Na~\textsc{i} $\lambda\lambda 5892, 5898$ doublets. Black lines show the total transmission fraction, while red and blue lines, respectively, show the effects of the shorter and longer wavelength transitions of the doublet by themselves. As indicated on the right, rows show a range of positions along the wind central axis, from $z=-2$ kpc (bottom) and increasing to $z=2$ kpc (top) in steps of 0.5 kpc (skipping $z=0$, where there is no wind). These positions are the same as those used for the H$\alpha$ emission spectra shown in \autoref{fig:m82_Ha}. Note that all Mg~\textsc{ii} spectra use the same $y$-axis scale, but for Na~\textsc{i} the scales are different in different panels. The central wavelengths for the two transitions of the doublets are marked by dotted vertical lines. The velocity scale shown on the top axis is relative to the lower frequency of the two transitions.
}
\end{figure}

In \autoref{fig:m82_odepth} we show absorption spectra for the Mg~\textsc{ii} $\lambda\lambda 2796, 2804$ and Na~\textsc{i} $\lambda\lambda 5892, 5898$ doublets, computed using the same model and measured at a range of distances along the wind axis. We use $f_w = \zeta_A$ for the wind filling fraction. For this computation, we use the value of $t_X$ (and thus the abundance, depletion factor, and ionisation correction) given for these doublets in \autoref{tab:tA}, and we assume the correlated case. We also assume that the source being absorbed is behind the wind, since in the geometry of M82 (and unlike the case for many high-redshift observations) the wind cone does not cover the disk of the galaxy.

We see that the Mg~\textsc{ii} doublet, due to its very large strength, produces deep absorption features broad enough that the two doublet transitions overlap. Moreover, because the optical depth along obscured lines of sight greatly exceeds unity even for the weaker doublet transition, the absorption features from the two transitions are of nearly equal strength. Absorption is not total only because the wind filling factor is less than unity. In contrast, the much weaker Na~\textsc{i} transition is not opaque along most lines of sight. This produces a clear difference in the strength of the two transitions, particularly at larger elevations where the overall optical depth is smaller. For both species the spectral shape is complex, as a result of the combination of there being two physical components to the wind (the near and far sides of the wind cone) as well as two transitions, with comparable velocity shifts between the physical components and the transitions.

\subsection{Variations: Dependence on Cloud Expansion, Potential, Mass Outflow Rate}
\label{ssec:variations}

Having considered the fiducial model, we now turn to the question of how the predictions of the model depend on three key parameters: the rate at which clouds expand as they are accelerated by the wind, the shape of the potential, and the overall mass flux of the wind. The latter is generally the quantity of greatest interest, while the other two quantities are essentially a nuisance parameters that, we as shall see, contributes significantly to our uncertainty in the outflow rate. This exercise will allow us to determine how well these quantities can be deduced from observations of the spectrum. (The geometric parameters will also of course affect the wind, but for simplicity we treat them as fixed here.)

For the purposes of this test, we consider isotropic mass outflow rates with a $\pm 1$ dex range about our fiducial estimate $\dot{M} = 100$ $M_\odot$ yr$^{-1}$, and we consider not just the fiducial case of constant solid angle and an isothermal potential, but also the cases of constant area and intermediate expansion, in both point and isothermal potentials. For simplicity we focus on spectra taken 1 kpc above the disk, along the central outflow axis; results for other positions yield qualitatively similar conclusions.

\begin{figure}
\includegraphics[width=\columnwidth]{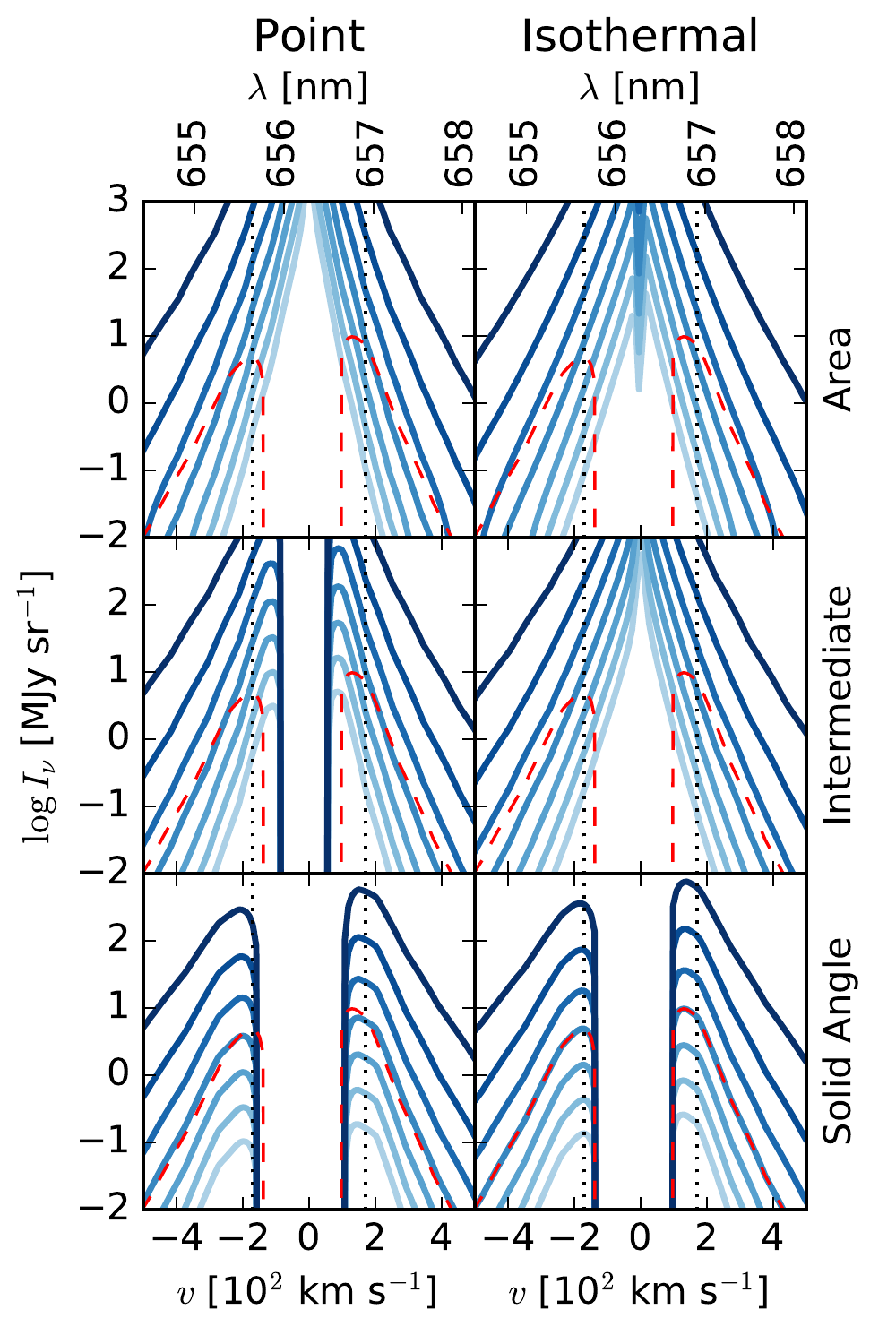}
\caption{
\label{fig:m82_Ha_var}
H$\alpha$ intensity measured at a position 1 kpc above the disc, along the outflow central axis, for our M82-like model. The left and right columns show the results for point and isothermal potentials, respectively, while the rows from top to bottom show the results for clouds with constant area, intermediate expansion, and constant solid angle. In each panel, solid lines show the intensity for isotropic mass outflow rates from $10 - 1000$ $M_\odot$ yr$^{-1}$, from lightest to darkest, in steps of $1/3$ of a dex; the actual mass outflow rate, considering our adopted geometry, is a factor of $0.22$ smaller. The red dashed line, which is the same in each panel, is the line profile for the fiducial case of $\dot{M} = 100$ $M_\odot$ yr$^{-2}$, isothermal potential, constant solid angle expansion. The black vertical dotted lines show velocities of $\pm v_0$.
}
\end{figure}

\begin{figure}
\includegraphics[width=\columnwidth]{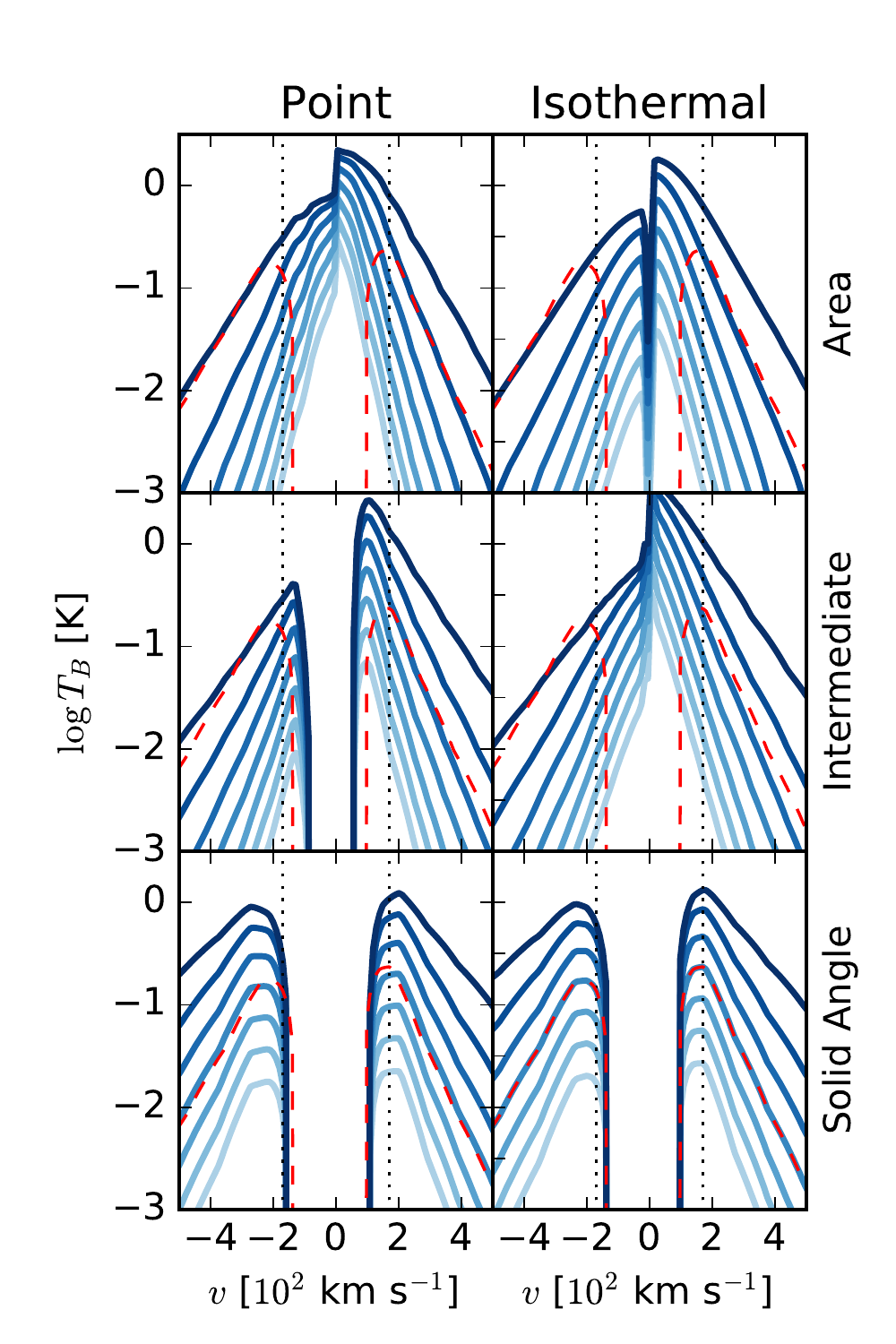}
\caption{
\label{fig:m82_CO_var}
Same as \autoref{fig:m82_Ha_var}, but showing CO rather than H$\alpha$ emission spectra.
}
\end{figure}

\begin{figure}
\includegraphics[width=\columnwidth]{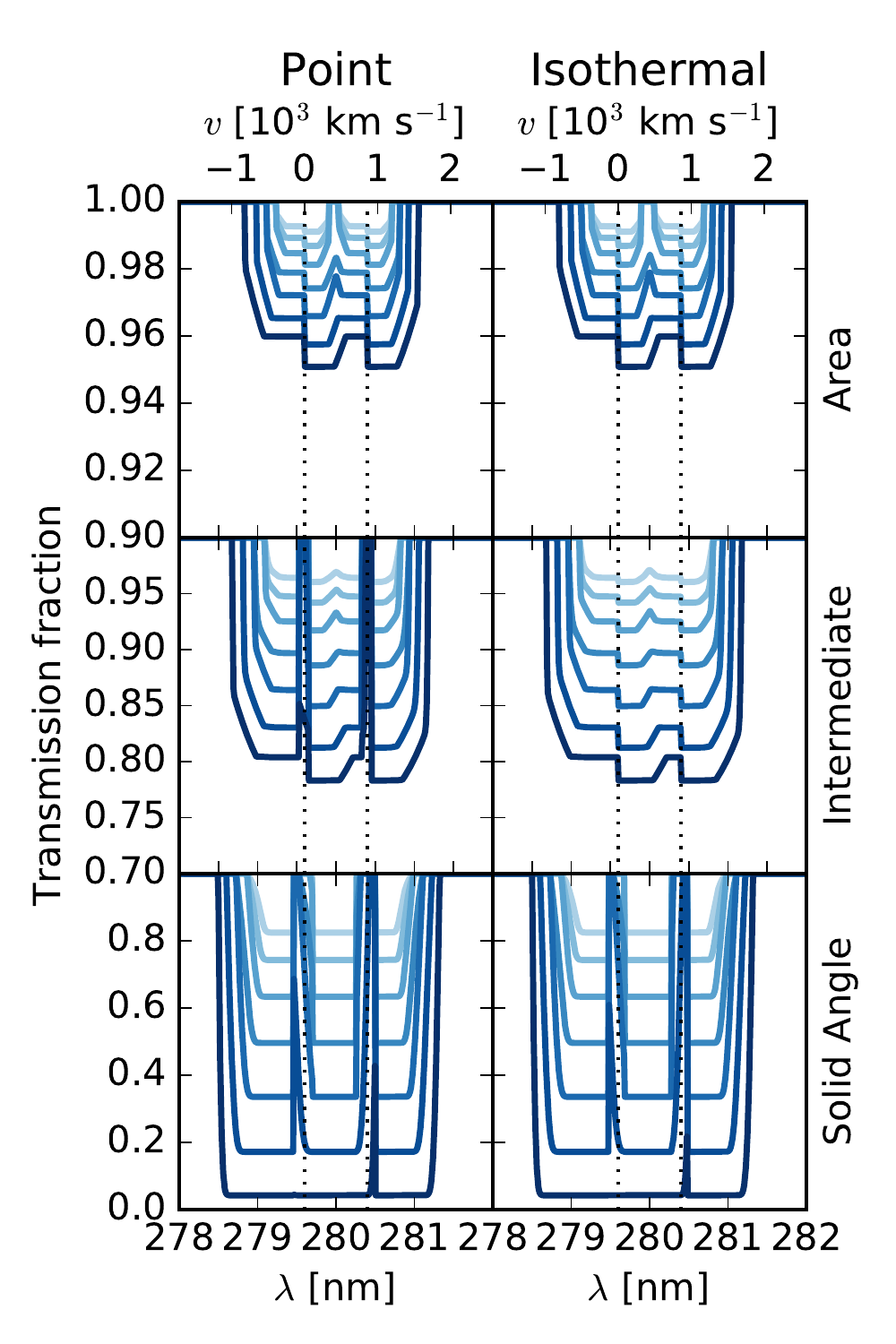}
\caption{
\label{fig:m82_MgII_var}
Same as \autoref{fig:m82_Ha_var}, but showing Mg~\textsc{ii} absorption rather than H$\alpha$ emission spectra. The velocities given on the top horizontal axis are relative to the shorter wavelength transition of the doublet, and the two vertical black dotted lines show the central wavelengths of the two double transitions. Note that the vertical axis range is different for each row.
}
\end{figure}

\begin{figure}
\includegraphics[width=\columnwidth]{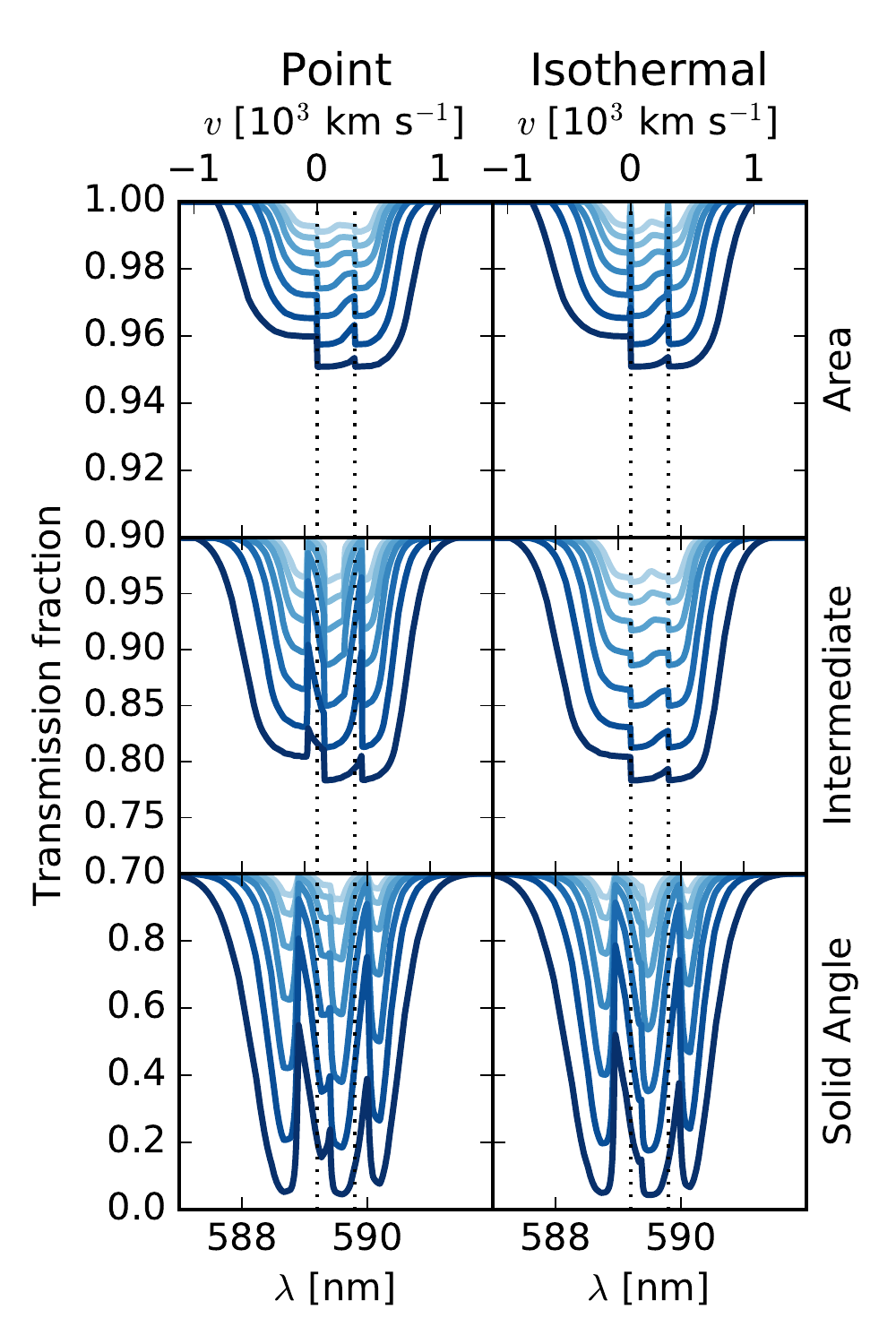}
\caption{
\label{fig:m82_NaI_var}
Same as \autoref{fig:m82_MgII_var}, but for Na~\textsc{i} instead of Mg~\textsc{ii}.
}
\end{figure}

\autoref{fig:m82_Ha_var}, \autoref{fig:m82_CO_var}, \autoref{fig:m82_MgII_var}, and \autoref{fig:m82_NaI_var} show the H$\alpha$ emission, CO emission, Mg~\textsc{ii} absorption, and Na~\textsc{i} absorption spectra, respectively, that we obtain. Clearly the expansion law can have as much or more impact on the observed spectrum than the outflow rate.  We can make a few remarks as to why. For H$\alpha$, we see line splitting only in the case of constant solid angle clouds or intermediate expansion in a point potential. For CO, the line splitting is much stronger for constant solid angle clouds or for intermediate expansion in a point potential than in the other cases. The reason is simple. Due to our adopted geometry there is no material whose direction of motion is purely in the plane of the sky, so to have emission near zero velocity there must be material in the wind with small radial velocity. Material at small radial velocity exists at large radii only if the potential $m$ grows as fast or faster than the cloud area $y$; in cases where this is not the case, there is a dip in emission near $v=0$. Absent this condition being met, emission continues all the way to $v=0$. (However, this is only true of material with density near $x_{\rm crit}$ actually enters the wind -- see \autoref{app:fcrit}.)

For the case of CO emission, we see a related phenomenon: the positive-negative velocity asymmetry is much larger for smaller expansion rates than for larger ones. The reason is the asymmetry pointed out in \autoref{ssec:LTE_lineprof}: at positive velocity, regardless of whether we are observing the portion of the outflow cone that is tipped slightly toward or slightly away from us, the light we see was emitted by that part of the outflow cone that lies on the far side of the plane of the sky. As this light travels to us, it can be absorbed only by gas that is at radii smaller than itself. At negative velocity, the converse is true: regardless of where on the plane of the sky we observe, the light at positive velocities must been emitted by gas that lies closer to us than the plane of the sky. As it travels to us it is therefore going from smaller to larger radii. These two situations are asymmetric: if clouds expand as they move out, then absorption by material at smaller radius blocks much less light than absorption by material at larger radius, because the small-radius absorbing clouds cover less area than the large-radius ones. Thus the positive velocity, far-side light is less absorbed than the negative velocity, near-side light, producing an asymmetry whereby the red side of the line is brighter than the blue one. It is worth noting that, while this is a phenomenon similar to p Cygni line profiles, the physics is somewhat different than in the classical p Cygni profile in that what drives the asymmetry is not density or temperature variation with radius. Instead, it is variations in covering factor with radius.

For H$\alpha$ emission and for both absorption measurements, we see that the effect of outflow rate on the emission profile is generally much smaller than the effects of expansion. For emission, the effect is not even in the same direction at all velocities. An increase in the outflow rate increases the value of the Eddington ratio $\Gamma$, which in turn produces less emission at small velocity and more at large velocity. For absorption, increasing the outflow rate decreases the transmission fraction at all wavelengths, but does so in a complex way that combines two different effects. As $\Gamma$ increases, the covering fraction of the outflow material goes up, and this manifests in the absorption spectra as an increase in the width and depth of the flat bottoms of the absorption features. However, the increasing mass outflow rate also increases the amount of material. This has no effect for velocities already covered by optically thick material, as is the case for most of the Mg~\textsc{ii} spectra, but it does increase the absorption feature depth for more optically thin velocities and tracers, for example across the Na~\textsc{i} spectra in the constant solid angle case. In any event, the takeaway conclusion from this complexity is that there is no simple way to deduce an outflow rate from an observation of subcritical emission or absorption without fitting the full spectrum. In contrast, for LTE emission, for example from CO, the strength of the wind features in the spectrum do correlate strongly with the outflow rate, particularly for positive velocities (i.e., emission from the far side) some distance from line centre, where geometric opacity effects are less important.

A third interesting point is that the potential matters significantly less than the expansion rate. Expansion rates always affect the result, but the potential matters only near the transition between $y > m$ and $y < m$., i.e., between a fountain and a wind. Because of the importance of the expansion rate, the often-used strategy of attempting to differentiate between wind and fountain material by dividing between emission at velocities above and below $v_0$ (the escape speed) clearly does not work. Even in a wind, some material may be at small velocity simply because it still accelerating and has not yet reached speeds above $v_0$, while even in a fountain some material may be above velocity $v_0$ at small radii, but be destined to decelerate and fall once it climbs to larger distances. All of the cases shown in \autoref{fig:m82_Ha_var} - \autoref{fig:m82_NaI_var} except for area expansion with an isothermal potential are winds with precisely the same total mass fluxes reaching infinity, while the isothermal-area case is a fountain where no mass flux reaches infinity. However, in terms of the observable emission and absorption, the point-area and isothermal-area cases (the former a wind, the latter a fountain) are clearly far more similar than the point-area and point-solid angle cases (both winds).

\section{Discussion}
\label{sec:discussion}

While we defer to paper II a full discussion of how to make use of the analytic model we present here for the purpose of analysing observations of winds, we can draw a few general conclusions based on the examples presented so far.

\subsection{Optical Depths of Winds}

One perhaps unexpected result of our analysis is that the optical depth of a particular wind tracer, either in emission or absorption, is controlled primarily by a single dimensionless ratio $t_X/t_w$, where $t_X$ is an intrinsic property of the transition that depends only on the abundance of the species in question and its wavelength and oscillator strength, and $t_w$ is the timescale over which the wind (if it were isotropic) would evacuate the wind launching region. This means that optical depth can, by itself, be used to obtain an upper or lower limit on a wind's mass flux.

Conversely, we can classify whether a particular transition is likely to be optically thin or thick for winds with astrophysically-interesting mass fluxes based solely on abundances and quantum mechanical transition properties. Lines such as Mg~\textsc{i} $\lambda 2853$ and Mg~\textsc{ii} $\lambda\lambda 2796, 2804$ have timescales of hundreds of Gyr, and thus are likely to be opaque for any galactic wind of interest. Weaker lines such as Na~\textsc{i} $\lambda\lambda 5892,5898$ and Fe~\textsc{ii} $\lambda 2383$ have timescales of $\lesssim 1$ Gyr, and thus may go between opaque and transparent depending on the outflow rate. In the radio, low $J$ lines of CO have timescales of $\sim 100$ Myr, and thus are likely to be opaque for strong starbursts or AGNs that are capable of evacuating their gas on timescales of $\lesssim 100$ Myr. They will be transparent in more gently mass-losing systems, Less abundant species such as HCN or CS have timescales of a few Myr, and thus will almost always be transparent for galactic winds, though they might become opaque for the winds driven by individual star clusters.

\subsection{Cloud Expansion as a Dominant Uncertainty in Mass Flux Measurements}

A second general conclusion we can draw is that, particularly for absorption and subcritical emission measurements, the observational signatures are at least as sensitive to the way that clouds of cool gas behave geometrically as they are carried out in a wind (as parameterised by their expansion laws) as they are to the overall rate of mass loss in the wind. For absorption measurements in strong lines, which include many of the commonly-used low ionisation transitions such as Mg~\textsc{ii} $\lambda\lambda 2796, 2804$, the optical depth along individual lines of sight is likely to be very high at any mass loss rate large enough to be of astrophysical interest. As a result, the amount of absorption is more closely related to the covering factor of the wind material than to its total amount, and this in turn is highly sensitive to how clouds of cool gas expand or do not expand as they propagate outward in the wind.

For subcritical emission tracers, the differences in expansion factor affect emission by changing the wind density and its scaling with velocity. If clouds do not expand substantially as they propagate outward, $y=1$, then most material accelerates slowly, and maintains high density. High velocities are only reached by rare gas that began with a column density much lower than the minimum required for entrainment and ejection. Consequently, emission is dominated by low velocities. On the other hand, if clouds expand rapidly and maintain constant solid angles as they move outward, $y=a^2$, then even material that is only marginally ejected at small radii will continue to be accelerated as it moves to large radii, shifting the emission to higher velocities.

Regardless of the ultimate source of the dependence on the cloud geometry, the dependence of the emission or absorption signature on the poorly known cloud expansion behaviour represents a dominant uncertainty in any attempt to infer mass loss rates from these techniques. A quick glance at \autoref{fig:m82_MgII_var} shows that the differences in the absorption spectra produced by a wind carrying a constant mass flux but with geometries of $y=1$ or $y=a^2$ are at least as large as the differences produced if we fix the geometry but vary the mass flux by a factor of 10 or more. The same is true for subcritical emission. In most phenomenological models used to compute wind mass fluxes, the flux is taken to be directly proportional to the absorption depth or emission intensity, so this translates directly into a factor of 10 variation in the mass flux that one would estimate.

In contrast, emission from tracers where we are confident that the gas is in local thermodynamic equilibrium are much less sensitive to this uncertainty, except at velocities near zero where the geometry of the absorbing material has large effects. In practice, however, this is unlikely to matter, since the signal near zero velocity will be dominated by the wind source rather than the wind itself. Converting an observed line profile in a line such as CO into a mass flux is by no means trivial, a point to be discussed momentarily, but at least the profile depends as or more strongly on the mass flux than on the poorly constrained cloud geometry.

\subsection{Winds versus Fountains}

A third important conclusion of our modelling is that the commonly-adopted method of estimating what gas will escape a galaxy versus what gas will fall back in a fountain based on a division at the escape velocity is potentially highly misleading. This picture is, at least implicitly, based on the assumption that the wind is launched ballistically, so that gas maintains its speed, with no additional forces applied at large distance from the source. While some material in winds is likely accelerated rapidly \citep[e.g., in superbubbles -- ][]{roy13a}, it is far from certain that this is the predominant source of cool gas in winds. Instead, entrainment by hot gas or pushing by radiation likely contribute, and both require some time and distance to accelerate material, with the bulk of the material ejected probably lying close to the low end of the range of possible velocities. Clouds that have a velocity below the escape speed at small radii may nonetheless be accelerated to the escape speed at larger radii.

This problem is compounded by biases in the mapping between mass and light. For absorption measurements, unless clouds maintain constant solid angle, measurements are most sensitive to the lowest velocity material. This is because clouds have the highest covering fractions, and thus absorb the most light, when they are close to the launch point and thus at low velocity. As they gain speed their covering fraction diminishes, and they are underrepresented in an effectively area-weighted absorption spectrum. For optically thin subcritical emission, such as that produced by H$\alpha$ or C~\textsc{ii}, there is another bias. As material moves outward and accelerates, its density drops, and the luminosity per unit mass drops proportionately. Again, the result is a strong bias toward low velocity emission, with high velocity material being underrepresented in the spectrum. Both biases are compounded by their dependence on the poorly-known wind geometry.

In contrast, for emission by gas in LTE the bias is substantially weaker because the emission scales linearly rather than quadratically with density. What bias there is tends to be in the opposite direction, in that gas at high velocities tends to have higher $X$ / $\alpha$ factors due to its lower optical depth. The lack of bias makes molecular lines a promising candidate for determining outflow rates with less bias, a topic to which we shall return in paper II. That said, even for LTE emission one still cannot neatly divide gas between fountain and escape based on its velocity. The low velocity material seen in molecular emission may be gas that has not yet accelerated up to escape speed but will do so as it continues to flow outward, and not counting it as part of the outflow can lead to a serious error. Only if one assumes that gas acceleration is purely impulsive does a division based on velocity work.

\subsection{Caveats and Limitations}

We conclude this section by discussing some of the limitations of our model. One prominent one that has already been mentioned several times is that we have presented a detailed model for the physical and kinematic properties of a wind, but not for its thermal or chemical properties. Thus we are forced to rely on simple assumptions about the abundances and chemical state of the gas, and, in cases where the wind temperature is important, to assume that the entire cool phase that we are modelling is characterised by a single temperature. In practice, this means that any inferences based on our models should be taken to apply to only on particular chemical and thermal phase of an outflow. It also means that our $t_X$ values that characterise transition strength are significantly uncertain. A related limitation is that we have only treated emission in the limiting cases of gas so dense that the emitting species can be assumed to be in LTE everywhere there is significant emission, or so diffuse that we can treat the entire population is being in the ground state. We have not attempted to treat the intermediate case of partial thermalisation, although in principle one could extend our model to include this effect, at least numerically.

A second significant limitation is that we have assumed that cool material is present at all radii in an outflow, and thus forms a continuous medium. A contrasting possibility is that originally-cool ISM gas is shock-heated to much higher temperature at small radii, e.g., as outer shells and embedded clouds in superbubbles \citep{kim17a}. While the denser material would cool rapidly and experience limited acceleration, other material may become part of a hot, diffuse flow that produces negligible absorption or emission in optical, infrared, or radio lines. The cool gas we observe appears only at large radii, where it condenses as the hot medium undergoes adiabatic cooling \citep{martin15a, thompson16b, roy16a}. Were this the primary source of cool wind gas, our model would not be applicable. While this is certainly possible, it remains an open question whether this mechanism is the primary source of cool gas, particularly molecular gas that requires very dust column densities to resist photodissociation. We intend to explore the observational signatures of such a re-condensation scenario in future work.

A final caveat to our work is that, for the absorption case, we have not included the effects of multiple scattering and fluorescent re-emission of photons, either those produced in the wind or from an AGN. There is some observational evidence for this effect operating on very strong lines, particularly the Mg~\textsc{ii} doublet \citep{weiner09a, erb12a, martin12a}. This means that, for sources with significant re-emission, we will overestimate the depths of the absorption features. All models of this effect published to date have treated the wind as a fully filled spherical shell \citep{prochaska11a, scarlata15a}, which in the context of our model is not realistic: at Mach numbers of $30-100$, we predict that the covering factor of a wind is always below 50\% (and even smaller if the areal PDF is not purely lognormal,  as discussed in \autoref{ssec:correlated}) unless the Eddington ratio $\Gamma > 1$, a condition which, if met, almost certainly leads to ejection of all the gas on a timescale comparable to the dynamical time. For more plausible wind mass fluxes, the covering fraction of the wind is typically tens of percent. In such a geometry, multiple scattering seems likely to be much less important than in the closed geometries that have been examined thus far. Clearly there is a need to revisit the question of multiple scattering in the context of the much more realistic geometry and density structure that we introduce here.

\section{Conclusion and Future Prospects}
\label{sec:conclusion}

In this paper we extend the simple but physically-consistent model of wind launching introduced by \citet{thompson16a} to consider the observable properties of the cool components of the flow. Because the model considers a distribution of initial properties for the clouds that are launching into a wind, it naturally gives rise to a spread of velocities and densities in the resulting wind. This produces line profiles that, when combined with simple but realistic geometries, produce relatively realistic line profiles for wind emission in both sub- and supercritical lines, and for absorption. We show that the strength of the emission and absorption features, and in particular whether a given line will be optically thin or thick, is well-characterised by the ratio between the timescale for the wind to evacuate the object from which it is launched, and a natural timescale that depends only on the abundance and quantum mechanical properties of the species in question.

Our model can accommodate both idealised winds and more realistic ones driven by either radiation pressure or the pressure of a volume-filling hot gas, It can be fully described in terms of a few free parameters, and is simple enough that we can use it to compute line profiles in fractions of a second even on a single processor. We provide software to carry out these computations as part of the \textsc{despotic} software suite \citep{krumholz14b}. This simplicity and speed makes our model ideal for the task of fitting observations and thereby deducing the physical properties of winds, particularly mass fluxes. We turn to the task of this fitting, and determining to what extend robust conclusions may be extracted from observations of various sorts, in Paper II.

\section*{Acknowledgements} 

MRK thanks J.~Bland-Hawthorn and S.~Veuilleux for helpful discussions, which took place during a visit to the University of Sydney supported by the Dick Hunstead Fund for Astrophysics. MRK's work on this project was supported under the Australian Research Council's \textit{Discovery Projects} funding scheme (project DP160100695). MRK and CLM thank the Kavli Institute for Theoretical Physics, where part of this work as performed, which is supported by the National Science Foundation under Grant No.~NSF PHY-1125915. TAT thanks E.~Quataert, N.~Murray, and D.~Zhang for discussions and collaboration. TAT is supported in part by NSF Grant AST-1516967. The work of ECO was supported by the NSF under grant AST-1312006. All authors thank the Simons Foundation for supporting the symposium ``Galactic Superwinds: Beyond Phenomenology", where this work was begun.

\bibliographystyle{mn2e}
\bibliography{refs}

\begin{appendix}

\section{Effects of Non-Escape of Material Near the Critical Surface Density}
\label{app:fcrit}

In the main text we assume that all material that has an outward acceleration at the wind launching radius $a=1$, i.e., all gas with $x<x_{\rm crit}$, enters the wind. However, for wind acceleration laws where the wind expansion factor $y$ grows with radius more slowly than the potential $m$, this results in a component of the wind with arbitrarily small velocities. It therefore might be reasonable to hypothesise that material only enters the wind if its surface density is small enough to produce an outward acceleration above some minimum value, so that the wind does not include material moving at arbitrarily small velocities.

\begin{figure}
\includegraphics[width=\columnwidth]{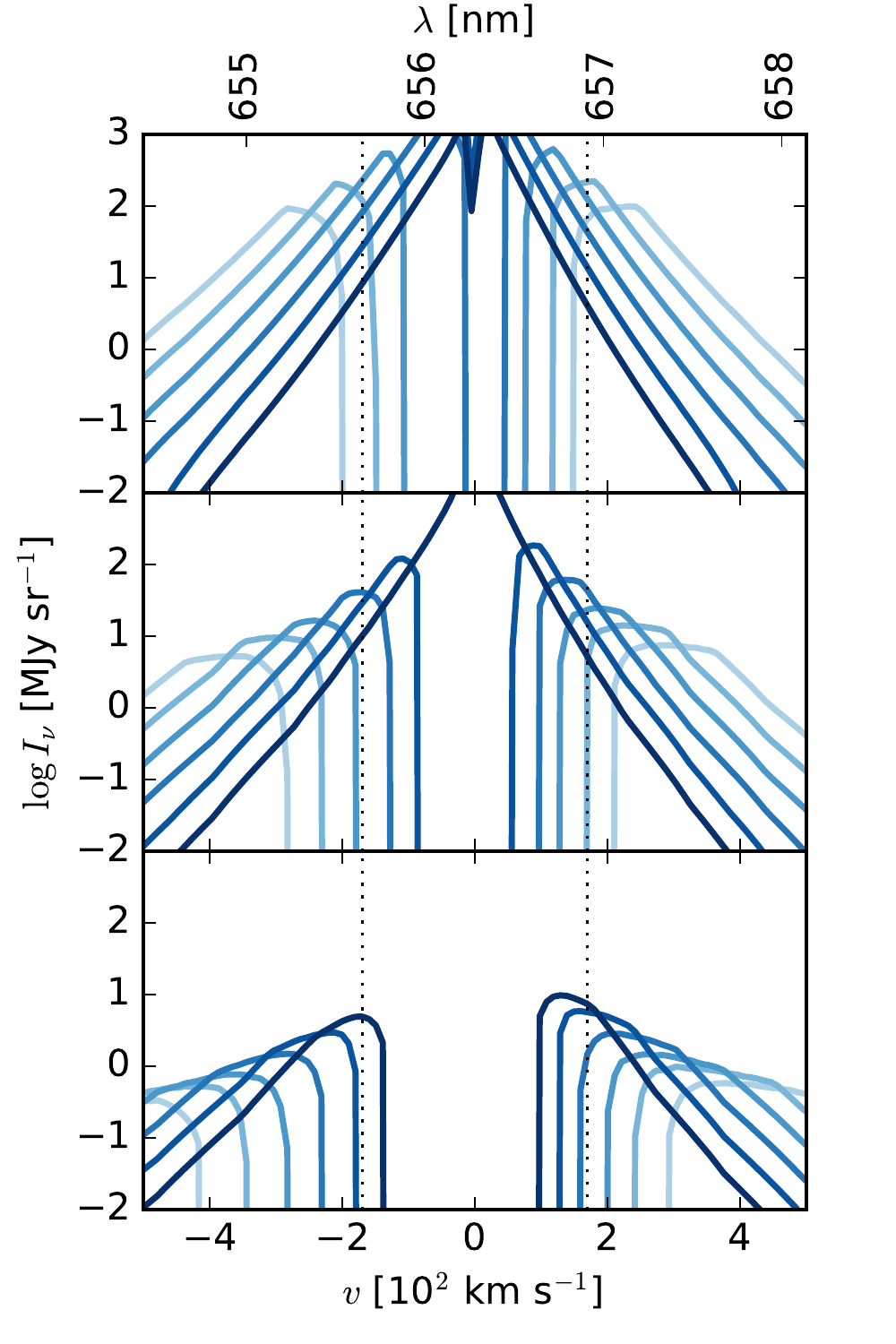}
\caption{
\label{fig:m82_Ha_fcrit}
Same as \autoref{fig:m82_Ha_var}, except that we only show the case for an isothermal potential, and all runs use the fiducial isotropic mass loss rate $100$ $M_\odot$ yr$^{-2}$. Curves show the results for varying $f_{\rm crit}$, from $f_{\rm crit} = 1$ (darkest line; fiducial case) to $f_{\rm crit} = 0.1$ (lightest line; only material with surface densities $10\times$ smaller than fiducial is ejected), in steps of 0.25 dex.
}
\end{figure}

\begin{figure}
\includegraphics[width=\columnwidth]{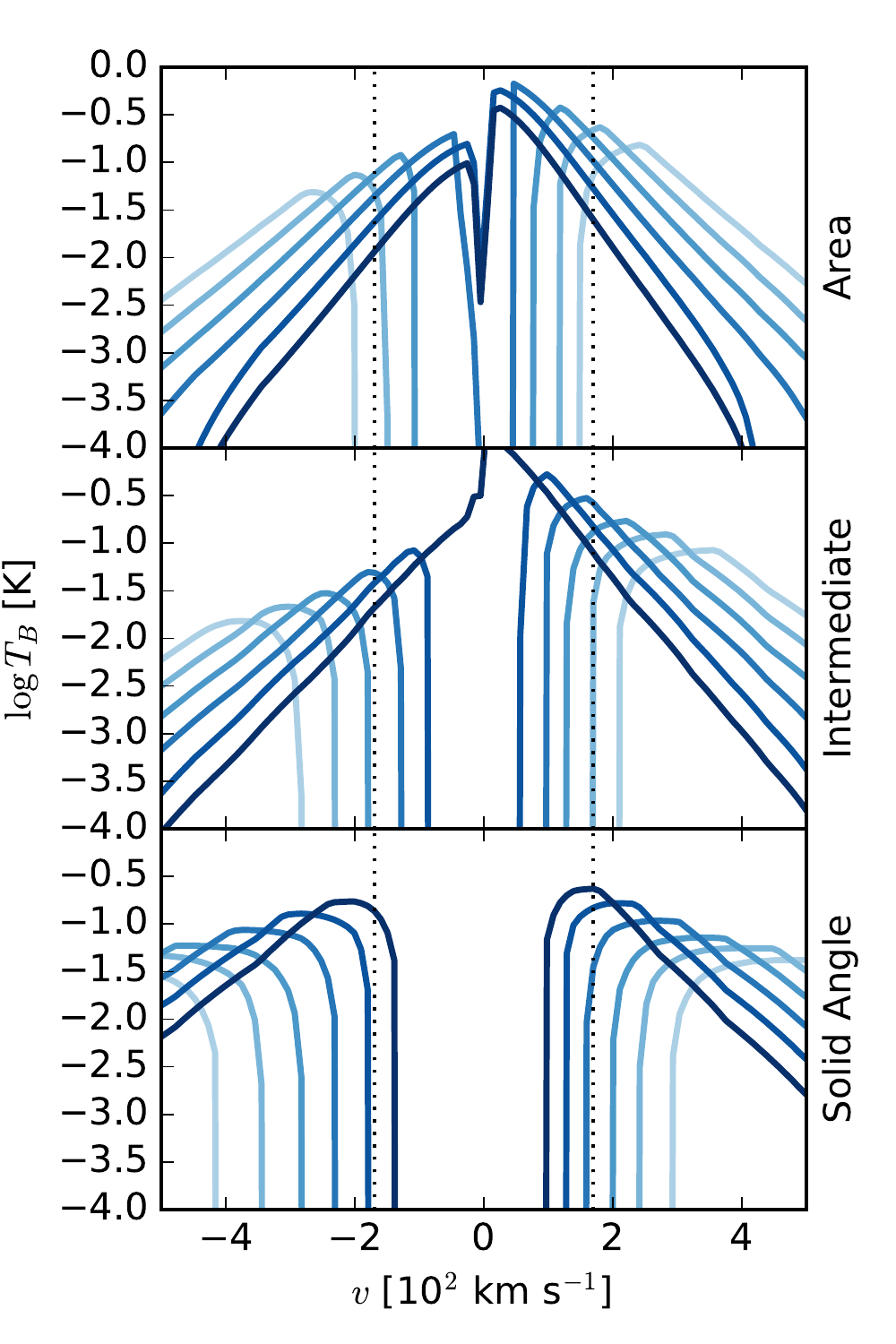}
\caption{
\label{fig:m82_CO_fcrit}
Same as \autoref{fig:m82_Ha_fcrit}, but showing the CO brightness temperature instead of the H$\alpha$ intensity.
}
\end{figure}

Specifically, suppose that we consider a wind consisting only of material with starting surface density $x < x_{\rm crit} + \log f_{\rm crit}$, with $f_{\rm crit} < 1$. The computational machinery described in the main text is unchanged, except that we set the contribution to emission or absorption to zero for any value of $x > x_{\rm crit} + \log f_{\rm crit}$. We use this machinery to compute the same emission and absorption profiles discussed in \autoref{ssec:variations} for cases with a range of values of $f_{\rm crit}$. To ensure that we are comparing like with like, when using $f_{\rm crit} < 1$, we change the value of $\Gamma$ in order to keep $\zeta_M$ and thus the overall mass flux fixed. Specifically, we compute $\Gamma$ by solving a modified version of \autoref{eq:eta},
\begin{equation}
\eta = \frac{f_A \dot{M}}{\dot{M}_*} = \frac{\zeta_M(f_{\rm crit})}{\epsilon_{\rm ff}},
\end{equation}
where $\zeta_M(f_{\rm crit})$ means $\zeta_M$ evaluated using $x = \ln f_{\rm crit} \Gamma$.

We show the results for H$\alpha$ and CO emission in \autoref{fig:m82_Ha_fcrit} and \autoref{fig:m82_CO_fcrit}, respectively; we focus on these rather than absorption because the results are easier to interpret. It is clear that removal of gas with $x$ close to $x_{\rm crit}$ from the flow has an effect that is qualitatively similar to increasing the expansion factor of the flow: it shifts emission to higher velocities, and causes line splitting to appear. This is not surprising: removing gas near $x_{\rm crit}$ from the flow increases the velocity of that gas that does remain, and, to keep the overall mass flow rate fixed, requires that $\Gamma$ increase as well, further raising the mass flux. Both effects put more mass at high velocity, exactly as happens for larger cloud expansion functions $y$.

\end{appendix}

\end{document}